\documentclass[preprint,showpacs,preprintnumbers,amsmath,amssymb,aps,floats,amsfonts,superscriptaddress,aps,prd,nofootinbib,11pt]{revtex4-2}

\selectfont

\usepackage[utf8]{inputenc}
\usepackage[]{graphicx}
\usepackage{hyperref}
\usepackage{url}
\usepackage{color}
\usepackage[usenames,dvipsnames,svgnames,table]{xcolor}
\usepackage{multirow}
\usepackage{ulem}
\usepackage[scr=boondoxo,scrscaled=1.05]{mathalfa}%script fonts
\usepackage{tensor}
\usepackage{booktabs}
\usepackage{array}
\usepackage{grffile}
\usepackage{amsmath}
\usepackage{amsfonts}
\usepackage{calc}
\usepackage{mathrsfs}
\usepackage{amssymb}
\usepackage{array}
\usepackage{tensor}
\usepackage{upgreek}

\newcommand{\npDel}{\triangle}

\newcommand{\beq}{\begin{equation}}
\newcommand{\eeq}{\end{equation}}

\newcommand{\nn}{\nonumber}

\newcommand{\mbar}{\overline{m}}

\newcommand{\dbar}{\overline{\delta}}

\newcommand{\cD}{\mathcal{D}}
\newcommand{\cL}{\mathcal{L}}
\newcommand{\cLh}{\hat{\cL}}

\newcommand{\etal}{{\it et al.}}

\newcommand{\cU}{\mathcal{U}}
\newcommand{\cV}{\mathcal{V}}
\newcommand{\cW}{\mathcal{W}}

\newcommand{\cH}{\mathcal{H}}

\newcommand{\rhoc}{\overline{\rho}}
\newcommand{\lambdah}{\hat{\lambda}}
\newcommand{\Lambdah}{\hat{\Lambda}}
\newcommand{\chiim}{\chi_\ell^{\text{im}}}
\newcommand{\chire}{\chi_\ell^{\text{re}}}
\newcommand{\Sim}{S_\ell^{\text{im}}}
\newcommand{\Sre}{S_\ell^{\text{re}}}
\newcommand{\calP}{\mathcal{P}}
\newcommand{\calS}{\mathcal{S}}
\newcommand{\hirg}{h_{\text{IRG}}}
\newcommand{\horg}{h_{\text{ORG}}}
\newcommand{\hirgup}{h^{\text{IRG}}}
\newcommand{\horgup}{h^{\text{ORG}}}
\newcommand{\barh}{\bar{h}}
\newcommand{\lmin}{\ell_{\text{min}}}
\newcommand{\hcomp}{h^{\text{comp.}}}

\newcommand{\hb}[1]{\bar{h}^{(#1)}}

\newcommand{\hrm}{\mathrm{h}} % ^{(lm)}}
\newcommand{\hrmlm}{\hrm^{\ell m}}

\newcommand{\LNW}{\hat{\mathcal{L}}_r}
\begin{document}

\title{Metric perturbations of Kerr spacetime in Lorenz gauge: \\ Circular equatorial orbits}

\author{Sam R.~Dolan}
\affiliation{
Consortium for Fundamental Physics,
School of Mathematics and Statistics,
University of Sheffield,
Hicks Building,
Hounsfield Road,
Sheffield S3 7RH, 
United Kingdom.}

\author{Leanne Durkan}
\affiliation{
Center for Gravitational Physics, University of Texas at Austin, Austin, TX 78712, USA. 
}
\affiliation{
School of Mathematics and Statistics,
University College Dublin,
Belfield,
Dublin 4,
Ireland.
}

\author{Chris Kavanagh}
\affiliation{
Max Planck Institute for Gravitational Physics (Albert Einstein Institute),
Am M\"uhlenberg 1,
Potsdam 14476,
Germany.}

\author{Barry Wardell}
\affiliation{
School of Mathematics and Statistics,
University College Dublin,
Belfield,
Dublin 4,
Ireland.
}

\begin{abstract}
We construct the metric perturbation in Lorenz gauge for a compact body on a circular equatorial orbit of a rotating black hole (Kerr) spacetime, using a newly-developed method of separation of variables. The metric perturbation is formed from a linear sum of differential operators acting on Teukolsky mode functions, and certain auxiliary scalars, which are solutions to {\it ordinary} differential equations in the frequency domain. For radiative modes, the solution is uniquely determined by the $s=\pm2$ Weyl scalars, the $s=0$ trace, and $s=0,1$ gauge scalars whose amplitudes are determined by imposing continuity conditions on the metric perturbation at the orbital radius. The static (zero-frequency) part of the metric perturbation, which is handled separately, also includes mass and angular momentum completion pieces. The metric perturbation is validated against the independent results of a 2+1D time domain code, and we demonstrate agreement at the expected level in all components, and the absence of gauge discontinuities. In principle, the new method can be used to determine the Lorenz-gauge metric perturbation at a sufficiently high precision to enable accurate second-order self-force calculations on Kerr spacetime in future. We conclude with a discussion of extensions of the method to eccentric and non-equatorial orbits.
\end{abstract}

\maketitle
\newpage

\tableofcontents

\section{Introduction\label{sec:introduction}}

In 1916, Schwarzschild showed that the vacuum field equations of Einstein's general relativity admit a special solution: a spacetime geometry with a trapped central region from which nothing, not even light, can escape. Over the course of the 20th century, it became clear that black holes -- objects with this key property -- are more than just theoretical curiosities. Almost every massive galaxy hosts a supermassive black hole ($M \sim 10^6$--$10^{10} M_{\odot}$) at its core, including the Milky Way \cite{EventHorizonTelescope:2022wkp}, and supermassive black holes are thought to play a key role in galaxy formation and evolution \cite{Cattaneo:2009ub}. A galaxy such as ours is likely to be replete with $10^7$--$10^9$ stellar-mass black holes \cite{OGLE:2022gdj, Lam:2022vuq}. Every quiescent black hole is essentially characterised by just two parameters, mass and angular momentum. Its spacetime geometry is described by Kerr's 1963 solution of the vacuum Einstein field equations \cite{Kerr:1963ud}. In the words of Chandrasekhar \cite{Chandrasekhar2013truth}, ``the most shattering experience has been the realisation that [Kerr's solution] provides the absolutely exact representation of untold numbers of massive black holes that populate the universe."

Exact solutions to the Einstein field equations are notoriously difficult to find in the absence of symmetries, and consequently several approaches have been developed to model the dynamics of black holes, including Numerical Relativity, post-Newtonian/post-Minkowskian expansions, and perturbation theory. 
The topic of \emph{black hole perturbation theory} is certainly one of deep and enduring interest. In 1957, Regge and Wheeler made a pioneering study of the odd-parity gravitational perturbations of Schwarzschild spacetime, and Zerilli later analysed the even-parity perturbations \cite{Zerilli:1970se,Zerilli:1971wd}, showing isospectrality. For perturbations of a charged (Reissner-Nordstrom) black hole, Moncrief showed that the coupled electromagnetic and gravitational field equations can be reduced to certain master equations which are separable \cite{Moncrief-1974a,Moncrief-1974b,Moncrief-1975}. In 1972, Teukolsky made a breakthrough in the analysis of the perturbations of a rotating (Kerr) black hole \cite{Teukolsky:1972my,Teukolsky:1973ha,Press:1973zz,Teukolsky:1974yv}, by showing that two Weyl scalars (that is, two scalars obtained by projecting the perturbed Weyl tensor onto the principal null tetrad) satisfy decoupled equations which are separable in the frequency domain. In fact, Teukolsky's master equations encapsulate perturbations of all spins on Kerr spacetime: the massless scalar ($s=0$) and spinor ($s=1/2$) fields, the electromagnetic field ($s=1$) and the linearised gravitational field ($s=2$).

The Einstein field equations are a \emph{nonlinear} system of ten coupled partial differential equations, for the ten components of the metric tensor that describes the spacetime geometry. Black hole perturbation theory generates a hierarchical system of \emph{linear} equations by expanding the metric tensor in a small parameter, $\epsilon$, as
\begin{equation}
  g^{\rm exact}_{\mu\nu} = g_{\mu\nu} + \epsilon h^{(1)}_{\mu\nu} + \epsilon^2 h^{(2)}_{\mu\nu} + O(\epsilon^3).\label{g expansion}
\end{equation}
The zeroth-order metric $g_{\mu\nu}$ is chosen to be an exact solution of Einstein's equations --- typically a black hole solution such as the Kerr metric. The metric perturbations (MPs) $h^{(1)}_{\mu\nu}$, $h^{(2)}_{\mu\nu}$, \ldots, 
all satisfy {\it linear} systems of partial differential equations that take the form
\begin{equation}
\Box \barh^{(i)}_{\mu \nu} + 2 \tensor{R}{^\alpha _\mu ^\beta _\nu} \barh^{(i)}_{\alpha \beta} + g_{\mu \nu} \nabla_\sigma Z_{(i)}^\sigma - 
2 \nabla_{(\mu} Z^{(i)}_{\nu)} = S^{(i)}_{\mu\nu}[h^{(i-1)}, \cdots, h^{(1)}, T_{\mu\nu}].  \label{eq:LEE}
\end{equation}
with $Z_\mu^{(i)} \equiv \nabla^\nu \barh^{(i)}_{\mu \nu}$, where $\barh^{(i)}_{\mu \nu} \equiv h^{(i)}_{\mu \nu} - \tfrac{1}{2} g_{\mu \nu} h^{(i)}$ is the trace-reversed metric perturbation. Here the covariant derivative $\nabla_\mu$, the d'Alembertian $\Box \equiv \nabla^\mu \nabla_\mu$ and the Riemann tensor are defined on the background geometry $g_{\mu \nu}$. The linearised Einstein equations (\ref{eq:LEE}) have two notable features: the differential operator on the left hand side is the same for all orders; and the source on the right hand side depends on the the stress-energy tensor $T_{\mu\nu}$ and on all lower-order metric perturbations, $h^{(i-1)}, \cdots, h^{(1)}$.

The diffeomorphism invariance of the Einstein field equations (i.e.~coordinate freedom) translates into \emph{gauge freedom} in the perturbation equations (\ref{eq:LEE}). As in electromagnetism, one may select a convenient gauge to suit the calculation at hand. \emph{Lorenz gauge} (also known as De Donder gauge, or harmonic gauge) is defined by $Z_\mu^{(i)} = 0$. In Lorenz gauge, Eq.~(\ref{eq:LEE}) reduces to a wave equation in manifestly hyperbolic form, which is a desirable feature for many applications. The question of how to find a metric perturbation in Lorenz gauge on Kerr spacetime in a way that builds on the approach of Teukolsky was recently addressed in Ref.~\cite{Dolan:2021ijg}; here we extend and build on that work to develop a fully-fledged calculational scheme. 

Although simpler than the fully nonlinear Einstein equations, the linearised equations (\ref{eq:LEE}) for the metric perturbation remain challenging to solve. There is no known method to decouple the ten equations for the ten components of the metric perturbation and, more significantly, in Kerr spacetime there has not been known a separation-of-variables ansatz that reduces the partial differential equations to a decoupled set of ordinary differential equations.
Despite being extremely useful, Teukolsky's method yields only two (of five) Weyl scalars, which on their own are insufficient for calculation of the full metric perturbation. Calculations that require access to the full metric perturbation have become more commonplace in recent years, with two of the most important applications being the calculation of Gravitational Self-Force to model Extreme Mass-Ratio Inspirals \cite{Poisson:2011nh,Wardell:2015kea,Barack:2018yvs,Pound:2021qin} --- where the self-force is expressed in terms of first derivatives of the metric perturbation --- and in non-linear perturbation theory where, as mentioned above, the source terms for higher-order perturbations necessarily depend on lower-order metric perturbations. The latter has played an important role in recent years, with non-linear perturbation theory being used to produce post-adiabatic waveforms for binary black hole inspirals \cite{Wardell:2021fyy}, to study the quasinormal mode ringdown following black hole mergers \cite{Ioka:2007ak,Ripley:2020xby,Sberna:2021eui,Mitman:2022qdl,Lagos:2022otp,Cheung:2022rbm}, in proofs of nonlinear stability of the Kerr solution \cite{Dafermos:2021cbw,Klainerman:2021qzy}, and in searches for signatures of turbulent dynamics \cite{Yang:2014tla,Green:2019nam,Green:2022htq}. In this paper our focus is on the former --- solving the gravitational self-force equations in Kerr spacetime --- but we anticipate that the approach we have developed will provide a useful tool for the latter too.

Shortly after Teukolsky's initial result, Chrzanowski \cite{Chrzanowski:1975wv} and Cohen and Kegeles \cite{Kegeles:1979an} (CCK) showed that the metric perturbation can be constructed from a Hertz potential that satisfies a separable differential equation sourced by the Weyl scalars that emerge from Teukolsky's equations (see also Wald \cite{Wald:1978vm} and Stewart \cite{Stewart:1978tm}). Strictly, the CCK reconstruction procedure only applies in vacuum. Nevertheless, it has been applied with some success to reconstruct metric perturbation in vacuum regions on either side of a particle's worldline. The CCK construction has three deficiencies that make it less suitable for modern applications: (i) the metric perturbation is constructed in a \emph{radiation gauge}, which means that (without remedy) it can't represent a full solution to a sourced equation unless certain components of the stress-energy tensor happen to be zero; (ii) the ``inversion'' relation between the Hertz potential and the Weyl tensor requires the solution of a fourth-order equation, which adds technical complexity; (iii) the reconstructed metric perturbation typically has extended string-like gauge singularities \cite{Barack:2001ph,Ori:2002uv,Keidl:2010pm,Pound:2013faa}, i.e., discontinuities or distributional terms in the metric perturbation. While gauge discontinuities in the first-order metric perturbation do not significantly impede self-force calculations at first order (see Refs.~\cite{Keidl:2006wk,Keidl:2010pm,Shah:2010bi,Shah:2012gu,vandeMeent:2015lxa,vandeMeent:2017bcc}), they become   problematic at second order, because they generate highly singular (non-distributional) terms in the second-order source $S^{(2)}_{\mu\nu}$. 

Recently, there has been significant progress on upgrading the CCK metric reconstruction procedure in the presence of sources. Green, Hollands and Zimmerman (GHZ) \cite{Green:2019nam} have shown that the CCK procedure can be extended to the sourced (i.e.~non-vacuum) case by augmenting the metric perturbation with a so-called corrector tensor, which is determined by solving certain decoupled ODEs by integrating over the outgoing Kerr–Newman radial coordinate. Toomani \emph{et al.}~\cite{Toomani:2021jlo} have shown that the gauge singularities arising in the GHZ approach can be softened by moving to a `shadowless' gauge, in order to obtain a metric perturbation suitable as an input for second-order calculations. This approach is certainly promising, and is under active development.
 
There are several other approaches to calculating metric perturbations for rotating black holes, which are in various states of progress:
\begin{itemize}
  \item Barack, Dolan and Wardell \cite{Dolan-Barack-Wardell} have developed a 2+1D time-domain code for calculating the (azimuthal) $m$-modes of the Lorenz-gauge metric perturbation for a particle on a circular, equatorial orbit, based on the puncture/effective source scheme developed in Refs.~\cite{Barack:2007jh,Barack:2007we, Dolan:2010mt,Dolan:2011dx,Dolan:2012jg}. (See also the work in this area by Thornburg \cite{Thornburg:2016msc}).
  \item Franchini \cite{Franchini:2023xhd} has recently introduced a slow-spin expansion which yields extended Regge-Wheeler and Zerilli equations at second order in the spin;
  \item Ripley and Loutrel \etal~\cite{Loutrel:2020wbw,Ripley:2020xby} have implemented an approach that circumvents the use of Hertz potentials altogether, by using a formulation rooted in early work by Chandrasekhar \cite{Chandrasekhar:1985kt}; 
  \item Osburn and Nishimura \cite{Osburn:2022bby} have used a scalar-field toy model to propose that it may be feasible to directly compute the metric perturbation by solving a set of two-dimension \emph{elliptic} partial differential equations in the $(r,\theta)$ domain;
  \item Aksteiner, Anderson and Backdahl (AAB) \cite{Aksteiner:2016pjt} have shown that, by combining the \emph{two} maximum-spin components of the Weyl tensor, a Hertz-potential approach can be used in the presence of sources without introducing a corrector tensor;
  \item Dolan, Kavanagh and Wardell \cite{Dolan:2021ijg} showed that, in vacuum regions, the radiation-gauge metric perturbation arising from the CCK procedure can be transformed into the Lorenz gauge by using a gauge vector that is straightforwardly obtained from solutions of the Teukolsky equation. 
  \end{itemize}

In this paper, we continue the development of the last approach \cite{Dolan:2021ijg}, by seeking to construct a Lorenz-gauge metric perturbation for a canonical non-vacuum case. More specifically, in the following sections we will calculate the linearised metric perturbation $h_{\mu \nu}^{(1)}$ (henceforth omitting the superscript) for the special case of a particle on a circular equatorial orbit of Kerr spacetime, by solving the system
\beq
\Box \barh_{\mu \nu} + 2 \tensor{R}{^\alpha _\mu ^\beta _\nu} \barh_{\alpha \beta} = - 16 \pi T_{\mu \nu}, \quad \quad \nabla^\nu \barh_{\mu \nu} = 0 , \label{eq:hLorenz-field-equation}
\eeq
with the appropriate stress-energy tensor in Eq.~(\ref{eq:stress-energy}). 

The adoption of Lorenz gauge brings several immediate benefits. First, the metric perturbation in Lorenz gauge should be free from extended discontinuties or distributions, and so it is a suitable input for second-order calculations. Second, 
the asymptotic behaviour of the Lorenz gauge MP towards the horizon and infinity is well-understood. Third, the existing Lorenz-gauge set-up has been successfully used in the Schwarzschild case for second-order calculations \cite{Warburton:2021kwk,Wardell:2021fyy, Durkan:2022fvm, Albertini:2022dmc,Albertini:2022rfe, vandeMeent:2023ols}. Fourth, the Lorenz-gauge MP has been computed by other means (e.g.~by a 2+1D time-domain code) and Kerr comparison data is available at the level of the metric perturbation, which we make use of here. Fifth, the Schwarzschild case has been well-studied and there are several comparison data sets available \cite{Barack:2005nr,Barack:2007tm, Berndtson:2007gsc, Barack:2008ms,Barack:2010tm, Akcay:2010dx,Hopper:2010uv,Hopper:2012ty,Akcay:2013wfa,Hopper:2015jxa}. Sixth, the regularization parameters for the self-force in Lorenz gauge have already been calculated \cite{Heffernan:2012vj, Heffernan:2022zgc}.

We can also identify three specific benefits of the scheme developed here. First, we construct the metric perturbation entirely from differential operators acting on linear combinations of single-variable functions of $r$ and $\theta$, such as the Teukolsky mode functions, that satisfy decoupled \emph{ordinary} differential equations. This brings clear advantages in accuracy and computational efficiency. Second, in the static sector, we are able to obtain closed-form solutions for \emph{all} of the mode functions used, including the completion pieces, in terms of elementary functions. Third, in the non-static sector, by constructing the $s=2$ part of the MP from the \emph{difference} between the ingoing and outgoing radiation-gauge perturbations (each transformed to Lorenz gauge), we can replace the Hertz potentials directly with the Weyl scalars in a straightforward way, evading a technically-complex ``inversion'' step (see also \cite{Aksteiner:2016pjt,Dolan:2021ijg}). The main drawbacks of the approach are that, first, since the formulation is in the frequency domain, it is not so well suited to particles on very eccentric, or unbound, trajectories. Second, the (inevitable) truncation of the $\ell$-mode sum leads to a loss of accuracy near the particle (see discussion in Sec.~\ref{sec:conclusions}).

In overview, the article is organised as follows. In Sec.~\ref{sec:vacuum}, we expand on the approach of Ref.~\cite{Dolan:2021ijg} to derive vacuum modes in Lorenz gauge in the radiative sector ($\omega \neq 0$). After a review of the Teukolsky formalism (Sec.~\ref{subsec:teukolsky}), and the CCK construction as clarified by Wald (Sec.~\ref{sec:radiation-gauge}), and various preliminaries (Sec.~\ref{subsec:preliminaries}), the transformation from radiation gauge is described in Sec.~\ref{subsec:toLorenz}. The key result of Sec.~\ref{sec:vacuum} is a set of $s=2$, $s=1$ and $s=0$ vacuum solutions to the Lorenz gauge equations (i.e.~Eq.~(\ref{eq:hLorenz-field-equation}) with $T_{\mu \nu} = 0$), whose metric components are listed in Eqs.~(\ref{eq:mp-s2}), (\ref{eq:mp-s1}) and (\ref{eq:mp-s0}) of Sec.~\ref{sec:mp}. We can conceive of these as jigsaw pieces to be fitted together correctly at the particle orbit radius $r_0$ in order to produce a regular Lorenz-gauge metric perturbation. In Sec.~\ref{sec:radiative} we describe how that jigsaw is assembled: we project the metric components onto an appropriate \emph{spherical} basis, and demand that each $\ell m$ mode is continuous at $r=r_0$. A range of numerical results are presented in Secs.~\ref{subsec:results-lm} and \ref{subsec:results-m}, where the results of the new method are compared with data from the 2+1D time domain code of Ref.~\cite{Dolan-Barack-Wardell}. In Sec.~\ref{sec:static} we address the question of how to transform the \emph{static} modes ($\omega=0$) to Lorenz gauge (Sec.~\ref{subsec:static-s2}), and how to construct a Lorenz-gauge MP with a trace (Sec.~\ref{subsec:static-s0}). In Sec.~\ref{subsec:completion} we present the non-radiative mass, angular momentum and gauge ``completion'' pieces in Lorenz gauge. In Sec.~\ref{subsec:Berndtson} we consider the boundary conditions imposed the metric perturbation, and  subtleties relating to mass and angular momentum. Numerical results for the static sector are presented in Sec~\ref{subsec:results-m0}, where we show a good agreement with the comparison data set. We conclude with a discussion of next steps in Sec.~\ref{sec:conclusions}. The appendices \ref{sec:spherical-harmonics}---\ref{sec:BL} give further details and useful expressions, including complete expressions for the projection of the metric perturbation onto the spherical basis in Appendix \ref{appendix:projections}.

Throughout this work we follow the conventions of Misner, Thorne and Wheeler
\cite{Misner:1973prb}: a ``mostly positive'' metric signature, $(-,+,+,+)$, is
used for the spacetime metric; the connection coefficients are defined by
$\Gamma^{\lambda}_{\mu\nu}=\frac{1}{2}g^{\lambda\sigma}(g_{\sigma\mu,\nu}
+g_{\sigma\nu,\mu}-g_{\mu\nu,\sigma})$; the Riemann tensor is
$R^{\tau}{}_{\!\lambda\mu\nu}=\Gamma^{\tau}_{\lambda\nu,\mu}
-\Gamma^{\tau}_{\lambda\mu,\nu}+\Gamma^{\tau}_{\sigma\mu}\Gamma^{\sigma}_{\lambda\nu} -\Gamma^{\tau}_{\sigma\nu}\Gamma^{\sigma}_{\lambda\mu}$, the Ricci
tensor and scalar are $R_{\mu\nu}=R^{\tau}{}_{\!\mu\tau\nu}$ and
$R=R_{\mu}{}^{\!\mu}$, and the Einstein equations are
$G_{\mu\nu}=R_{\mu\nu}-\frac{1}{2}g_{\mu\nu}R=8\pi
T_{\mu\nu}$. Standard geometrised units are used, with $c=G=1$.
We use Greek letters for spacetime indices, denote symmetrisation
of indices using round brackets [e.g. $T_{(\alpha \beta)} = \tfrac12 (T_{\alpha \beta}+T_{\beta
\alpha})$] and anti-symmetrisation using square brackets [e.g. $T_{[\alpha \beta]} = \tfrac12
(T_{\alpha \beta}-T_{\beta \alpha})$], and exclude indices from symmetrisation by surrounding them by
vertical bars [e.g. $T_{(\alpha | \beta | \gamma)} = \tfrac12 (T_{\alpha \beta \gamma}+T_{\gamma \beta
\alpha})$].

\section{Vacuum metric perturbations in Lorenz gauge\label{sec:vacuum}}
 \subsection{Preliminaries\label{subsec:preliminaries}}
 
\subsubsection{Metric and null tetrad\label{subsec:metric}}

A Kerr black hole of mass $M$ and angular momentum $J = aM$ is described by the metric $g_{\mu \nu}$ in Boyer-Lindquist coordinates $x^\mu = \{t, r, \theta, \phi \}$. It is standard to express the inverse metric in terms of a null tetrad $\{ l^\mu, n^\mu, m^\mu, \mbar^\mu \}$, 
\beq
g^{\mu \nu} = - 2 l^{(\mu} n^{\nu)} + 2 m^{(\mu} \mbar^{\nu)}. 
\eeq
Here $l^\mu$ and $n^\mu$ are real vectors, and $m^\mu$ is complex and $\mbar^\mu$ is its complex conjugate, normalised such that $m \cdot \bar{m} = - \ell \cdot n = 1$, with all other inner products equal to zero (where $a \cdot b \equiv g_{\mu \nu} a^\mu b^\nu$).
Kerr spacetime is algebraically special and of Petrov type D, and many simplifications arise by aligning $l^\mu$ and $n^\mu$ with the repeated principal null directions. A common choice of tetrad is that of Kinnersley,
\begin{align}
l^\mu = l_+^\mu, \quad n^\mu = - \frac{\Delta}{2 \Sigma} l_-^\mu, \quad m^\mu = \frac{1}{\sqrt{2} \, \rho} m_+^\mu , \quad \mbar^\mu = \frac{1}{\sqrt{2} \, \rhoc} m_-^\mu , \label{eq:Kinnersley}
\end{align}
where $\Delta \equiv r^2 - 2Mr + a^2$ and $\Sigma \equiv \rho \rhoc = r^2 + a^2 \cos^2 \theta$ with
\beq 
\rho \equiv r + i a \cos \theta.
\eeq
Here we have introduced an unnormalized tetrad $\{l_\pm^\mu, m_\pm^\mu\}$,
\begin{align}
l^\mu_\pm &\equiv \left[ \pm (r^2+a^2) / \Delta, 1, 0, \pm a/\Delta \right] , \\
m^\mu_{\pm} &\equiv \left[ \pm i a \sin \theta, 0, 1, \pm i \csc \theta \right] .
\label{eq:tetrad2}
\end{align}
Note that $l_\pm^\mu$ and $m^\mu_{\pm}$ are functions of $r$ only and $\theta$ only, respectively. 
The inverse metric can be written in terms of this unnormalized tetrad as
\begin{align}
g^{\mu \nu}
 &= \frac{1}{\Sigma} \left( \Delta l_{+}^{(\mu} l_{-}^{\nu)} + m_{+}^{(\mu} m_{-}^{\nu)} \right) . \label{eq:inverse-metric}
\end{align}
The metric determinant is $g = - \Sigma^2 \sin^2 \theta$.

 \subsubsection{Directional derivatives}
Following Chandrasekhar \cite{Chandrasekhar:1985kt}, and others, we now introduce directional derivatives along the null directions.
The directional derivatives along $\{ l^\mu, n^\mu, m^\mu, \mbar^\mu \}$ are denoted by $\{ D, \npDel, \delta, \dbar \}$, respectively. The directional derivatives along $\{ l_+^\mu, l_-^\mu, m_+^\mu, m_-^\mu \}$ are denoted by
 $\{ \cD, \cD^\dagger, \cL^\dagger, \cL \}$ (note ordering).

Throughout this work, we shall assume that the differential operators always act on functions $\mathcal{X}$ with harmonic dependence on time and azimuthal angle, $\mathcal{X}(x^\mu) =\mathcal{X}(r, \theta) e^{- i \omega t + i m \phi}$. In this prescription, the $\partial_t$ and $\partial_\phi$ derivatives are replaced by factors of $- i \omega$ and $i m$, respectively, and we may write the operators in their Chandrasekhar form \cite{Chandrasekhar:1985kt},
\begin{subequations}
\begin{align}
\cD \equiv l_+^\mu \partial_\mu &= \partial_r - \frac{i K}{\Delta},  & 
\cL^\dagger \equiv m_+^\mu \partial_\mu =  \partial_\theta - Q ,
 \\
\cD^\dagger  \equiv l_-^\mu \partial_\mu &= \partial_r + \frac{i K}{\Delta} , 
& \cL \equiv m_-^\mu \partial_\mu = \partial_\theta + Q ,
\end{align}
\label{eq:DLoperators}
\end{subequations}
where $K \equiv \omega (r^2 + a^2) - a m$ and $Q \equiv m \csc \theta - a \omega \sin \theta$. 
In a standard way, we also introduce $\cD_n = \cD + n \Delta' / \Delta$, $\cD^\dagger_n = \cD^\dagger + n \Delta' / \Delta$, $\cL_n = \cL + n \cot\theta$, and $\cL^\dagger_n = \cL^\dagger + n \cot\theta$ where $\Delta' = \partial_r \Delta = 2 (r - M)$.

Later we consider the projection of spheroidal harmonics onto a spherical basis, and it is helpful to decompose the angular operators into
\beq
\cL_n = \cLh_n - a\omega \sin \theta, \quad \quad \cL^\dagger_n = \cLh^\dagger_n + a\omega \sin \theta,
\eeq
where $\hat{\cL}_n$ and $\hat{\cL}^\dagger_n$ are ladder operators that lower and raise the spin-weight of the \emph{spherical} harmonics, as described in Appendix \ref{sec:spherical-harmonics}. 
 
 \subsubsection{Bivectors and the principal tensor}
In the following sections, we make use of the following null bivectors (rank-two antisymmetric tensors):
\begin{subequations} \label{UVWdef}
\begin{align}
\cU^{\mu \nu} &= 2 l_+^{[\mu} m_+^{\nu]}     \\
\cV^{\mu \nu} &= 2 l_-^{[\mu} m_-^{\nu]}  \\
\cW^{\mu \nu} &= 2 \Delta l_+^{[\mu} l_-^{\nu]} + 2 m_+^{[\mu} m_-^{\nu]}  .
\end{align}
\end{subequations}
These bivectors are self-dual, in the sense that $({}^\star \cU)^{\mu \nu} = i \, \cU^{\mu \nu}$, where $i$ is the unit imaginary, and ${}^\star$ denotes the Hodge-dual operation,
\beq
({}^\star X)_{\mu \nu} \equiv \frac{1}{2} \epsilon_{\mu \nu \sigma \lambda} X^{\sigma \lambda} , 
\eeq
where $\epsilon_{\mu \nu \sigma \lambda}$ is the Levi-Civita tensor. 
It follows that the complex-conjugate bivectors $\{ \overline{\cU}^{\mu \nu}, \overline{\cV}^{\mu \nu}, \overline{\cW}^{\mu \nu} \}$ are anti-self-dual, $({}^\star \overline{\cU})^{\mu \nu} = - i \, \overline{\cU}^{\mu \nu}$. 

The inner products of the bivector set $\{ \cU, \cV, \cW, \overline{\cU}, \overline{\cV}, \overline{\cW} \}$ are $\cU \cdot \cV = \overline{\cU} \cdot \overline{\cV} = 2  (2\Sigma)^2 / \Delta$, $\cW \cdot \cW = \overline{\cW} \cdot \overline{\cW} = - 4  (2\Sigma)^2$, with all other inner products zero, where $\cU \cdot \cV \equiv \cU_{\mu \nu} \cV^{\mu \nu}$.

The following bivectors are
divergence-free on Kerr spacetime:
\beq
\nabla_\mu \left( \frac{1}{\Sigma \sin \theta} \cU^{\mu \nu} \right) =  \nabla_\mu \left( \frac{1}{\Sigma \sin \theta} \cV^{\mu \nu} \right) = \nabla_\mu \left( \frac{1}{\Sigma \rhoc^2} \cW^{\mu \nu} \right) =   0 .
\eeq

The \emph{principal tensor} $\tilde{f}_{\mu \nu}$ is used in the construction of the $s=0$ trace modes of the metric perturbation in Sec.~\ref{subsec:scalar}. It is a conformal Killing-Yano tensor, that is, a two-form satisfying the equation $\nabla_{\sigma} \tilde{f}_{\mu \nu} = g_{\nu \sigma} X_{\mu} - g_{\mu \sigma} X_{\nu}$, where $X^\mu = \delta^{\mu}_{(t)}$ is the time-translation Killing vector field. On Kerr spacetime it takes the form
\begin{align}
\tilde{f}^{\mu \nu}
 &= \frac{r \Delta}{\Sigma} l_+^{[\mu} l_-^{\nu]} - \frac{i a \cos \theta}{\Sigma} m_+^{[\mu} m_-^{\nu]} . \label{eq:cky}
\end{align}

\subsubsection{Weyl scalars}
With a null tetrad (\ref{eq:Kinnersley}), the Weyl tensor $C_{\mu \nu \sigma \lambda}$ can be decomposed into five complex Weyl scalars $\Psi_i$ ($i=0\ldots 4$). On the background Kerr spacetime, the scalars take the values
\begin{align}
\Psi_0 = \Psi_1 = \Psi_3 = \Psi_4 = 0, \quad \Psi_2 = - M / \rhoc^3 . 
\end{align}
Of particular importance in the following are the Weyl scalars of maximal spin weight,
\begin{subequations}
\begin{align}
\Psi_0 &=  C_{\mu \nu \sigma \lambda} l^\mu m^\nu l^\sigma m^\lambda \\
\Psi_4 &=  C_{\mu \nu \sigma \lambda} n^\mu \mbar^\nu n^\sigma \mbar^\lambda 
\end{align}
\end{subequations}
At perturbative order they are invariant under changes of gauge and tetrad. Moreover, as shown by Teukolsky \cite{Teukolsky:1972my}, $\Psi_0$ and $\Psi_4$ satisfy decoupled second-order PDEs that are separable in the frequency domain.

For future reference, we also introduce a rescaled Weyl scalar
\begin{align}
\widetilde{\Psi}_4 \equiv \frac{4 \rhoc^4 \Psi_4}{\Delta^2} ,
\end{align}
and the `primed' scalars
\begin{subequations}
\begin{align}
\Psi_0^\prime &= C_{\mu \nu \sigma \lambda} l^\mu \overline{m}^\nu l^\sigma \overline{m}^\lambda \\
\widetilde{\Psi}_4^\prime &=  \frac{4 \rho^4}{\Delta^2} C_{\mu \nu \sigma \lambda} n^\mu m^\nu n^\sigma m^\lambda .
\end{align}
\end{subequations}

\subsubsection{Forms\label{subsec:forms}}
The language of forms is used in Sec.~\ref{sec:hLorenz}.
A $p$-form is equivalent to a fully antisymmetric tensor of rank $(0,p)$. The exterior derivative $d$ maps a $p$-form $\beta$ to a $(p+1)$-form $d \beta$, and the coderivative $\delta \equiv {}^\star d {}^\star$ (where ${}^\star$ denotes the Hodge-dual operation) maps a $p$-form $\beta$ to a $(p-1)$-form $\delta \beta$, according to the rules
\begin{align}
(d \beta)_{\mu_0 \cdots \mu_p} &= (p+1) \nabla_{[\mu_0} \beta_{\mu_1 \cdots \mu_p]} , \\
(\delta \beta)_{\mu_2 \cdots \mu_p} &= - \nabla^{\mu_1} \beta_{\mu_1 \cdots \mu_p} . 
\end{align}
For a fuller summary, see e.g. Appendix A.2 in Ref.~\cite{Frolov:2017kze}. 
 
 \subsection{The Teukolsky formalism\label{subsec:teukolsky}}
 
  \subsubsection{Gravitational fields $s=2$}
In Refs.~\cite{Teukolsky:1972my,Teukolsky:1973ha}, Teukolsky derived separable equations for the Weyl scalars of maximal spin-weight ($\mathfrak{s} = \pm 2$), which can be written in the Chandrasekhar form \cite{Chandrasekhar:1985kt}
\begin{subequations}
\begin{align}
\left[ \Delta \cD_1 \cD_2^{\dagger} + \cL_{-1}^\dagger \cL_2 + 6 i \omega \rho  \right] \Psi_0 &= -8 \pi \Sigma T_0 , \label{eq:sourced-psi0} \\
\left[ \Delta \cD_1^\dagger \cD_2 + \cL_{-1} \cL_2^\dagger - 6 i \omega \rho \right] \widetilde{\Psi}_4 &= -8 \pi \Sigma \widetilde{T}_4 . \label{eq:sourced-psi4}
\end{align}
\label{eq:sourced-teukolsky}
\end{subequations}
These equations are separable with the mode ansatz
\begin{subequations}
\begin{align}
\Psi_0 = \Delta^{-2} P_{+2}(r) S_{+2}(\theta) e^{-i \omega t + i m \phi} , \\
\widetilde{\Psi}_4 = \Delta^{-2} P_{-2}(r) S_{-2}(\theta) e^{-i \omega t + i m \phi} , 
\end{align}
\label{eq:Psi04-modes}
\end{subequations}
Here $P_{\pm2}(r)$ are related to the Teukolsky radial functions $R_{\pm2}(r)$ by $P_{+2} = \Delta^2 R_{+2}$ and $P_{-2}(r) = R_{-2}(r)$.
In vacuum ($T_0 = 0 = \widetilde{T}_4$) one obtains ordinary differential equations in the form
\begin{subequations}
\begin{align}
\left(\Delta \cD_{-1} \cD^\dagger + 6 i \omega r - \Lambda \right) P_{+2} &= 0 , &
\left(\cL_{-1}^\dagger \cL_2 - 6 a \omega \cos \theta + \Lambda \right) S_{+2} &= 0 ,  \label{eq:TeukSp} \\ 
\left(\Delta \cD_{-1}^\dagger \cD - 6 i \omega r - \Lambda \right) P_{-2} &= 0 ,   &
\left(\cL_{-1} \cL_2^\dagger + 6 a \omega \cos \theta + \Lambda \right) S_{-2} &= 0 .   \label{eq:TeukSm}  
\end{align}
\label{eq:teukolsky-eqns2}
\end{subequations}
The separation constant $\Lambda$ is the Teukolsky constant for $s=-2$, and in the Schwarzschild case ($a=0$) it takes the value $\Lambda = (l-1)(l+2)$. The functions $S_{\pm2}(\theta)$ are spheroidal harmonics of spin-weight $\pm 2$.

Using the vacuum Teukolsky equations (\ref{eq:teukolsky-eqns2}), it is quick to establish the following identities,
\begin{subequations}  \label{key-identities-1}
\begin{align}
\Delta \cD^\dagger \cD (\cD P_{-2}) &= +(\Lambda + 2 + 2 i \omega r) \cD P_{-2} + 6 i \omega P_{-2} , \\
\cL \cL_1^\dagger (\cL_2^\dagger S_{-2}) &= - (\Lambda + 2 + 2 a \omega \cos \theta) \cL_2^\dagger S_{-2} + 6 a \omega \sin \theta S_{-2} ,
\end{align}
\end{subequations}
and, taking an additional derivative,
\begin{subequations}  \label{key-identities-2}
\begin{align}
\cD \Delta \cD^\dagger (\cD \cD P_{-2}) &= +(\Lambda + 2 + 2 i \omega r) \cD \cD P_{-2} + 8 i \omega \cD P_{-2} , \\
\cL_1^\dagger \cL (\cL_1^\dagger \cL_2^\dagger S_{-2}) &= - (\Lambda + 2 + 2 a \omega \cos \theta) \cL_1^\dagger \cL_2^\dagger S_{-2} + 8 a \omega \sin \theta \cL_2^\dagger S_{-2} . 
\end{align}
\end{subequations}
These equations will be used in Sec.~\ref{sec:hLorenz}. 
Similar identities hold for $P_{+2}$ and $S_{+2}$, making the swaps $\cD \leftrightarrow \cD^\dagger$ and $\cL_n \leftrightarrow \cL^\dagger_n$ and $P_{+2} \leftrightarrow P_{-2}$ and $S_{+2} \leftrightarrow S_{-2}$ and $\omega \leftrightarrow - \omega$. 

To represent a valid solution to the linearized vacuum Einstein equations satisfying the Bianchi identities, the Weyl scalars must be related by the \emph{Teukolsky-Starobinskii identities}, which in mode form read
\begin{subequations}
\label{eq:TS-identities}
\begin{align}
\Delta^2 \cD \, \cD \, \cD \, \cD \, P_{-2} &= \mathcal{C} \, P_{+2} , \label{TS:Rm} \\
\Delta^2 \cD^\dagger \cD^\dagger \cD^\dagger \cD^\dagger P_{+2} &= \mathcal{C}^\ast \, P_{-2} , \label{TS:Rp}  \\
\cL_{-1}^\dagger \cL_0^\dagger \cL_1^\dagger \cL_2^\dagger S_{-2} &= A \, S_{+2} ,  \label{TS:Sm}  \\
\cL_{-1} \cL_0 \cL_1 \cL_2 S_{+2} &= A \, S_{-2} , \label{TS:Sp} 
\end{align}
\end{subequations}
with 
\beq
A^2 = \Lambda^2 (\Lambda+2)^2 + 8 a \omega \Lambda \left[ ( m - a \omega) (5\Lambda+6) + 12 a \omega \right] + 144 a^2 \omega^2 (m - a \omega)^2 , 
\label{eq:Adef}
\eeq
and
\beq
\mathcal{C} = A + (-1)^{\ell + m} 12 i \omega M .
\eeq
The sign $(-1)^{\ell + m}$ here is in agreement with Eq.~(104) of Ref.~\cite{Pound:2021qin}), and it is necessary to recover the even-parity ($\ell + m$ even) and odd-parity ($\ell + m$ odd) modes in the Schwarzschild limit.

  \subsubsection{Electromagnetic fields $s=1$}
The Maxwell equations on Kerr spacetime are also amenable to a separation of variables \cite{Teukolsky:1972my}. The Maxwell scalars $\Phi_0 = F_{\mu \nu} l^\mu m^\nu$ and $\widetilde{\Phi}_2 = (2 \rhoc^2 / \Delta) F_{\mu \nu} \overline{m}^\mu n^\nu$ satisfy the sourced equations
\begin{align}
\left( \Delta \cD_1 \cD_1^\dagger + \cL_0^\dagger \cL_1 + 2 i \omega \rho \right) \Phi_0 &= 4 \pi \Sigma J_{0} , \\
\left( \Delta \cD_1 \cD_1^\dagger + \cL_0 \cL_1^\dagger - 2 i \omega \rho \right) \widetilde{\Phi}_2 &= 4 \pi \Sigma \widetilde{J}_{2} .  
\end{align}
These equations are separable with the mode ansatz
\begin{subequations}
\begin{align}
\Phi_0 = \Delta^{-1} P_{+1}(r) S_{+1}(\theta) e^{-i \omega t + i m \phi} , \\
\widetilde{\Phi}_2 = \Delta^{-1} P_{-1}(r) S_{-1}(\theta) e^{-i \omega t + i m \phi} . 
\end{align}
\end{subequations}
Here $P_{\pm1}(r)$ are related to the Teukolsky radial functions $R_{\pm1}(r)$ by $P_{+1} = \Delta R_{+1}$, and $P_{-1} = R_{-1}$. In vacuum ($J_0 = 0 = \tilde{J}_2$) one obtains ordinary differential equations in the Chandrasekhar form
\begin{subequations}
\begin{align}
\left(\Delta \cD \cD^\dagger + 2 i \omega r - \lambda \right) P_{+1} &= 0 , &
\left(\cL^\dagger \cL_1 - 2 a \omega \cos \theta + \lambda \right) S_{+1} &= 0 , \\
\left(\Delta \cD^\dagger \cD - 2 i \omega r - \lambda \right) P_{-1} &= 0 , &
\left(\cL \cL_1^\dagger + 2 a \omega \cos \theta + \lambda \right) S_{-1} &= 0 , 
\end{align}
\label{eq:teukolsky-eqns1}
\end{subequations}
with $P_{-1}=R_{-1}$, $P_{+1} = \Delta R_{+1}$, where $R_{\pm1}(r)$ are the radial functions of Teukolsky. The separation constant $\lambda$ is that for $s=-1$ (in the Teukolsky equations), and in the Schwarzschild limit it is $\lambda = l(l+1)$. 

Using the vacuum Teukolsky equations, it is quick to establish identities such as
\begin{subequations}  \label{key-identities-vector}
\begin{align}
\cD \Delta \cD^\dagger (\cD P_{-1}) &= +(\lambda + 2 i \omega r) \cD P_{-1} + 2 i \omega P_{-1} , \\
\cL_1^\dagger \cL (\cL_1^\dagger S_{-1}) &= - (\lambda + 2 a \omega \cos \theta) \cL_1^\dagger S_{-1} + 2 a \omega \sin \theta S_{-1} .
\end{align}
\end{subequations}
It follows also that
\beq
\Box (\cD \cL_1^\dagger \psi) = \frac{1}{\Sigma} \left( 
\cD \Delta \cD^\dagger + \cL^\dagger_1 \cL - 2 i \omega \rho 
\right) (\cD \cL_1^\dagger \psi) = \frac{2 i \omega}{\Sigma} \left(\cL_1^\dagger - i a \sin \theta \cD \right) \psi
\eeq
where $\psi = P_{-1}(r) S_{-1}(\theta)$. This equation will be used in the following section.

The functions $P_{\pm1}(r)$ and $S_{\pm1}(\theta)$ are not independent; to represent a valid solution of the vacuum Maxwell equations, the Maxwell scalars must also satisfy the Teukolsky-Starobinskii equations. In mode form, these are 
\begin{subequations}
\begin{align}
\Delta \cD \cD P_{-1} &= \mathcal{B} \, P_{+1} , &
\cL_0^\dagger \cL_1^\dagger S_{-1} &= \mathcal{B} \, S_{+1} ,   \\
\Delta \cD^\dagger \cD^\dagger P_{+1} &= \mathcal{B} \, P_{-1} ,  &
\cL_0 \cL_1 S_{+1} &= \mathcal{B} \, S_{-1} , 
\end{align}
\label{eq:TSs1}
\end{subequations}
with 
\beq
\mathcal{B} \equiv \sqrt{\lambda^2 + 4 a m \omega - 4 a^2 \omega^2} . \label{eq:Bteuk}
\eeq

 \subsubsection{Scalar fields $s=0$}
The scalar field equation $\Box \chi = 0$ is straightforwardly separable on Kerr spacetime \cite{Brill:1972xj}. The d'Alembertian of a scalar field $\chi$ may be written as
\begin{subequations}
\begin{align}
\Box \chi \equiv \nabla_\mu \nabla^\mu \chi &= \frac{1}{\sqrt{-g}} \partial_\mu \left( \sqrt{-g} g^{\mu \nu} \partial_\nu \chi  \right) , \nn \\
&= \frac{1}{2 \Sigma \sin \theta} \partial_\mu \left( \sin \theta (\Delta \cD^\dagger \chi \, l_+^\mu + \Delta \cD \chi \, l_-^\mu + \cL \chi \, m_+^\mu + \cL^\dagger \chi \, m_-^\mu )  \right) , \nn \\
&= \frac{1}{2\Sigma} \left( \cD \Delta \cD^\dagger + \cD^\dagger \Delta \cD + \cL_1^\dagger \cL + \cL_1 \cL^\dagger \right) \chi .
\end{align}
\end{subequations}
Here we have made use of the inverse metric (\ref{eq:inverse-metric}), the metric determinant $\sqrt{-g} = \Sigma \sin \theta$, and directional derivatives (\ref{eq:DLoperators}). Alternative forms of the d'Alembertian include
\begin{subequations}   \label{eq:Box-alt}
\begin{align}
\Box \chi &= \frac{1}{\Sigma} \left( 
\cD \Delta \cD^\dagger + \cL_1 \cL^\dagger - 2 i \omega \rhoc 
\right) \chi , \\
&=  \frac{1}{\Sigma} \left( 
\cD \Delta \cD^\dagger + \cL^\dagger_1 \cL - 2 i \omega \rho 
\right) \chi .
\end{align} \label{eq:dAlembertian-alt}
\end{subequations}
The scalar field equation $\Box \chi = 0$ is separable with the ansatz $\chi = P_0(r) S_0(\theta) e^{- i \omega t + i m\phi}$. One obtains a pair of second-order ODEs, viz.,
\begin{subequations}
\begin{align}
\left( \cD \Delta \cD^\dagger - 2 i \omega r - \lambda_0 \right) P_0 &= 0 , \\
\left( \cL_1 \cL^\dagger - 2 a \omega \cos \theta + \lambda_0 \right) S_0 &= 0 . 
\end{align}
\end{subequations}
Here $S_0(\theta)$ is a spheroidal harmonic (of spin-weight zero), and $\lambda_0$ is the angular eigenvalue, which in the Schwarzschild case ($a=0$) takes the value $\lambda_0 = \ell(\ell+1)$.

 \subsection{Metric perturbations in radiation gauges\label{sec:radiation-gauge}}

In vacuum, a metric perturbation (or a vector potential) in a radiation gauge may be obtained from scalar potentials by applying the method of adjoints introduced by Wald \cite{Wald:1978vm}, which we recap below. 

Abstractly, suppose we wish to solve a field equation of the form $\mathcal{E}(h) = 0$, where $\mathcal{E}$ is a linear partial differential operator, and $h$ is a tensor field of interest (with indices omitted). Suppose also that a decoupled equation has been found, $\mathcal{O} \psi = 0$, with a scalar variable $\psi$  that is derived from the field, $\psi = \mathcal{T} h$ (here $\mathcal{T}$ and $\mathcal{O}$ are also linear differential operators). There then exists an operator $\mathcal{S}$ to complete the operator identity $\mathcal{S} \mathcal{E} = \mathcal{O} \mathcal{T}$. Taking the adjoint (see Ref.~\cite{Wald:1978vm} for the definition) of this identity yields $\mathcal{E}^\dagger \mathcal{S}^\dagger = \mathcal{T}^\dagger \mathcal{O}^\dagger$. In typical cases of interest, the field operator is self-adjoint ($\mathcal{E}^\dagger = \mathcal{E}$).  Hence it follows that, if $\hat{\psi}$ satisfies $\mathcal{O}^\dagger \hat{\psi} = 0$, then $\hat{h} \equiv \mathcal{S}^\dagger \hat{\psi}$ satisfies the field equation $\mathcal{E} \hat{h} = 0$. This gives us a method for constructing a tensorial perturbation (such as the metric perturbation) from a scalar potential by the application of a (tensorial) differential operator. 

Application of this method in the electromagnetic case recovers the expressions for the vector potential first found by Cohen and Kegeles \cite{Cohen:1974cm}. Application of this method in the gravitational case recovers expressions for the metric perturbation first found by Chrzanowski \cite{Chrzanowski:1975wv}. In the section below, we quote these expressions without derivation.

The Ingoing Radiation Gauge (IRG) metric perturbation in vacuum, satisfying $h^{\text{IRG}}_{\mu \nu} l_+^\nu = 0$, is
\begin{align}
\hirg^{\mu\nu}
&= -\frac{1}{2 \rho^2} \left[ l_+^\mu l_+^\nu \left( \cL_1^\dagger - \frac{2 \rhoc_{,\theta}}{\rhoc} \right)  \cL_2^\dagger + m_+^\mu m_+^\nu \left(\cD - \frac{2}{\rhoc} \right) \cD  \right. \nn \\
& \left. \quad \quad \quad \quad - 2 \, l_+^{(\mu} m_+^{\nu)}  
\left( \cD \cL_2^\dagger  - \frac{1}{\rho} \left( \cL_2^\dagger + \rho_{,\theta} \cD \right)  - \frac{1}{\rhoc} \left( \cL_2^\dagger + \rhoc_{,\theta} \cD \right) \right) 
 \right] \psi_{-} . \label{h-irg}
\end{align} 
This is the second part of Chrzanowski's IRG perturbation (Ref.~\cite{Chrzanowski:1975wv}, Table I). The metric perturbation (\ref{h-irg}) is complex; to obtain a real metric perturbation one should add the complex conjugate. Here $\psi_{-}$ is a \emph{Hertz potential}, and $\Delta^{-2} \psi_{-}$ must satisfy the vacuum $s=-2$ Teukolsky equation (Eq.~(\ref{eq:sourced-psi4}) with $\tilde{T}_4=0$). In mode form,
\beq
\psi_{-} = \beta_{-} \, P_{-2}(r) S_{-2}(\theta) e^{i (m \phi - \omega t)} ,  \label{eq:psi-minus}
\eeq 
where $\beta_{-}$ is a constant whose value will be ascertained in the next section. 
For the GHP version of this construction, see Eq.~(60a) and (67) in Ref.~\cite{Pound:2021qin}.

The vacuum metric perturbation in outgoing radiation gauge (ORG), satisfying $h_{\mu \nu}^{\text{ORG}} l_-^\nu = 0$, takes a similar form, 
\begin{align}
\horg^{\mu \nu} &= -\frac{1}{2 \rho^2} \left[ l_-^\mu l_-^\nu \left( \cL_1 - \frac{2 \rhoc_{,\theta}}{\rhoc} \right)  \cL_2 + m_-^\mu m_-^\nu \left(\cD^\dagger - \frac{2}{\rhoc} \right) \cD^\dagger  \right. \nn \\
& \left. \quad \quad \quad \quad - 2 \, l_-^{(\mu} m_-^{\nu)}  
\left( \cD^\dagger \cL_2  - \frac{1}{\rho} \left( \cL_2 + \rho_{,\theta} \cD^\dagger \right)  - \frac{1}{\rhoc} \left( \cL_2 + \rhoc_{,\theta} \cD^\dagger \right) \right) 
 \right] \psi_{+} , \label{h-org}
\end{align}
where $\Delta^{-2} \psi_{+}$ satisfies the vacuum $s=+2$ Teukolsky equation (Eq.~(\ref{eq:sourced-psi0}) with $T_0 = 0$). In mode form,
\beq
 \psi_{+} = \beta_{+} \, P_{+2}(r) S_{+2}(\theta) e^{i (m \phi - \omega t)} .  \label{eq:psi-plus}
\eeq 
where $\beta_+$ is a constant.

These metric perturbations (\ref{h-irg}) and (\ref{h-org}) may also be written in a manifestly covariant form \cite{Aksteiner:2016pjt},
\beq
h_{\text{IRG/ORG}}^{\mu \nu} = \nabla_{(\lambda}  \rhoc^{4}\nabla_{\sigma)}  \rhoc^{-4} \cH_{\text{IRG/ORG}}^{\mu \sigma \nu \lambda}  ,
\eeq
where
\begin{align}
\cH_{\text{IRG}}^{\mu \sigma \nu \lambda} &= - \frac{1}{2 \rho^2} \psi_{-} \,  \cU^{\mu \nu} \cU^{\sigma \lambda} , \\
\cH_{\text{ORG}}^{\mu \sigma \nu \lambda} &= - \frac{1}{2 \rho^2} \psi_{+} \,  \cV^{\mu \nu} \cV^{\sigma \lambda} , 
\end{align}
and $\cU^{\mu \nu}$ and $\cV^{\mu \nu}$ are bivectors defined in Eqs.~(\ref{UVWdef}). 

\subsection{Key properties of the radiation-gauge solutions\label{subsec:rad-gauge-properties}}

By construction, the IRG/ORG metric perturbations are traceless ($h \equiv g^{\mu \nu} h_{\mu \nu} =0$).  A metric perturbation $h_{\mu \nu}$ which satisfies the vacuum field equations, and which is traceless, must necessarily also satisfy $\tensor{h}{^{\mu \nu} _{; \mu \nu}} = 0$. This is because the perturbed Ricci scalar is $\delta R = \tfrac{1}{2} \Box h + \tensor{\bar{h}}{^{\mu \nu} _{; \mu \nu}}$, and in vacuum $\delta R = 0$. 
It follows from Poincar\'e's lemma that there exists (locally) two-forms $J_{\mu \nu} = J_{[\mu \nu]}$ such that $\tensor{h}{^{\mu \nu} _{; \nu}} = - \tensor{J}{^{\mu \nu} _{; \nu}}$ (the sign is introduced here for later convenience). Below we obtain explicit forms for $J^{\mu \nu}$.
 
By direct calculation, the divergence of the IRG metric perturbation is
\beq
\nabla_\nu h^{\mu \nu}_{\text{IRG}} =  \alpha l_{+}^\mu + \beta m_{+}^\mu
\eeq
where
\begin{align}
\alpha &= -\frac{1}{4 \rho^2} \left[ 2 \left( \cD - \frac{1}{\rho} + \frac{1}{\rhoc} \right)\left( \cL_1^\dagger - \frac{2 \rhoc_{,\theta}}{\rhoc} \right)  \cL_2^\dagger \right. \nn \\
& \left. \quad\quad\quad \quad - 2 \left( \cL_1^\dagger + \frac{\rho_{,\theta}}{\rho} + \frac{\rhoc_{,\theta}}{\rhoc} \right) 
\left\{ 
\cD \cL_2^\dagger  - \frac{1}{\rho} \left( \cL_2^\dagger + \rho_{,\theta} \cD \right)  - \frac{1}{\rhoc} \left( \cL_2^\dagger + \rhoc_{,\theta} \cD \right)
\right\}
\right] \psi_- , \\
\beta &= -\frac{1}{4 \rho^2} \left[ 2 \left( \cL_2^\dagger - \frac{\rho_{,\theta}}{\rho} + \frac{\rhoc_{,\theta}}{\rhoc} \right) \left( \cD - \frac{2}{\rhoc} \right)  \cD  \right. \nn \\
 & \quad \quad \quad \quad \left. - 2 \left( \cD + \frac{1}{\rho} + \frac{1}{\rhoc} \right)
 \left\{
\cD \cL_2^\dagger  - \frac{1}{\rho} \left( \cL_2^\dagger + \rho_{,\theta} \cD \right)  - \frac{1}{\rhoc} \left( \cL_2^\dagger + \rhoc_{,\theta} \cD \right)
\right\}
 \right] 
 \psi_- .
\end{align}
After some manipulation using commutation relations for the operators, it is straightforward to show that
\begin{align}
\alpha &= \frac{\cL^\dagger_1 \gamma}{\Sigma} , & 
\beta &= - \frac{\cD \gamma}{\Sigma} , 
\end{align}
where
\beq
\gamma \equiv - \frac{1}{\rho} \left( \cL_2^\dagger - i a \sin \theta \cD \right) \psi_- .
\eeq
Hence the divergence of the ingoing radiation-gauge metric perturbation is equal to (minus) the divergence of a two-form,
$
\nabla_\nu h^{\mu \nu}_{\text{IRG}} = - \nabla_\nu J^{\mu \nu}_{\text{IRG}}
$, 
where
\beq
J^{\mu \nu}_{\text{IRG}} = \frac{\cU^{\mu \nu}}{\rho \Sigma} \left(\cL_2^\dagger - i a \sin \theta \cD \right) \psi_{-}  , \label{eq:J-irg}
\eeq
where $\cU^{\mu \nu}$ is defined in Eq.~(\ref{UVWdef}). 
By similar steps, the divergence of the outgoing radiation-gauge metric perturbation is equal to (minus) the divergence of the two-form
\beq
J^{\mu \nu}_{\text{ORG}} =  \frac{\cV^{\mu \nu}}{\rho \Sigma} \left( \cL_2 - i a \sin \theta \cD^\dagger \right) \psi_{+} .
\label{eq:J-org}
\eeq

 \subsection{Transforming the radiation-gauge vector potential to Lorenz gauge\label{subsec:toLorenz}}
Before proceeding to consider the metric perturbation, we first examine how the transformation to Lorenz gauge proceeds in the spin-one case. A vector potential $A^{\mu}_L$ that satisfies the EM field equation and the (vector) Lorenz gauge condition $\nabla_\mu A^\mu_L = 0$ generates a traceless, pure-gauge metric perturbation in (tensor) Lorenz gauge, via $h_{\mu \nu} = - \mathcal{L}_{A_L} g_{\mu \nu} = -2 A^{L}_{(\mu ; \nu)}$. The components of that metric perturbation are needed in our construction, and are listed in Sec.~\ref{sec:mp}. 

Following the adjoint construction (Sec.~\ref{sec:radiation-gauge}), a vector potential in Ingoing Radiation Gauge (IRG) is given by
\begin{align}
A_{\text{IRG}}^\mu
&= - \frac{\rhoc^2}{\sqrt{2}} \nabla_{\nu} \left( \frac{\varphi_{-}}{\rhoc \Sigma} \, \cU^{\mu \nu} \right) ,  \label{eq:A1eqnalt}
\end{align}
where $\varphi_{-}$ is a Hertz potential satisfying the vacuum $s=-1$ Teukolsky equation; in mode form, $\varphi_{-} = P_{-1}(r) S_{-1}(\theta) e^{i m \phi - i \omega t}$. This solution, first found in Ref.~\cite{Cohen:1974cm}, is listed in Table I of Ref.~\cite{Chrzanowski:1975wv}. It satisfies the vacuum electromagnetic field equation $\nabla_{\nu} ( A^{[\nu ; \mu]} ) = 0$, and the ingoing radiation-gauge condition $A_\mu l_+^\mu = 0$.  
 
 The Lorenz-gauge vector potential $A^\mu_{L}$ is found by applying a gauge transformation,
\beq
A^\mu_{L} = A^\mu_{\text{IRG}} - \nabla^\mu \upchi ,  \label{eq:AL}
\eeq
where $\chi$ is a scalar field. By taking the divergence of Eq.~(\ref{eq:AL}), the scalar field $\upchi$ must satisfy
\beq
\Box \upchi = \nabla_\mu A^\mu_{{IRG}} .
\eeq
Using (\ref{eq:A1eqnalt}), it is straightforward to show that this equation is equivalent to
\beq
\frac{1}{\Sigma} \left( 
\cD \Delta \cD^\dagger + \cL^\dagger_1 \cL - 2 i \omega \rho 
\right) \upchi =  -\frac{\sqrt{2}}{\Sigma} \left( \cL_1^\dagger - i a \sin \theta \cD \right) \varphi_{-} . \label{eq:gaugesourceEM}
\eeq
With the identities (\ref{key-identities-vector}), one can then verify that Eq.~(\ref{eq:gaugesourceEM}) has an elementary solution,
\beq
\upchi = -\frac{1}{\sqrt{2} \, i \omega} \cD \cL_1^\dagger \varphi_{-} . \label{eq:chiEM}
\eeq
Note that the gauge transformation is invalid in the static case $\omega = 0$, due to the factor of $\omega$ in the denominator of Eq.~(\ref{eq:chiEM}).

The components of the Lorenz-gauge vector $A^{\mu}_L$ are found by combining Eqs.~(\ref{eq:A1eqnalt}), (\ref{eq:AL}) and (\ref{eq:chiEM}).
A similar procedure can be applied to transform the \emph{outgoing} radiation gauge potential to Lorenz gauge.

Complementary to the above, the vector potential in Lorenz gauge has also been obtained from an ansatz in Ref.~\cite{Lunin:2017drx}, and from the Proca field in the massless limit (see Refs.~\cite{Frolov:2018ezx,Krtous:2018bvk}), and (for $\omega \neq 0$) directly from the divergence of a two-form (see Refs.~\cite{Dolan:2019hcw, Wardell:2020naz}).

 \subsection{Transforming the radiation-gauge metric perturbation to Lorenz gauge\label{sec:hLorenz}}
In this section we describe how the IRG/ORG metric perturbations of Sec.~\ref{sec:radiation-gauge} are transformed to Lorenz gauge, adding some detail to the presentation in Ref.~\cite{Dolan:2021ijg}.

We apply a gauge transformation, such that
\beq
{h}_{L}^{\mu \nu} = h_{\text{IRG}}^{\mu \nu} - \xi^{\mu ; \nu} - \xi^{\nu ; \mu} ,  \label{toLorenz}
\eeq
and we demand that the new metric perturbation is also traceless ($g_{\mu \nu} h_{L}^{\mu \nu} = 0$) and is in Lorenz gauge ($\nabla_\nu h_{L}^{\mu \nu} =0$).
The traceless condition implies that $\nabla_\mu \xi^\mu = 0$. 
Taking the divergence of (\ref{toLorenz}) yields
\beq
\Box \xi^{\mu} + \tensor{R}{^\mu _\nu} \xi^\nu = - j^\mu .  \label{eq:Boxxi}
\eeq
Here we have defined a `current' $j^\mu \equiv - \nabla_\nu h^{\mu \nu}_{\text{IRG}}$ which is conserved by the results of Sec.~\ref{subsec:rad-gauge-properties} ($\nabla_\mu j^\mu = - h^{\mu \nu}_{\text{IRG}; \mu \nu} = 0$), and which is the divergence of a two-form: $j^\mu = \nabla_\nu J^{\mu \nu}$ (see Eqs.~(\ref{eq:J-irg})). 

Now consider Maxwell's equations for a vector potential $\xi^\mu$ in vector-Lorenz gauge ($\nabla_\mu \xi^\mu = 0$),
that is,
\begin{align}
\Box \xi^\mu - \tensor{R}{^\mu _\nu} \xi^\nu &= - j^\mu . \label{eq:Maxwell}
\end{align}
In a Ricci-flat spacetime ($R_{\mu \nu} = 0$) such as Kerr spacetime, Eqs.~(\ref{eq:Boxxi}) and (\ref{eq:Maxwell}) are identical, and so we may work with Maxwell's equations, expressed in the language of forms, to seek the gauge vector $\xi^\mu$.

Informed by the work of Green and Toomani \cite{Green-talk-2021,Green-paper,Green-Toomani}, and adopting the language of forms (Sec.~\ref{subsec:forms}), we make the ansatz $\xi = \rhoc^2 \delta H - d \chi$, that is,
\beq
\xi^\mu = \rhoc^2 \tensor{H}{^{\mu \nu}}_{; \nu} - \nabla^\mu \chi .  \label{eq:xi-ansatz}
\eeq
where $H_{\mu \nu} = H_{[\mu \nu]}$ is a two-form, and $\chi$ is a scalar field (which is distinct from $\upchi$ introduced in the previous section, though it is playing an analogous role). The gradient term is included to enforce vector Lorenz gauge ($\nabla_\mu \xi^\mu = 0$). 
Then Maxwell's equations read
\beq
\delta \left( d \rhoc^2 \delta H - J + {}^\star d Y \right) = 0 ,
\eeq
where $Y$ is an arbitrary vector field, known as a gauge vector of the third kind. We will make use of the freedom afforded by $Y$ to seek to solve
\beq
d \rhoc^2 \delta H - J + {}^\star d Y = 0 .
\eeq
Following the approach in Ref.~\cite{Mustafa:1987hertz}, we choose  $Y$ to cancel out the anti-self-dual part of the equation, i.e.,
$
Y = - i \rhoc^2 \delta H 
$. 
Then the equation becomes
\beq
(1 - i \, {}^\star) d \rhoc^2 \delta H = J. \label{eq:JHertzLorenz}
\eeq

At this point, we note that $J^{\mu \nu}_{\text{IRG}}$ in Eq.~(\ref{eq:J-irg}) is self-dual. Hence we choose $H$ be self-dual also (${}^\star H = i H$), and in the general form
\beq
H^{\mu \nu} = \frac{1}{\Sigma} \left( H_U \cU^{\mu \nu} + H_W \cW^{\mu \nu} + H_V \cV^{\mu \nu} \right),
\eeq
where $H_U$, $H_W$ and $H_V$ are scalar functions to be determined. 
Note that $(1 - i \, {}^\star) \cU = 2 \cU$ and $(1 - i {}^\star) \overline{\cU} = 0$, etc. By direct calculation, the left-hand side of the equation is
\begin{align}
((1 - i {}^\star) \mathrm{d} \rhoc^2 \delta H)^{\mu \nu} &= 
\frac{\Psi}{\Sigma^2} \left( - \cU^{\mu \nu} \left\{ \Delta \cD^\dagger \rhoc^2 \cD + \cL \rhoc^2 \cL_1^\dagger \right\} H_U
- \cV^{\mu \nu} \left\{ \Delta \cD \rhoc^2 \cD^\dagger + \cL^\dagger \rhoc^2 \cL_1 \right\} H_V
 \right. \nn \\
& \left. \quad \quad \quad  - \frac{\rhoc^2}{2} \cW^{\mu \nu} \left[ \cD \rhoc^{-2} \Delta \cD^\dagger + \cD^\dagger \rhoc^{-2} \Delta \cD
 + \cL_1^\dagger \rhoc^{-2} \cL + \cL_1 \rhoc^{-2} \cL^\dagger \right]  (\rhoc^2 H_W) \right). 
\end{align}
Here, the $\cU$ and $\cV$ parts yield the $s=-1$ and $s=+1$ Teukolsky equations, and the $\cW$ part yields the Fackerell-Ipser equation \cite{Fackerell:1972hg}.
Noting that $J_{\text{IRG}}^{\mu \nu}$ in Eq.~(\ref{eq:J-irg}) has a $\cU$ part only, we can proceed by setting $H_V = H_W = 0$ and seeking a solution to
\beq
\left( \Delta \cD^\dagger \rhoc^2 \cD + \cL \rhoc^2 \cL_1^\dagger \right) H_U = - \rhoc \left( \cL_2^\dagger - i a \sin \theta \cD \right) \psi_{-}
\eeq
for $H_U$. Manipulating the left-hand side in the standard way yields a Teukolsky equation with a source term derived from the Hertz potential, viz.,
\beq
\left( \Delta \cD^\dagger \cD + \cL \cL_1^\dagger - 2 i \omega \rho \right) (\rhoc H_U) = - \left( \cL_2^\dagger - i a \sin \theta \cD \right) \psi_{-} . \label{eq:HUode}
\eeq
By use of the identities (\ref{key-identities-1}), one can show that Eq.~(\ref{eq:HUode}) has an elementary solution:
\beq
H_U = - \frac{1}{6 i \omega \rhoc } \cD \cL_2^\dagger \psi_{-} .  \label{eq:HUsolution}
\eeq

It now remains to seek the gradient piece in Eq.~(\ref{eq:xi-ansatz}) that restores $\xi^\mu$ to (vector) Lorenz gauge ($\nabla_\mu \xi^\mu = 0$). Taking the divergence of Eq.~(\ref{eq:xi-ansatz}), 
\begin{align}
\Box \chi &= 2 \rhoc \, \rhoc_{, \mu} \nabla_\nu H_U^{\mu \nu} 
=  \frac{2 \rhoc \, \rhoc_{, \mu}}{\Sigma} \left( l_+^\mu \cL_1^\dagger - m_+^\mu \cD \right) H_U , \nn\\
&=  \frac{2 \rhoc}{\Sigma} \left( \cL_1^\dagger - i a \sin \theta \cD \right) H_U  , \label{eq:Boxchi-rhs}
\end{align}
and hence
\begin{align}
 \Box \chi &=  \frac{1}{\Sigma} \left( 
\cD \Delta \cD^\dagger + \cL^\dagger_1 \cL - 2 i \omega \rho 
\right) \chi  = - \frac{1}{3 i \omega \Sigma} \left[ \cL_1^\dagger - i a \sin \theta \cD \right] \cD \cL_2^\dagger \psi_- \label{eq:Boxchi}
\end{align}
(and here using $(\cL_1^\dagger - i a \sin \theta \cD) \rhoc = 0$). 

Using the identities (\ref{key-identities-2}), it is straightforward to show that Eq.~(\ref{eq:Boxchi}) also admits an elementary closed-form solution, viz.,
\beq
\chi = \frac{1}{24 \omega^2} \cD \cD \cL_1^\dagger \cL_2^\dagger \, \psi_- . \label{eq:chi}
\eeq
Again, the vacuum Teukolsky equations (but not the Teukolsky-Starobinskii identities) were required to establish this. 

In summary, the gauge vector in (\ref{toLorenz}) and (\ref{eq:xi-ansatz}) that transforms from ingoing radiation gauge to Lorenz gauge (in vacuum), via Eq.~(\ref{toLorenz}), can be written in the form
\beq
\xi^\mu = - \rhoc^2 \nabla_{\nu} \left( \frac{ \cU^{\mu \nu}}{6 i \omega \rhoc \Sigma} \cD \cL_2^\dagger \psi_{-} \right)
 - g^{\mu \nu} \partial_\nu \left( \frac{1}{24 \omega^2}  \cD \cD \cL_1^\dagger \cL_2^\dagger \psi_{-} \right) , \label{eq:gauge-vec-irg}
\eeq
where (we remind that) $\psi_-$ satisfies the $s=-2$ vacuum Teukolsky equations and $\cU^{ab}$ was defined in Eq.~(\ref{UVWdef}). 

By a similar series of steps, the gauge vector that transforms from \emph{outgoing} radiation gauge (\ref{h-org}) to Lorenz gauge is
\beq
\xi^\mu = - \rhoc^2 \nabla_{\nu} \left( \frac{ \cV^{\mu \nu}}{6 i \omega \rhoc \Sigma} \cD^\dagger \cL_2 \psi_{+} \right)
 - g^{\mu \nu} \partial_\nu \left( \frac{1}{24 \omega^2}  \cD^\dagger \cD^\dagger \cL_1 \cL_2 \psi_{+} \right) . \label{eq:gauge-vec-org}
\eeq

 \subsection{Weyl scalars: the inverse problem and its solution\label{sec:weyl}}
In the preceding construction, it is important to note that the Hertz potentials and the Weyl scalars are distinct entities, even though in vacuum $\Delta^{-2} \psi_+$ satisfies the same equation as $\Psi_0$, and $\Delta^{-2} \psi_-$ satisfies the same equation as $\widetilde{\Psi}_4$. If one works with an IRG perturbation only, or an ORG perturbation only, then the process of deducing the correct Hertz potential from a solution for the Weyl scalars is rather involved. Moreover, strict imposition of either of the two radiation gauge conditions leads to gauge singularities in the metric perturbation. Fortunately, there is a simpler solution to this inversion problem (for $\omega \neq 0$).

The Weyl scalars associated with the IRG and ORG perturbations in Eq.~(\ref{h-irg}) and (\ref{h-org}), and hence also with their Lorenz-gauge counterparts, can be found by direct calculation, as shown in Appendix \ref{appendix:weyl}. This leads to the results summarized in Table \ref{tbl:weyl}.
\begin{table}
\begin{center}
\begin{tabular}{|c|c|c|}
\hline
& IRG  &  ORG \\
\hline
$\Psi_0 =$  & $0$ & $+6 i M \omega \Delta^{-2} \psi_+$  \\
$\widetilde{\Psi}_4 =$ & $-6 i M \omega \Delta^{-2} \psi_- $ & $0$  \\
$\Psi^\prime_0 =$  & $-\frac{1}{2} \cD \cD \cD \cD \psi_- $ & $-\frac{1}{2} \Delta^{-2} \cL_{-1} \cL \cL_1 \cL_2 \psi_+$ 
 \\
$\widetilde{\Psi}^\prime_4 =$ & $-\frac{1}{2} \Delta^{-2} \cL^\dagger_{-1} \cL^\dagger \cL^\dagger_1 \cL^\dagger_2 \psi_- $  & $-\frac{1}{2} \cD^\dagger \cD^\dagger \cD^\dagger \cD^\dagger \psi_+$ 
\\
\hline
\end{tabular}
\caption{The Weyl scalars associated with the metric perturbations (\ref{h-irg}) and (\ref{h-org}) generated by Hertz potentials $\psi_-$ and $\psi_+$. \label{tbl:weyl}} 
\end{center}
\end{table}

A simplification occurs when one considers the \emph{difference} between the IRG and ORG metric perturbations (or their Lorenz-gauge equivalents). The metric perturbation 
\beq
h_{\mu \nu}^{(-)} \equiv h_{\mu \nu}^{IRG} - h_{\mu \nu}^{ORG} \label{eq:h-minus}
\eeq 
generates the Weyl scalars
\begin{subequations}
\begin{align}
\Psi_0 &= (- 6 i M \omega \beta_+) \Delta^{-2} P_{+2} S_{+2}, \\
\widetilde{\Psi}_4 &= (- 6 i M \omega \beta_-) \Delta^{-2} P_{-2} S_{-2}. 
\end{align}
\end{subequations}
Hence, with the choice 
\beq
\beta_{\pm} = \beta \equiv 1/ (- 6 i M \omega), \label{eq:beta}
\eeq 
the Weyl scalars of Eqs.~(\ref{eq:Psi04-modes}) are recovered. In other words, with the construction (\ref{eq:h-minus}) we choose the Hertz potentials to be directly proportional to the Weyl scalars themselves: $\psi_+ = (-6 i M \omega)^{-1} \Delta^2 \Psi_0$ and $\psi_- = (-6 i M \omega)^{-1} \Delta^2 \widetilde{\Psi}_4$.

To check consistency, we can also calculate the `primed' Weyl scalars. After application of the Teukolsky-Starobinskii identities,
\begin{subequations}
\begin{align}
\Psi_0^\prime &= (-1)^{\ell + m} \Delta^{-2}  P_{+2} S_{-2}, \\
\tilde{\Psi}_4^\prime &= (-1)^{\ell + m} \Delta^{-2} P_{-2} S_{+2} .
\end{align}
\end{subequations}

It is notable that this construction breaks down in two cases: for modes of zero frequency ($\omega = 0$), and in Minkowski spacetime ($M = 0$). In this case, the gauge-invariant Weyl scalars generated by the IRG and ORG metric perturbations are identical: from Table \ref{tbl:weyl} we learn that $\Psi^{\text{IRG}}_0 = \widetilde{\Psi}^{\text{IRG}}_4 =  0 = \Psi^{\text{ORG}}_0 = \widetilde{\Psi}^{\text{ORG}}_4$ and (with $\beta_\pm = 1$ and after application of the Teukolsky-Starobinskii identities) $\Psi^{\prime \text{IRG}}_0 = \Psi^{\prime \text{ORG}}_0 = - \tfrac{1}{2 \Delta^2} A P_{+2} S_{-2}$ and $\widetilde{\Psi}^{\prime \text{IRG}}_4 = \widetilde{\Psi}^{\prime \text{ORG}}_4 = - \tfrac{1}{2 \Delta^2} A P_{-2} S_{+2}$, with $A$ as defined in Eq.~(\ref{eq:Adef}). In the case $M\omega = 0$, the metric perturbation $h^{(-)}_{\mu \nu}$ in Eq.~(\ref{eq:h-minus}) must be a pure-gauge solution (that is, generated from a gauge vector alone).

 \subsection{Modes of the Lorenz-gauge metric perturbation\label{sec:mp}}
In the previous sections, we have described how to find certain gauge vectors in a Lorenz-gauge construction. In this section we give explicit forms for the metric perturbation in terms of projections onto the unnormalised null tetrad basis.
 
  \subsubsection{Tensor modes ($s=2$)}

The gauge vector for transforming from IRG to Lorenz gauge, Eq.~(\ref{eq:gauge-vec-irg}), can be expressed as
\beq
\xi^\mu = - \frac{1}{2\Sigma} \left[(\rhoc^2 \cL_1^\dagger \alpha + \Delta \cD^\dagger \chi ) \, l_+^\mu + (\Delta \cD \chi) \, l_-^\mu  + (-\rhoc^2 \cD \alpha + \cL \chi) \, m_+^\mu + (\cL^\dagger \chi) \, m_-^\mu \right] ,
\eeq
where $\psi$ is the Hertz potential in Eq.~(\ref{eq:psi-minus}), $\chi$ was defined in Eq.~(\ref{eq:chi}), and  
\beq
\alpha =  \frac{1}{3 i \omega \rhoc} \cD \cL_2^\dagger \psi .
\eeq
(N.B.~Here the subscript $-$ associated with the IRG has been suppressed for clarity). 

The metric perturbation generated by this gauge vector is $h^{(\xi)}_{\mu \nu} = - \xi_{\mu ; \nu} - \xi_{\nu ; \mu}$, and its components can be calculated using 
\beq
{h}^{(\xi)}_{\mu \nu} l_+^\mu m_-^\nu = -l_+^\mu ( m_{-}^{\nu} \xi_\nu  )_{,\mu} -  m_-^\mu ( l_{+}^{\nu} \xi_\nu  )_{, \mu} + (l_+^\mu \nabla_\mu m_-^\nu + m_-^\mu \nabla_\mu l_+^\nu ) \xi_\nu ,
\eeq
and the spin coefficients
listed in Eq.~(\ref{eq:spin-coefficients}). The components of the metric perturbation generated by the gauge vector are listed in Eq.~(\ref{h-from-xi}) of Appendix \ref{sec:spin-coefficients}. 

The components of the (vacuum) spin-two Lorenz gauge metric perturbation derived from the IRG perturbation are
\begin{subequations}
\begin{align}
{h}_{l_+ l_+} &= 2 \cD \cD \chi , \\
{h}_{l_- l_-} &= 2 \left[ \cD^\dagger \cD^\dagger \chi + \cD^\dagger \left( \frac{\rhoc^2 \cL_1^\dagger \alpha}{\Delta} \right) - \frac{\rhoc^2}{\Delta^2} \left( \cL_1^\dagger - \frac{2 \rhoc_{,\theta}}{\rhoc} \right) \cL_2^\dagger \psi \right] , \\
{h}_{m_+ m_+} &= 2 \cL_{-1}^\dagger \cL^\dagger \chi , \\
{h}_{m_- m_-} &= 2 \left[  
\cL_{-1} \cL \chi - \cL_{-1} \left( \rhoc^2 \cD \alpha \right) - \rhoc^2 \left( \cD - \frac{2}{\rhoc} \right) \cD \psi
\right] \\
{h}_{l_+ m_+} &= \left(\cL^\dagger - \frac{2 \rho_{,\theta}}{\rho} \right) \cD \chi + \left( \cD - \frac{2}{\rho} \right) \cL^\dagger \chi , \\
{h}_{l_+ m_-} &=  - \rhoc^2 \cD \cD \alpha + 2 \cD \cL \chi - \frac{2}{\rhoc} \left( \cL + \rhoc_{,\theta} \cD \right) \chi  \\
{h}_{l_- m_+} &=  \frac{\rhoc^2}{\Delta} \cL^\dagger \cL_1^\dagger \alpha + 2 \cD^\dagger \cL^\dagger \chi - \frac{2}{\rhoc} \left(\cL^\dagger + \rhoc_{,\theta} \cD^\dagger \right) \chi  \\
{h}_{l_- m_-} &=  
2 \cL \cD^\dagger \chi - \frac{2}{\rho} \left( \cL + \rho_{,\theta} \cD^\dagger \right) \chi 
 \nn \\
&  \quad \quad \quad  +\frac{1}{\Delta} \left(\cL - \frac{2 \rho_{,\theta}}{\rho} \right) \left( \rhoc^2 \cL_1^\dagger \alpha \right) - \left(\cD^\dagger - \frac{2}{\rho} \right) \left( \rhoc^2 \cD \alpha \right) \nn \\
&   \quad \quad \quad  + \frac{2 \rhoc^2}{\Delta} \left\{   \cD \cL_2^\dagger  - \frac{1}{\rho} \left( \cL_2^\dagger + \rho_{,\theta} \cD \right)  - \frac{1}{\rhoc} \left( \cL_2^\dagger + \rhoc_{,\theta} \cD \right)  \right\} \psi  \\
\Delta {h}_{l_+ l_-} &=  (\cD^\dagger \Delta \cD + \cD \Delta \cD^\dagger) \chi - \frac{\Sigma_{,r}}{\Sigma} \Delta (\cD + \cD^\dagger) \chi + \frac{\Sigma_{,\theta}}{\Sigma} (\cL + \cL^\dagger) \chi   \nn \\
& \quad \quad \quad \quad 
+ \cD ( \rhoc^2 \cL_1^\dagger \alpha ) - \frac{\rhoc^2}{\Sigma} \left( \Sigma_{,r} \cL_1^\dagger + \Sigma_{,\theta} \cD \right) \alpha 
 , \\
{h}_{m_+ m_-} &=  (\cL_1^\dagger \cL + \cL_1 \cL^\dagger) \chi + \frac{\Sigma_{,r}}{\Sigma} \Delta (\cD + \cD^\dagger) \chi - \frac{\Sigma_{,\theta}}{\Sigma} (\cL + \cL^\dagger) \chi  \nn \\
& \quad \quad \quad  - \cL_1^\dagger (\rhoc^2 \cD \alpha) + \frac{\rhoc^2}{\Sigma} \left( \Sigma_{,r} \cL_1^\dagger + \Sigma_{,\theta} \cD \right) \alpha , 
\end{align}
\label{eq:s2-metric-components}
\end{subequations}
where $h_{l_+ l_+} \equiv h_{\mu \nu} l^\mu_+ l^\nu_+$, etc.
It is straightforward to verify that its trace is zero,
\begin{align}
h = \frac{1}{\Sigma} ( \Delta h_{l_+ l_-} + h_{m_+ m_-} ) &= \frac{1}{\Sigma} \left[ (\cD^\dagger \Delta \cD + \cD \Delta \cD^\dagger  + \cL_1^\dagger \cL + \cL_1 \cL^\dagger) \chi + \cD \left(\rhoc^2 \cL_1^\dagger \alpha \right) - \cL_1^\dagger \left( \rhoc^2 \cD \alpha \right) \right] \nn \\
&= 2 \Box \chi + \frac{2}{\rho} \left( \cL_1^\dagger - \rhoc_{,\theta} \cD \right) \alpha  = 0 ,
\end{align}
as expected.

In a similar way, one can write down the metric components of the ORG solution transformed to Lorenz gauge. As motivated in Sec.~\ref{sec:weyl}, we now construct $h_{\mu \nu}^{L(-)}$, which is defined as the difference of the IRG and ORG solutions after transforming each to Lorenz gauge separately (cf.~Eq.~(\ref{eq:h-minus})). This metric perturbation has components
\begin{subequations}
\begin{align}
 h^{L(-)}_{l_+ l_+} &=  + \frac{2 \rhoc^2}{\Delta^2} \left(\cL_1 - \frac{2 \rhoc_{,\theta}}{\rhoc} \right) \cL_2 \psi_+ + 2 \cD \cD \left( \chi_- - \chi_+ \right) - 2 \cD \rhoc^2 \Delta^{-1} \cL_1 \alpha_{+}  , \\
 h^{L(-)}_{l_- l_-} &=  - \frac{2 \rhoc^2}{\Delta^2} \left(\cL_1^\dagger - \frac{2 \rhoc_{,\theta}}{\rhoc} \right) \cL_2^\dagger \psi_-  + 2 \cD^\dagger \cD^\dagger \left( \chi_- - \chi_+ \right) + 2 \cD^\dagger \rhoc^2 \Delta^{-1} \cL^\dagger_1 \alpha_{-}  , \\
 h^{L(-)}_{m_+ m_+} &=  + 2 \rhoc^2 \left( \cD^\dagger - \frac{2}{\rhoc} \right) \cD^\dagger \psi_+  + 2 \cL_{-1}^\dagger \cL^\dagger \left(\chi_- - \chi_+ \right) + 2 \cL^\dagger_{-1} \rhoc^2 \cD^\dagger \alpha_+  , \\
 h^{L(-)}_{m_- m_-} &=  - 2 \rhoc^2 \left( \cD - \frac{2}{\rhoc} \right) \cD \psi_-  + 2 \cL_{-1} \cL \left(\chi_- - \chi_+ \right)  - 2 \cL_{-1} \rhoc^2 \cD \alpha_- , \\
 h^{L(-)}_{l_+ m_+} &=  - \frac{2 \rhoc^2}{\Delta} \left\{ \cD^\dagger \cL_2 - \tfrac{\Sigma_{,r}}{\Sigma} \cL_2 - \tfrac{\Sigma_{,\theta}}{\Sigma} \cD^\dagger  \right\} \psi_{+}  + 
\left( \cD - \frac{2}{\rho} \right) \left( \cL^\dagger (\chi_{-}  -  \chi_+) + \rhoc^2 \cD^\dagger \alpha_+ \right) \nn \\ 
& \quad \quad + \left( \cL^\dagger - \frac{2 \rho_{,\theta}}{\rho} \right)  \left(  \cD (\chi_{-}  - \chi_+) - \rhoc^2 \Delta^{-1} \cL_1 \alpha_+  \right) 
 , \\
 h^{L(-)}_{l_- m_-} &= \frac{2 \rhoc^2}{\Delta} \left\{ \cD \cL_2^\dagger - \tfrac{\Sigma_{,r}}{\Sigma} \cL_2^\dagger - \tfrac{\Sigma_{,\theta}}{\Sigma} \cD  \right\} \psi_{-}  + 
\left( \cD^\dagger - \frac{2}{\rho} \right) \left( -\rhoc^2 \cD \alpha_{-} + \cL (\chi_{-} - \chi_{+}) \right) \nn \\
& \quad \quad + \left( \cL - \frac{2 \rho_{,\theta}}{\rho} \right) \left( \rhoc^2 \Delta^{-1} \cL_1^\dagger \alpha_{-} +  \cD^\dagger (\chi_{-}  -  \chi_+) \right)
, \\
 h^{L(-)}_{l_+ m_-} &=  \left(\cD - \frac{2}{\rhoc} \right) \left(\cL (\chi_- - \chi_+) - \rhoc^2 \cD \alpha_- \right) + \left( \cL - \frac{2 \rhoc_{,\theta}}{\rhoc} \right) \left( \cD (\chi_- - \chi_+) - \rhoc^2 \Delta^{-1} \cL_1 \alpha_+ \right) , \\
 h^{L(-)}_{l_- m_+} &=  \left(\cD^\dagger - \frac{2}{\rhoc} \right) \left( \cL^\dagger (\chi_- - \chi_+) + \rhoc^2 \cD^\dagger \alpha_+ \right)  + \left( \cL^\dagger - \frac{2\rhoc_{,\theta}}{\rhoc} \right) \left( \cD^\dagger(\chi_- - \chi_+) + \rhoc^2 \Delta^{-1} \cL_1^\dagger \alpha_- \right) , \\
\Sigma \Delta h^{L(-)}_{l_+ l_-} &=  \left( \Sigma (\cD \Delta \cD^\dagger + \cD^\dagger \Delta \cD) - 2 \Delta \Sigma_{,r} \partial_r + 2 \Sigma_{,\theta} \partial_\theta \right) (\chi_- - \chi_+) \nn \\
 & \quad +   \Sigma (\cD \rhoc^2 \cL_1^\dagger \alpha_- - \cD^\dagger \rhoc^2 \cL_1 \alpha_+ ) - \rhoc^2 \Sigma_{,r} (\cL_1^\dagger \alpha_- - \cL_1 \alpha_+ ) + \rhoc^2 \Sigma_{,\theta} ( \cD^\dagger \alpha_+ - \cD \alpha_- ) , \\
 h^{L(-)}_{m_+ m_-} &= - \Delta h^{L(-)}_{l_+ l_-}   .
\end{align}
\end{subequations}
where $\psi_{\pm}$ are defined in Eq.~(\ref{eq:psi-minus}) and (\ref{eq:psi-plus}) with Eq.~(\ref{eq:beta}) and 
\begin{align}
\alpha_- &= \frac{1}{3 i \omega \rhoc} \cD \cL_2^\dagger \psi_{-} , &
\chi_- &= \frac{1}{24 \omega^2} \cD \cD \cL_1^\dagger \cL_2^\dagger \psi_{-} ,  \\
\alpha_+ &= \frac{1}{- 3 i \omega \rhoc} \cD^\dagger \cL_2 \psi_{+} , &
\chi_+ &= \frac{1}{24 \omega^2} \cD^\dagger \cD^\dagger \cL_1 \cL_2 \psi_{+} .
\end{align}

After applying the Teukolsky-Starobinskii identities (\ref{eq:teukolsky-eqns2}), we can write these components in a more compact and symmetric form,
\begin{subequations}
\begin{align}
h^{L(-)}_{l_+ l_+} &= \frac{-1}{6 \omega^2 \Delta^2} P_{+2} \left((-1)^{\ell + m} \cL_1^\dagger \cL_2^\dagger S_{-2} + \cL_1 \cL_2 S_{+2} \right)  \\
h^{L(-)}_{l_- l_-} &= \frac{-1}{6 \omega^2 \Delta^2} P_{-2} \left(\cL_1^\dagger \cL_2^\dagger S_{-2} + (-1)^{\ell + m} \cL_1 \cL_2 S_{+2} \right)  \\
h^{L(-)}_{m_+ m_+} &= \frac{-1}{6 \omega^2} \left( (-1)^{\ell + m} \cD \cD P_{-2} + \cD^{\dagger} \cD^{\dagger} P_{+2} \right) S_{+2}  \\
h^{L(-)}_{m_- m_-} &= \frac{-1}{6 \omega^2} \left( \cD \cD P_{-2} + (-1)^{\ell + m} \cD^{\dagger} \cD^{\dagger} P_{+2} \right) S_{-2}  \\
h^{L(-)}_{l_+ m_+} &= (-1)^{\ell + m} \left\{ 2 \left(\cD \cL^\dagger - \frac{1}{\rho} \cL^\dagger - \frac{\rho_{,\theta}}{\rho} \cD \right) \left( \chi_-^\prime - \chi_+^\prime \right) - \frac{\rho^2}{\Delta} \left( \Delta \cD \cD \alpha_-^\prime + \cL^\dagger \cL_1^\dagger \alpha_+^\prime \right) \right\} ,  \\
h^{L(-)}_{l_- m_-} &= (-1)^{\ell + m} \left\{ 2 \left(\cD^\dagger \cL - \frac{1}{\rho} \cL - \frac{\rho_{,\theta}}{\rho} \cD^\dagger \right) \left( \chi_-^\prime - \chi_+^\prime \right) + \frac{\rho^2}{\Delta} \left( \Delta \cD^\dagger \cD^\dagger \alpha_+^\prime + \cL \cL_1 \alpha_-^\prime \right) \right\}  \\
h^{L(-)}_{l_+ m_-} &=  \left(\cD - \frac{2}{\rhoc} \right) \left(\cL (\chi_- - \chi_+) - \rhoc^2 \cD \alpha_- \right) + \left( \cL - \frac{2 \rhoc_{,\theta}}{\rhoc} \right) \left( \cD (\chi_- - \chi_+) - \rhoc^2 \Delta^{-1} \cL_1 \alpha_+ \right) ,  \\
h^{L(-)}_{l_- m_+} &=  \left(\cD^\dagger - \frac{2}{\rhoc} \right) \left( \cL^\dagger (\chi_- - \chi_+) + \rhoc^2 \cD^\dagger \alpha_+ \right)  + \left( \cL^\dagger - \frac{2\rhoc_{,\theta}}{\rhoc} \right) \left( \cD^\dagger(\chi_- - \chi_+) + \rhoc^2 \Delta^{-1} \cL_1^\dagger \alpha_- \right) ,  \\
h^{L(-)}_{l_+ l_-} &= \frac{1}{\Delta \Sigma} \left( \Sigma (\cD \Delta \cD^\dagger + \cD^\dagger \Delta \cD) - 2 \Delta \Sigma_{,r} \partial_r + 2 \Sigma_{,\theta} \partial_\theta \right) (\chi_- - \chi_+)  \\
 & \quad + \frac{1}{\Delta \Sigma} \left( \Sigma (\cD \rhoc^2 \cL_1^\dagger \alpha_- - \cD^\dagger \rhoc^2 \cL_1 \alpha_+ ) - \rhoc^2 \Sigma_{,r} (\cL_1^\dagger \alpha_- - \cL_1 \alpha_+ ) + \rhoc^2 \Sigma_{,\theta} ( \cD^\dagger \alpha_+ - \cD \alpha_- ) \right),  \\
h^{L(-)}_{m_+ m_-} &= - \, \Delta h^{L(-)}_{l_+ l_-} ,
\end{align}
\label{eq:mp-s2}
\end{subequations}
where
\begin{align}
\psi_-^\prime &= \beta P_{-2}(r) S_{+2}(\theta) , &
\alpha_-^\prime &= \frac{1}{3 i \omega \rho} \cD \cL_2 \psi_{-}^\prime, &
\chi_-^\prime &= \frac{1}{24 \omega^2} \cD \cD \cL_1 \cL_2 \psi_{-}^\prime, \\
\psi_+^\prime &= \beta P_{+2}(r) S_{-2}(\theta) , &
\alpha_+^\prime &= \frac{1}{- 3 i \omega \rho} \cD^\dagger \cL_2^\dagger \psi_{+}^\prime, &
\chi_+^\prime &= \frac{1}{24 \omega^2} \cD^\dagger \cD^\dagger \cL_1^\dagger \cL_2^\dagger \psi_{+}^\prime . 
\end{align}
and the harmonic factor $e^{-i \omega t + i m \phi}$ is omitted for brevity. 
This is the form we shall use in the following sections.

  \subsubsection{Vector modes ($s=1$)}
A gauge vector generating a traceless spin-one Lorenz-gauge metric perturbation is
\beq
\xi^\mu_{(s=1)} = - \frac{1}{2\Sigma} \left( P_{-1} \calS \, l_+^\mu \pm P_{+1} \calS \, l_-^\mu - \calP S_{-1} \, m_+^\mu \pm \calP S_{+1} \, m_-^\mu \right)
\eeq 
where the upper (lower) sign is for even parity (odd parity), and 
\begin{align}
\calP &\equiv \cD P_{-1} \pm \cD^\dagger P_{+1} \\
\calS &\equiv \cL_1^\dagger S_{-1} \mp \cL_1 S_{+1} . 
\end{align}
The Teukolsky functions $P_{+1}(r)$, $P_{-1}(r)$, $S_{+1}(\theta)$ and $S_{-1}(\theta)$ satisfy the vacuum equations, Eqs.~(\ref{eq:teukolsky-eqns1}), and the Teukolsky-Starobinskii identities, Eqs.~(\ref{eq:TSs1}). Using Eq.~(\ref{h-from-xi}), the metric perturbation $h^{(s=1)}_{\mu \nu} = - 2 \xi^{(s=1)}_{(\mu ; \nu)}$ has the components
\begin{subequations}
\begin{align}
h^{(s=1)}_{l_+ l_+} &= \pm 2 \Delta^{-1} \cD_{-1} P_{+1} \calS  , \\
h^{(s=1)}_{l_- l_-} &= \; \; \, 2 \Delta^{-1} \cD^\dagger_{-1} P_{-1} \calS ,  \label{eq:vector-lmlm} \\
h^{(s=1)}_{m_+ m_+} &= \pm 2 \calP \cL_{-1}^\dagger S_{+1} , \\
h^{(s=1)}_{m_- m_-} &=  - 2 \calP \cL_{-1} S_{-1},  \\
h^{(s=1)}_{l_+ m_+} &= \pm \left(\cD - \frac{2}{\rho} \right) \calP S_{+1} \pm \Delta^{-1} P_{+1}  \left( \cL^\dagger -  \frac{2 \rho_{,\theta}}{\rho} \right) \calS , \\
h^{(s=1)}_{l_+ m_-} &= - \left(\cD - \frac{2}{\rhoc} \right) \calP S_{-1} \pm \Delta^{-1} P_{+1}  \left(\cL - \frac{2 \rhoc_{,\theta}}{\rhoc} \right) \calS , \\
h^{(s=1)}_{l_- m_+} &= \pm \left( \cD^\dagger - \frac{2}{\rhoc} \right) \calP S_{+1} + \Delta^{-1} P_{-1} \left(\cL^\dagger - \frac{2 \rhoc_{,\theta}}{\rhoc} \right) \calS  , \\
h^{(s=1)}_{l_- m_-} &=  - \left(\cD^\dagger - \frac{2}{\rho} \right) \calP S_{-1} + \Delta^{-1} P_{-1} \left( \cL - \frac{2 \rho_{,\theta}}{\rho} \right) \calS , \\
\Delta h^{(s=1)}_{l_+ l_-} &= \calP \calS - \frac{\Sigma_{,r}}{\Sigma} \left( P_{-1} \pm P_{+1} \right) \calS \pm \calP \frac{\Sigma_{,\theta}}{\Sigma} \left(S_{+1} \mp S_{-1} \right), \\
h^{(s=1)}_{m_+ m_-} &= -\Delta h^{(s=1)}_{l_+ l_-} . 
\end{align}
\label{eq:mp-s1}
\end{subequations}

  \subsubsection{Scalar gauge modes ($s=0$) and the trace\label{subsec:scalar}}
As shown in Ref.~\cite{Dolan:2021ijg}, a Lorenz-gauge mode with trace $h$ is generated by the gauge vector 
\beq
\xi^\mu_{(s=0)} = -\frac{1}{2 i \omega} \tilde{f}^{\mu \nu} \nabla_\nu h + 2 \nabla^\mu \kappa 
\eeq
where, in vacuum,
\begin{subequations}
\begin{align}
\Box h &= 0, \\
\Box \kappa &= \tfrac{1}{2} h ,  \label{eq:Box-kappa}
\end{align}
\end{subequations}
and the conformal Killing-Yano tensor is stated in Eq.~(\ref{eq:cky}).
This gauge vector generates a metric perturbation $h^{(s=0)}_{\mu \nu} \equiv -2 \xi_{(\mu ; \nu)}^{(s=0)}$ with metric components 
\begin{subequations}
\begin{align}
h_{l_+ l_+}^{(s=0)} &= - \left( \frac{1}{i \omega} \cD r \cD h  + 4 \cD \cD \kappa \right) , \\
h_{l_- l_-}^{(s=0)} &= -  \left( - \frac{1}{i \omega} \cD^\dagger r \cD^\dagger h  + 4 \cD^\dagger \cD^\dagger \kappa \right) , \\
h_{m_+ m_+}^{(s=0)} &= - \cL^\dagger_{-1} \left( - \frac{a \cos \theta}{\omega} \cL^\dagger h + 4 \cL^\dagger \kappa \right) , \\
h_{m_- m_-}^{(s=0)} &= - \cL_{-1} \left( + \frac{a \cos \theta}{\omega} \cL h + 4 \cL \kappa \right) , 
\\
\rho \, h_{l_+ m_+}^{(s=0)} &= - \left[ \frac{\Sigma}{2i \omega} \cD \cL^\dagger h + \frac{a}{\omega} \left(r \sin \theta \cD + \cos \theta \cL^\dagger \right) h + 4 \rho \cD \cL^\dagger \kappa - 4 \left(\cL^\dagger - i a \sin \theta \cD \right) \kappa \right] , \\
\rhoc \, h_{l_+ m_-}^{(s=0)} &= - \left[ \frac{\Sigma}{2i \omega} \cD \cL h - \frac{a}{\omega} \left(r \sin \theta \cD + \cos \theta \cL \right) h + 4 \rhoc \cD \cL \kappa - 4 \left(\cL + i a \sin \theta \cD \right) \kappa  \right] , \\
\rhoc \, h_{l_- m_+}^{(s=0)} &= - \left[ -\frac{\Sigma}{2i \omega} \cD^\dagger \cL^\dagger h + \frac{a}{\omega} \left(r \sin \theta \cD^\dagger + \cos \theta \cL^\dagger \right) h + 4 \rhoc \cD^\dagger \cL^\dagger \kappa - 4 \left(\cL^\dagger + i a \sin \theta \cD^\dagger \right) \kappa \right] , \\
\rho \, h_{l_- m_-}^{(s=0)} &= - \left[ -\frac{\Sigma}{2i \omega} \cD^\dagger \cL h - \frac{a}{\omega} \left(r \sin \theta \cD^\dagger + \cos \theta \cL \right) h + 4 \rho \cD^\dagger \cL \kappa - 4 \left(\cL - i a \sin \theta \cD^\dagger \right) \kappa \right] , \\
h_{m_+ m_-}^{(s=0)} &= - \left( \frac{a \sin \theta Q}{\omega} - 2\Sigma \right) h - 2   
\left(\cL_1 \cL^\dagger + \cL_1^\dagger \cL \right) \kappa + \frac{4}{\Sigma} \left( -\Sigma_{,r} \Delta \kappa_{,r} + \Sigma_{,\theta} \kappa_{,\theta}  \right) ,   \\
\Delta h_{l_+ l_-}^{(s=0)} &= -\left(\frac{K_r}{\omega} - 2\Sigma \right) h - 2   
\left(\cD \Delta \cD^\dagger + \cD^\dagger \Delta \cD \right) \kappa - \frac{4}{\Sigma} \left( -\Sigma_{,r} \Delta \kappa_{,r} + \Sigma_{,\theta} \kappa_{,\theta}  \right) .
\end{align}
\label{eq:mp-s0}
\end{subequations}
Combining the $l_+ l_-$ and $m_+ m_-$ components and using $K_r + a \sin \theta Q = \omega \Sigma$ and $\Box \kappa = \frac{1}{2} h$ yields
\beq
\frac{1}{\Sigma} \left( \Delta h^{(s=0)}_{l_+ l_-} + h^{(s=0)}_{m_+ m_-} \right) = h ,
\eeq
as expected.

\section{The metric perturbation for a particle on an equatorial circular orbit: radiative modes}\label{sec:radiative}

\subsection{Method}
In the preceding section, we have established that, in vacuum regions, Lorenz-gauge metric perturbations can be constructed directly from scalar potentials. More precisely, the spin-two perturbation is made from Hertz potentials $\psi_{\pm}$ that are in proportion to the required Weyl scalars $\Psi_0$ and $\widetilde{\Psi}_4$; the spin-one perturbation is made from spin-1 Hertz potentials $\varphi_{\pm}$ satisfying vacuum Maxwell equations; and the $s=0$ perturbation that is made from the trace of the metric pertubation $h$, and an auxiliary scalar $\kappa$ that is sourced by the trace. Moreover, each of these scalars $\{\psi_{\pm}, \varphi_{\pm}, h, \kappa\}$ satisfies a decoupled and separable equation (the special case of $\kappa$ is described below). Consequently, we can express these scalars in terms of mode sums of products of radial functions $P_{\pm2}(r)$, $P_{\pm1}(r)$, $h(r)$ and spin-weighted spheroidal harmonics $S_{\pm2}(\theta)$, $S_{\pm 1}(\theta)$ and $S_0(\theta)$.

In this section we seek to construct the Lorenz gauge metric perturbation in the presence of a source, specifically, a particle on a circular orbit at $r=r_0$ in the equatorial plane ($\theta_0 = \pi/2$). This particle has a worldline $x_0^\mu(\tau)$ and a tangent covector $u_{\mu} \equiv g_{\mu \nu} \tfrac{dx_0^\nu}{d\tau} = [-E, 0, 0, L]$, where $E$ and $L$ are the specific energy and angular momentum of the particle's orbit. We seek a solution to Eq.~(\ref{eq:LEE}) with the stress-energy tensor 
\begin{subequations}
\begin{align}
T_{\mu \nu} &= \mu \int_{-\infty}^\infty (-g)^{-1/2} \delta^4 (x^\mu - x_0^\mu(\tau) ) u_{\mu} u_{\nu} d\tau   \\
   &= \frac{\mu}{r_0^2 u^t} u_{\mu} u_{\nu} \delta(r - r_0) \delta(\theta - \pi/2) \delta(\phi - \Omega t) \\ 
   &= \frac{\mu}{2 \pi r_0^2 u^t} u_{\mu} u_{\nu} \sum_{m} \delta(r - r_0) \delta(\theta - \pi/2) e^{- i \omega_m t + i m \phi} ,  \label{eq:stress-energy}
\end{align}
\end{subequations}
where $\omega_m = m \Omega$,  and $\Omega$ is the angular frequency of the circular orbit. The metric perturbation can also be decomposed into $m$-modes, as
\begin{align}
h_{\mu \nu}(x) = \sum_{m} h^{(m)}_{\mu \nu}(r,\theta) e^{- i \omega_m t + i m \phi} ,
\end{align}
and henceforth we shall omit the harmonic factor for brevity.

Our central assumption is that in the vacuum regions $x^\mu \neq x^\mu_0(\tau)$ the metric perturbation can be written as a linear sum of the vacuum solutions assembled in Sec.~\ref{sec:vacuum}. Moreover, we shall assume that in Lorenz gauge a unique solution exists which satisfies the physical boundary conditions, and which is smooth everywhere except on the particle worldline.
These considerations lead us to construct a solution by `glueing' IN and UP solutions at the sphere at $r=r_0$, 
\beq
h_{\mu \nu}^{(m)}(r,\theta) = \Theta(r - r_0) h_{\mu \nu}^{\text{up}}(r,\theta) + \Theta(r_0 - r) h_{\mu \nu}^{\text{in}}(r,\theta) ,
\eeq
where $\Theta(\cdot)$ is the Heaviside step function. 
The UP (IN) solution is constructed from scalar potentials, and consequently from radial functions, that satisfy the outgoing-wave (ingoing-wave) boundary conditions at spatial infinity (at the outer horizon). The IN and UP solutions are constructed from a sum of spin-2, spin-1 and spin-0 pieces, each of which can be expressed as a mode sum. Schematically,
\beq
h_{\mu \nu}^{(m)\text{in/up}}(r,\theta) = \sum_{\ell} h^{L(-)}_{\mu \nu} [P_{\pm2}^{\Lambda}(r)] + h^{(s=1)}_{\mu \nu} [P_{\pm1}^{\Lambda}(r)] + h^{(s=0)}_{\mu \nu} [h^{\Lambda}(r), \kappa^{\Lambda}(r)] , 
\eeq
where $\Lambda = \{ \text{in/up}, \ell, m\}$, and the dependence on the spin-weighted spheroidal harmonics $S^{\Lambda}_{s}(\theta)$ has been omitted for brevity.

In the vacuum regions ($r>r_0$ and $r<r_0$), the spin-two part of the metric perturbation is determined uniquely by (the modal expansion of) the Weyl scalars $\Psi_0$ and $\widetilde{\Psi}_4$, and the spin-zero piece is partially determined by the trace $h$. To  determine the remaining degrees of freedom, we shall demand that the metric perturbation is smooth across the (topological) sphere $r=r_0$ except at the particle position ($\theta=\pi/2$, $\phi = \Omega t$), and that the field equation is satisfied. After projecting the metric perturbation on to a common basis of (spin-weighted) spherical harmonics, these conditions translate into a set of linear equations to determine the `jumps' in the radial functions. The jumps are defined by
\begin{subequations}
\begin{align}
J_0^{(\mathcal{R})} &= \lim_{\epsilon \rightarrow 0^+} \left[ \mathcal{R}^{\text{up}}(r_0 + \epsilon) - \mathcal{R}^{\text{in}}(r_0 - \epsilon) \right] ,  \label{eq:Jump0} \\
J_1^{(\mathcal{R})} &= \lim_{\epsilon \rightarrow 0^+} \left[ \partial_r \mathcal{R}^{\text{up}}(r_0 + \epsilon) - \partial_r \mathcal{R}^{\text{in}}(r_0 - \epsilon) \right] , \label{eq:Jump1}
\end{align}
\label{eq:Jumps}
\end{subequations}
where $\mathcal{R}$ is any radial function in the set $\{ P^{\Lambda}_{\pm2}(r), P^{\Lambda}_{\pm1}(r), h^\Lambda(r), \kappa^{\Lambda}(r) \}$. (N.B.~The higher-derivative jumps are not linearly independent; for $n>1$, $J_{n}^{(\mathcal{R})}$ can be written as linear combination of $J_0^{(\mathcal{R})} $ and $J_1^{(\mathcal{R})} $ by using the second-order vacuum equations.)

Given a pair $J_0^{(\mathcal{R})}$ and $J_1^{(\mathcal{R})}$, one can find unique solutions for corresponding radial functions $\mathcal{R}^{\text{up}}(r)$ and $\mathcal{R}^{\text{in}}(r)$ that satisfy the vacuum equations and the boundary conditions. Hence the jumps at $r=r_0$ are sufficient to determine the global solution for the metric perturbation.

The jump conditions for $P_{\pm2}^{\Lambda}$ are uniquely determined by the sourced Teukolsky equations. The jump conditions for $h^{\Lambda}$ are uniquely determined by the decoupled trace equation. The jump conditions for $P_{\pm1}^{\Lambda}$ and $\kappa^{\Lambda}$ are determined by the regularity requirement imposed on the metric perturbation. As we shall see, for each $\ell$-value, we have four unknowns (i.e.~the jumps in $P_{-1}$ and $\kappa$ and their first derivative, with $P_{+1}$ replaced using the vacuum Teukolsky-Starobinskii identities) but 10 continuity equations, and 10 equations for the first derivatives arising from the field equation. Consequently, the linear system for the jumps is significantly overdetermined. This means we can use a subset of 4 equations to determine the unknowns, and the remaining 16 equations as consistency checks.

We now turn attention to obtain modal solutions for the scalars that satisfy decoupled sourced equations, namely, the trace and the Teukolsky scalars.

\subsection{The trace}
In Lorenz gauge, by contracting the linearized field equation (\ref{eq:LEE}), the trace $h = g^{\mu \nu} h_{\mu \nu}$ satisfies the scalar wave equation
\beq
\Box h = 16 \pi T, \quad \quad \quad \eeq
with the source 
\beq
T \equiv \tensor{T}{^\mu _{\mu}} 
= -\frac{\mu}{2 \pi r_0^2 u^t} \sum_{m} \delta(r - r_0) \delta(\theta - \pi/2) e^{- i \omega_m t + i m \phi} . \label{eq:Ttrace}
\eeq
The trace can be expanded into a sum of $m$-modes, and $l$-modes, as follows:
\begin{subequations}
\begin{align}
h &= \sum_{m} h^{(m)}(r,\theta) e^{- i \omega_m t + i m \phi} , \label{eq:m-mode-expansion}\\
h^{(m)}(r,\theta) &= \sum_{\lmin}^\infty h_{lm}(r) S_{0lm}(\theta)  ,
\end{align} \label{eq:lmmodes}
\end{subequations}
$S_{0lm}(\theta)$ is a scalar spheroidal harmonic and $\lmin \equiv \text{max}(|m|, |s|)$.
Henceforth we drop the label $m$ for brevity. Inserting (\ref{eq:Ttrace}) and (\ref{eq:lmmodes}) into the d'Alembertian (\ref{eq:dAlembertian-alt}), one obtains a sourced radial equation,
\begin{align}
\left[ \cD \Delta \cD^\dagger - 2 i \omega r - \lambda_0 \right] h_{l}(r) &= - \frac{16 \pi \mu}{u^t} S_{0l}(\tfrac{\pi}{2}) \delta(r-r_0) ,  \label{eq:radial-h}
\end{align}
with the solution
\beq
h_{l}(r) = - \frac{16 \pi \mu}{u^t} S_{0l}(\tfrac{\pi}{2}) \frac{1}{W_0} \begin{cases}
P_{0l}^{\text{in}}(r_0) P^{\text{up}}_{0l}(r), & r \ge r_0 , \\
P_{0l}^{\text{up}}(r_0) P^{\text{in}}_{0l}(r), & r < r_0 ,
\end{cases} \label{eq:hsol}
\eeq
where $P^{\text{in}}_{0l}(r)$ and $P^{\text{up}}_{0l}(r)$ are the IN and UP homogeneous solutions that satisfy the boundary conditions are the horizon and infinity, respectively. Here, the (constant) Wronskian is
\beq
W_0 \equiv \Delta \left( P_{l}^{\text{in}} \frac{dP_{l}^{\text{up}}}{dr} - P_{l}^{\text{up}} \frac{dP_{l}^{\text{in}}}{dr} \right) .
\eeq 
Note that $S_{lm}(\pi/2) = 0$ for $l+m$ odd; hence the trace is made up of even-parity modes only.

The jumps in the radial function $h_{l}(r)$ can be deduced from Eq.~(\ref{eq:hsol}), or from the radial equation (\ref{eq:radial-h}), and they are
\begin{align}
J_0^{(h)} &= 0,  & 
J_1^{(h)} &= - \frac{16 \pi \mu}{u^t \Delta_0} S_{0\ell m}(\pi/2) .
\label{eq:jumps-h}
\end{align}
where $\Delta_0 \equiv \Delta(r_0) = r_0^2 - 2Mr_0 + a^2$. 
Conversely, from these jumps one can derive Eq.~(\ref{eq:hsol}). 

\subsubsection{Expansion of the $\kappa$ function}
Now we turn our attention to the auxiliary scalar function $\kappa$, which satisfies an equation sourced by the trace, Eq.~(\ref{eq:Box-kappa}), in the vacuum regions. As before, the expansion of $\kappa$ in $m$-modes $\kappa^{(m)}(r,\theta)$ is straightforward, following the template in Eq.~(\ref{eq:m-mode-expansion}). However, the expansion in $\ell$ modes needs a little more care. We start by writing
\begin{align}
\kappa^{(m)}(r,\theta) &= \sum_{l = 0}^\infty \kappa_{\ell}(r,\theta)  ,
\end{align}
where $\kappa_{\ell}(r,\theta)$ is the function sourced by an $\ell$-mode of the trace, satisfying 
\begin{align}
\Box \kappa_{\ell}(r,\theta) =  \tfrac{1}{2} h_{\ell}(r) S_{0\ell}(\theta) ,
\end{align}
Now we expand $\kappa_{\ell}(r,\theta)$ in the basis of scalar spheroidal harmonics, 
\beq
\kappa_\ell (r,\theta) = \sum_{j} \kappa_{\ell j}(r) S_{0j}(\theta) . \label{eq:kappa-lj-def}
\eeq
Inserting into Eq.~(\ref{eq:Box-kappa}), and moving the operator inside the mode sum, leads to
 \beq
 \sum_{j} \left(\cD \Delta \cD^\dagger - 2 i \omega r - \lambda_{(j)} \right) \kappa_{\ell j}(r) S_{0j}(\theta) = \tfrac{1}{2} (r^2 + a^2\cos^2\theta) h_{\ell}(r) S_{0\ell}(\theta) ,
\eeq
where $\lambda_{(j)}$ denotes the $s=0$ eigenvalue for mode $j$.

Now we can decompose $\cos^2 \theta S_{0\ell}$ into spheroidal harmonics as follows,
\beq
\cos^2 \theta S_{0\ell}(\theta) = \sum_{j} \gamma_{\ell j} S_{0j}(\theta) ,
\quad \quad
\gamma_{\ell j} = \oint  S_{0j}(\theta) \cos^2 \theta S_{0\ell}(\theta) d\Omega ,  \label{eq:gamma-lj}
\eeq
by making use of the orthonormality of the spheroidal basis. (Expressions for the mixing coefficients $\gamma_{\ell j}$ are given in the next section; see Eq.~(\ref{eq:gamma-lj2})). Hence we obtain a set of sourced ODEs, 
\beq
 \left(\cD \Delta \cD^\dagger - 2 i \omega r - \lambda_{(j)} \right)  \kappa_{\ell j}(r) = \tfrac{1}{2} \left(\delta_{\ell j} r^2 + a^2 \gamma_{\ell j} \right) h_{\ell}(r) . 
\eeq
For $\ell \neq j$, the equation is straightforward to solve with the ansatz $\kappa_{\ell j} = \alpha_{\ell j} h_{\ell}(r)$, where $\alpha_{\ell j}$ is a constant that is determined by inserting into the above, yielding
\begin{align}
\kappa_{\ell j} = \frac{a^2 \gamma_{\ell j}}{2 (\lambda_{(\ell)} - \lambda_{(j)})} h_{\ell}(r), \quad \quad \quad \ell \neq j. \label{eq:kappa-lj}
\end{align}
For $\ell = j$, the radial equation is
\beq
\left(\cD \Delta \cD^\dagger - 2 i \omega r  - \lambda_{(\ell)} \right) \kappa_{\ell \ell} (r) = \tfrac{1}{2} \left( r^2 + a^2 \gamma_{\ell \ell} \right) h_{\ell}(r) .  \label{eq:kappa-ll}
\eeq
Methods for numerically solving this ODE with an extended source are described in Appendix \ref{appendix:kappa}.

For $\ell \neq j$, the IN and UP modes of $\kappa_{\ell j}$ are in direct linear proportion to the IN and UP modes of $h_\ell$, by Eq.~(\ref{eq:kappa-lj}). For $\ell = j$, the IN and UP modes are determined only up to the addition of homogeneous modes. These degrees of freedom are fixed once we know the jumps in $\kappa_{\ell \ell}(r)$ and its radial derivative, denoted $J_0^{(\kappa)}$ and $J_1^{(\kappa)}$ (see Eq.~(\ref{eq:Jumps})), which are found via the numerical method in Sec.~\ref{sec:construction}. Sample data is given in Tables \ref{tbl:jumps-a06}, \ref{tbl:jumps-a0} and \ref{tbl:jumps-a099}.

 \subsection{The Teukolsky scalars}

Solutions to the sourced Teukolsky equation can be constructed from IN and UP homogeneous solutions. This was considered in e.g.~Refs.~\cite{Hughes:1999bq, Keidl:2010pm}, and an independent analysis is given in Appendix \ref{appendix:sourced-teuk}. The key result for our purposes is that the jumps in the radial function $P_{+2}(r)$ and its derivative across the particle radius $r=r_0$ 
are determined from the linear equations
\beq
\begin{pmatrix} J^{(P_{+2})}_{0} \\ J^{(P_{+2})}_{1} \end{pmatrix}
= \Delta_0 \mathfrak{D} \begin{pmatrix}
\mathfrak{C}_1 - \tfrac{\Delta_0'}{\Delta_0} \mathfrak{C}_2 \\ 
\mathfrak{C}_0 - \tfrac{(V_0 - 4)}{\Delta_0} \mathfrak{C}_2 
\end{pmatrix} . \label{eq:jumps}
\eeq
where
\begin{subequations}
\begin{align}
\mathfrak{D} &= -\frac{4\pi\mu}{r_0^2 u^t} \\
\mathfrak{C}_2 &= \mathcal{B}^2 S_0 \\
\mathfrak{C}_1 &= 2 \mathcal{B} \left[ i \mathcal{A} (S_0^\prime + Q_0 S_0) + \mathcal{B} \left(-iW_0 + \frac{1}{r_0}\right)S_0 \right] \\
\mathfrak{C}_0 &= 2 \mathcal{A}  \left[ i \mathcal{B} \left(\frac{2}{r_0} - i W_0\right) + \mathcal{A} \left(-Q_0 + \frac{ia}{r_0} \right) \right] \left(S_0^\prime + Q_0 S_0 \right) + \left[ \mathcal{A}^2 \Lambda + \mathcal{B}^2 \left(i W_0^\prime - \frac{2 i W_0}{r_0} - W_0^2 \right) \right] S_0
\end{align}
\end{subequations}
where $Q_0 = m (1 - a \Omega)$, $W_0 = m \left[ \Omega (r_0^2+a^2) - a \right] / \Delta_0$, $W_0^\prime = \partial_{r_0} W_0$, $S_0 = S_{+2lm}(\tfrac{\pi}{2})$, $S_0^\prime = \partial_\theta S_{+2lm}(\tfrac{\pi}{2})$, and
\beq
V_0 - 4 = \Delta_0 ( W_0^2 + i W_0') - i \Delta'_0 W_0 +  6 i m \Omega r - \Lambda ,
\eeq
and $\mathcal{A}$ and $\mathcal{B}$ are defined in Eq.~(\ref{eq:AB}).

To find the jumps for the $s=-2$ mode function $P_{-2}$, one can apply the vacuum Teukolsky-Starobinskii identities; in fact, it is sufficient to swap the sign of $m$ in all the terms above.

 \subsection{Construction of the metric perturbation\label{sec:construction}}
As shown above, the jumps in the radial functions for the spin-two and trace functions -- and thus the functions themselves -- are determined uniquely. On the other hand, at this point we have not determined the jumps in the radial functions $P_{\pm1 \ell}$ and $\kappa_{\ell \ell}$, and nor do we have decoupled \emph{sourced} equations for these pieces. To remedy this, we now project the metric perturbation and field equations onto a \emph{spherical} basis of spin-weighted harmonics.

To illustrate the approach, consider first one component of the stress-energy, $T_{l_+ l_+} \equiv T_{\mu \nu} l_+^\mu l_+^\nu$ in Eq.~(\ref{eq:stress-energy}), which can be straightforwardly expanded in scalar \emph{spherical} harmonics, as
\beq
\Sigma T_{l_+ l_+} = \frac{\mu}{u^t} (u \cdot l_+)^2 \sum_{\ell m} \delta(r - r_0) Y_{\ell m}(\tfrac{\pi}{2} ) Y_{\ell m}(\theta) e^{-i \omega_mt + i m\phi} , \label{eq:T-expansion}
\eeq
where $u \cdot l_+ \equiv u_{\mu} l_+^\mu$.  
In principle, the metric projection $h_{l_+ l_+}$ can also be expanded in the same basis,
\beq
h_{l_+ l_+} = \sum_{\ell m} \hrm^{\ell m}_{l_+ l_+}(r) Y_{\ell m}(\theta) e^{-i \omega_mt + i m\phi} . \label{eq:h-expansion}
\eeq
where $\hrm^{\ell m}_{l_+ l_+}(r)$ are $lm$-mode functions to be determined. 
Our first requirement is that $h_{l_+ l_+}$ is continuous on the sphere, away from the worldline; and free from distributions on the worldline (due to the hyperbolic form the field equation in Lorenz gauge). This implies that the $lm$-modes, in the spherical basis, are $C^0$ at $r=r_0$, that is,
\beq
\lim_{\epsilon \rightarrow 0^+} \left[ \hrm^{\ell m}_{l_+ l_+}(r) \right]_{r_0 - \epsilon}^{r_0 + \epsilon} = 0 . \label{eq:C0}
\eeq
A second condition is obtained from the corresponding component of the field equation. In Lorenz gauge, the component of the field equation can be written in the schematic form
\beq
\Delta \partial_r \partial_r h_{l_+ l_+} + \mathcal{O}(\partial_r) = - 16 \pi \Sigma T_{l_+ l_+} .   
\eeq
with  $\mathcal{O}(\partial_r)$ representing all first- and zero-derivative terms, as well as couplings to all other metric components. By inserting the expressions (\ref{eq:T-expansion}) and (\ref{eq:h-expansion}), and integrating from $r_0 - \epsilon$ to $r_0 + \epsilon$, we observe that
\beq
\lim_{\epsilon \rightarrow 0^+}
\left[ \partial_r \hrm^{\ell m}_{l_+ l_+}(r) \right]_{r_0 - \epsilon}^{r_0 + \epsilon} = -\frac{16 \pi \mu}{u^t \Delta_0} (u \cdot l_+)^2 Y_{\ell m}(\tfrac{\pi}{2}). \label{eq:C1}
\eeq
Similar expressions can be derived for other projected components. 

We shall consider a set of 10 tetrad components of the metric perturbation, namely,
\beq
\left[ h_{l_+ l_+}, h_{l_- l_-}, h_{m_+ m_+}, h_{m_- m_-}, \rho h_{l_+ m_+} , \rhoc h_{l_+ m_-}, \rhoc h_{l_- m_+}, \rho h_{l_- m_-},\Sigma \Delta h_{l_+ l_-}, h \right] , \label{eq:h-projections}
\eeq
where $\rho$, $\rhoc$, $\Sigma$ and $\Delta$ are as defined in Sec.~\ref{subsec:metric}.
Each tetrad component can be decomposed into $\ell m$ modes, as in Eq.~(\ref{eq:h-expansion}). The $\ell m$ modes are defined with respect to the spin-weighted \emph{spherical} basis, where the spin-weight is deduced by adding the number of $m_+$ projections and subtracting the number of $m_-$ projections; hence the spin-weights associated with Eq.~(\ref{eq:h-projections}) are $[0,0,2,-2,1,-1,1,-1,0,0]$, respectively. These $\ell m$ modes are denoted by
\begin{align}
\left[ 
\hrm_{l_+ l_+}^{\ell m}(r), \hrm_{l_- l_-}^{\ell m}(r), \hrm_{m_+ m_+}^{\ell m}(r), \hrm_{m_- m_-}^{\ell m}(r), \right. & \rho \hrm_{l_+ m_+}^{\ell m}(r), \rhoc \hrm_{l_+ m_-}^{\ell m}(r), \rhoc \hrm_{l_- m_+}^{\ell m}(r), \rho \hrm_{l_- m_-}^{\ell m}(r), \nn \\
& \left.\Sigma \Delta \hrm_{l_+ l_-}^{\ell m}(r), \hrm^{\ell m}(r) \right] . \label{eq:h-lm-modes}
\end{align}
Where necessary for clarity the superscripts $\ell m$ will be omitted.

To apply (\ref{eq:C0}) and (\ref{eq:C1}) to deduce the jumps in the radial functions $P_{\pm1}$ and $\kappa$ (for each $\ell m$), we now need a practical method for projecting metric components onto a spin-weighted spherical basis; this is addressed in the next section.
 
 \subsection{Projecting onto the spin-weighted spherical basis\label{subsec:projections}}
 
The vacuum metric perturbations of Sec.~\ref{sec:mp} are constructed from differential operators acting on radial functions and \emph{spheroidal} angular functions of spins 0, 1 and 2. Here we consider the projection of the metric perturbations onto a common basis of spin-weighted \emph{spherical} harmonics. The projection onto a spherical basis serves two purposes: first, it yields a system of linear equations for the unknown jumps. Second, it yields expressions for the $\ell m$ radial modes in Eq.~(\ref{eq:h-lm-modes}) for all 10 tetrad components of the metric perturbation.

To begin, it is well-known that the spin-weighted \emph{spheroidal} harmonics can be decomposed in a basis of spherical harmonics of the same spin-weight \cite{Hughes:1999bq}, viz.
\beq
S_{s \ell m}(\theta) = \sum_{j=\lmin}^{\infty} b^{(s)}_{ \ell j} Y_{s j m}(\theta) 
\eeq
where $\lmin \equiv \text{max}(|m|,|s|)$. Note that the spheroidal function $S_{s \ell m}(\theta)$ and the coefficients $b^{(s)}_{ \ell j}$ depend on spheroidicity parameter $a \omega$. The latter coefficients can be calculated using the {\tt SpinWeightedSpheroidalHarmonics} \cite{barry_wardell_2020_8091168} package of the Black Hole Perturbation Toolkit \cite{BHPToolkit}, using {\tt Method -> "SphericalExpansion"}. 

To determine the jumps in the $s=1$ and $s=0$ radial functions uniquely, it suffices to consider just \emph{two} tetrad components of the metric perturbation, and their radial derivatives. For $m \ge 2$, we use the $h_{l_+ l_+}$ and $h_{m_+ m_+}$ components (for $m=1$, one additional component is required, and we choose $\rho h_{l_+ m_+}$), made from a linear sum of the following: 
\begin{align}
h^{(s=2)}_{l_+ l_+} &= \frac{-1}{6 \omega^2 \Delta^2} P_{+2} \left(\pm \cL_1^\dagger \cL_2^\dagger S_{-2} + \cL_1 \cL_2 S_{+2} \right),  &
h^{(s=2)}_{m_+ m_+} &= \frac{-1}{6 \omega^2} \left( \pm \cD \cD P_{-2} + \cD^{\dagger} \cD^{\dagger} P_{+2} \right) S_{+2} . \nn \\
h^{(s=1)}_{l_+ l_+} &= \pm 2 \Delta^{-1} \cD_{-1} P_{+1} (\cL_1^\dagger S_{-1} \mp \cL_1 S_{+1})  , &
h^{(s=1)}_{m_+ m_+} &= \pm 2 \left(\cD P_{-1} \pm \cD^\dagger P_{+1} \right) \cL_{-1}^\dagger S_{+1} , \nn \\
h^{(s=0)}_{l_+ l_+} &= - \left( \frac{1}{i \omega} \cD r \cD h  + 4 \cD \cD \kappa \right) , &
h^{(s=0)}_{m_+ m_+} &= - \cL^\dagger_{-1} \left( - \frac{a \cos \theta}{\omega} \cL^\dagger h + 4 \cL^\dagger \kappa \right) . 
\label{eq:lplpmpmp}
\end{align}
Here the upper sign is for even parity, and the lower sign is for odd parity (that is, $\pm = (-1)^{\ell + m}$). The labels $\ell m$, and the mode sum, are suppressed for brevity.

Below we examine the projection of $h_{l_+ l_+}$ and $h_{m_+m_+}$ in some detail. Results for the other eight tetrad components are given in Appendix \ref{appendix:projections}.

To project modes defined in terms of spheroidal functions (of mixed spin-weights) onto a spherical basis, we make use of  \emph{mixing coefficients} with the following definitions:
\begin{subequations}
\begin{align}
(\mathfrak{s}^n)^{s_1 s_2}_{\ell_1 \ell_2} &\equiv \left< s_2 \ell_2 m \left| \sin^n (\theta) \right| s_1 \ell_1 m \right>  \equiv \oint Y_{s_2 \ell_2 m}^\ast (\theta,\phi) \, \sin^n(\theta) \, Y_{s_1 \ell_1 m} (\theta,\phi) d\Omega , \\
(\mathfrak{c}^n)^{s_1 s_2}_{\ell_1 \ell_2} &\equiv \left< s_2 \ell_2 m \left| \cos^n (\theta) \right| s_1 \ell_1 m \right>  \equiv \oint Y_{s_2 \ell_2 m}^\ast (\theta,\phi) \, \cos^n(\theta) \, Y_{s_1 \ell_1 m} (\theta,\phi) d\Omega .
\end{align} 
\label{eq:mixing-coeffs}
\end{subequations}
Note that the azimuthal number $m$ is omitted in the labelling, since there is no mixing between different $m$ modes. 
The inner products can be calculated in terms of Clebsch-Gordan coefficients $C(j_1 m_1, j_2 m_2 ; j_3 m_3)$ by following the method of Campbell and Morgan, Physica 53, 264 (1971) (see page 278). 
Below is a collection of projection coefficients that will be needed:
\begin{subequations}
\begin{align}
(\mathfrak{s})_{\ell , \ell'}^{\pm 1 , 0} \equiv \left< 0 \ell' m \left| \sin \theta \right| {\pm 1} \ell m \right>  &= \mp \sqrt{ \frac{2 (2 \ell + 1)}{2 \ell^\prime + 1} } C(\ell m,10 ; \ell^\prime m) C(\ell {\mp 1}, 1 {\pm 1} ; \ell^\prime 0) , \\
(\mathfrak{s})_{\ell , \ell'}^{\pm 2 , \pm 1} \equiv \left< \pm1 \ell' m \left| \sin \theta \right| {\pm 2} \ell m \right>  &= \mp \sqrt{ \frac{2 (2 \ell + 1)}{2 \ell^\prime + 1} } C(\ell m, 10 ; \ell^\prime m) C(\ell {\mp 2}, 1 {\pm 1} ; \ell^\prime {\mp1})   , \\
(\mathfrak{s}^2)_{\ell , \ell'}^{\pm 2 , 0} \equiv \left< 0 \ell' m \left| \sin^2 \theta \right| {\pm 2} \ell m \right>  &= \phantom{\pm} \sqrt{ \frac{8 (2 \ell + 1)}{3 (2 \ell^\prime + 1)} } C(\ell m, 20 ; \ell^\prime m) C(\ell {\mp 2}, 2 {\pm 2} ; \ell^\prime 0) , 
\end{align}
with $(\mathfrak{s}^n)^{s_1, s_2}_{l_1, l_2} = (\mathfrak{s}^n)^{s_2, s_1}_{l_2, l_1}$ (i.e.~for the matrix representation below, $\mathfrak{s}^{s_2,s_1} = (\mathfrak{s}^{s_1,s_2})^T$), 
and
\begin{align}
(\mathfrak{c})_{\ell , \ell'}^{s , s} \equiv \left< s \ell' m \left| \cos \theta \right| s \ell m \right> &= \sqrt{\frac{2 \ell + 1}{2 \ell' + 1}} C(\ell m, 1 0 ; \ell' m) C(\ell , -s, 1 0 ; \ell' , -s),  \\
(\mathfrak{c}^2)_{\ell , \ell'}^{s , s} \equiv \left< s \ell' m \left| \cos^2 \theta \right| s \ell m \right> &= \frac{1}{3} \delta_{\ell \ell'} + \frac{2}{3} \sqrt{\frac{2 \ell + 1}{2 \ell' + 1}} C(\ell m, 2 0 ; \ell' m) C(\ell , -s, 2 0 ; \ell' , -s).
\end{align}
\end{subequations}
In our usage, the $\mathfrak{s}$ coefficients couple harmonics of different spin weight, and the $\mathfrak{c}$ coefficients couple harmonics of the same spin weight. 

In this notation, the mixing coefficient in Eq.~(\ref{eq:gamma-lj}) is given by
\beq
\gamma_{\ell j} = \sum_{\ell' j'} b^{(0)}_{j j'} b^{(0)}_{\ell \ell'} (\mathfrak{c}^2)_{j' , \ell'}^{0 , 0} .  \label{eq:gamma-lj2}
\eeq
 
\subsubsection{Spin one}
As a first example, let us consider $h^{(s=1)}_{l_+ l_+}$ in Eq.~(\ref{eq:lplpmpmp}). The first task is to project the term $\cL^{\dagger}_1 S_{-1 \ell m}(\theta)$ onto the spin-zero basis. We can write the differential operators as $\cL_n = \widehat{\cL}_n - a\omega \sin \theta$ and $\cL_n^\dagger = \widehat{\cL}_n^\dagger + a\omega \sin \theta$, where $\widehat{\cL}_n$ and $\widehat{\cL}^\dagger_n$ are the spin-weight raising and lowering operators for the spin-weighted \emph{spherical} harmonics. Hence
\begin{align}
\cL^{\dagger}_1 S_{-1 \ell m}(\theta) &= \left(\widehat{\cL}^{\dagger}_1 + a \omega \sin \theta \right) \sum_{j} b^{({-1})}_{\ell j} Y_{-1jm}(\theta) \\
&= \sum_{j} \sum_{k} b^{({-1})}_{\ell j} \left( -\hat{\lambda}_{j} \delta_{j k}  + a \omega (\mathfrak{s})^{-10}_{jk} \right) Y_{0km}(\theta) .  
\end{align}
where $\hat{\lambda}_{j} = \sqrt{j (j+1)}$. This can be written in terms of matrix multiplication, 
\begin{align}
\cL^{\dagger}_1 S_{-1}(\theta) &\rightarrow \mathbf{b}_{-1} \left(- \hat{\boldsymbol{\lambda}} + a \omega (\mathfrak{s})_{-10} \right) \mathbf{Y}_0 , \\
\cL_1 S_{+1}(\theta) &\rightarrow \mathbf{b}_{+1} \left(+ \hat{\boldsymbol{\lambda}} - a \omega (\mathfrak{s})_{+10} \right) \mathbf{Y}_0 ,
\end{align}
where $\mathbf{b}_{s}$ and $\hat{\boldsymbol{\lambda}}$ and $(\mathfrak{s})_{s_1 s_2}$ are matrices with components $(\mathbf{b}_{s})_{jk} = b^{({s})}_{j k}$ and $\hat{\boldsymbol{\lambda}} =  \hat{\lambda}_{j} \delta_{jk} = \sqrt{j (j+1)}$ and $((\mathfrak{s})_{s_1 s_2})_{jk} = (\mathfrak{s})^{s_1s_2}_{jk}$, and $\mathbf{Y}_0$ is a column vector of spin-zero spherical harmonics.
Hence
\begin{align}
h^{(s=1)}_{l_+ l_+} &= \pm 2 \Delta^{-1} \sum_{\ell j k} \cD_{-1} P_{+1\ell m} 
\left[ 
 b_{\ell j}^{-1} \left( -\hat{\lambda}_{j} \delta_{j k}  + a \omega (\mathfrak{s})^{-10}_{jk} \right) 
 \mp 
 b_{\ell j}^{+1} \left( \hat{\lambda}_{j} \delta_{j k}  - a \omega (\mathfrak{s})^{+10}_{jk} \right)
  \right] 
  Y_{0km}(\theta) 
\end{align}
or, in matrix form, 
\begin{subequations}
\begin{align}
h^{(s=1)}_{l_+ l_+} &\rightarrow  2 \Delta^{-1} \left( \pm \cD_{-1} P_{+1}  \right)^T  \mathbf{S}^{(s=1)}_{l_+ l_+} \mathbf{Y}_0 , \\
\mathbf{S}^{(s=1)}_{l_+ l_+} &\equiv  - ( \mathbf{b}_{-1} \pm \mathbf{b}_{+1} ) \boldsymbol{\hat{\lambda}} + a \omega \left(\mathbf{b}_{-1} \mathfrak{s}_{-10} \pm \mathbf{b}_{+1} \mathfrak{s}_{+10}  \right)  , 
\end{align}
\end{subequations}
with the upper (lower) sign for even (odd) parity.
Schematically, this is an expression in the form
$
h =  \mathbf{R}^T \mathbf{M} \mathbf{Y}
$
where $\mathbf{R}^T$ is a row vector, $\mathbf{M}$ is a matrix, and $\mathbf{Y}$ is a column vector. The projection onto the $k$th harmonic is found from the entry in the $k$th column of the row vector $\mathbf{R}^T \mathbf{M}$. Henceforth we will adopt the matrix notation exclusively.

Following similar steps for the $m_+ m_+$ component in Eq.~(\ref{eq:lplpmpmp}), we obtain
\begin{subequations}
\begin{align}
h^{(s=1)}_{m_+ m_+} &\rightarrow \left[ \pm 2 \left(\cD P_{-1} \pm \cD^\dagger P_{+1} \right) \right]^T \mathbf{S}_{m_+m_+}^{(s=1)} \mathbf{Y}_{+2} , \\ 
\mathbf{S}_{m_+m_+}^{(s=1)}  &\equiv  \mathbf{b}_{+1} \left( - \mathbf{\hat{\boldsymbol{\lambda}}_2} + a \omega  (\mathfrak{s})_{12} \right) ,
\end{align}
\end{subequations}
where $\mathbf{Y}_{+2}$ is a column vector of spin-2 spherical harmonics.

\subsubsection{Spin two}
Following the approach in the previous section, it is straightforward to establish that
\begin{subequations}
\begin{align}
h_{l_+ l_+}^{(s=2)} &\rightarrow - \frac{1}{6 \omega^2 \Delta^2} (P_{+2})^T \mathbf{S}_{l_+ l_+}^{(s=2)} \mathbf{Y}_0 , \\
h_{m_+ m_+}^{(s=2)} &\rightarrow \frac{-1}{6 \omega^2} \left( \pm \cD \cD P_{-2} + \cD^{\dagger} \cD^{\dagger} P_{+2} \right)^T \mathbf{S}_{m_+m_+}^{(s=2)} \mathbf{Y}_{+2} ,
\end{align}
where
\begin{align}
\mathbf{S}_{l_+ l_+}^{(s=2)} &\equiv 
\mathbf{b}_{+2} \left( \boldsymbol{\hat{\Lambda}}  - 2 a \omega \hat{\boldsymbol{\lambda}}_2 \mathfrak{s}_{10} + a^2 \omega^2 (\mathfrak{s}^2)_{20} \right) 
\pm
\mathbf{b}_{-2} \left(  \boldsymbol{\hat{\Lambda}} - 2 a \omega \hat{\boldsymbol{\lambda}}_2 \mathfrak{s}_{-10} + a^2 \omega^2 (\mathfrak{s}^2)_{-20} \right)
 , \\
 \mathbf{S}_{m_+m_+}^{(s=2)} &\equiv \mathbf{b}_{+2} . 
\end{align}
\end{subequations}
with $(\hat{\boldsymbol{\lambda}}_2)_{jk} \equiv \sqrt{(j-1)(j+2)} \delta_{jk}$ and $(\hat{\boldsymbol{\Lambda}})_{jk} \equiv \sqrt{(j-1) j (j+1) (j+2)} \delta_{jk}$. 

\subsubsection{Spin zero}
Continuing in this manner,
\begin{subequations}
\begin{align}
h^{(s=0)}_{l_+ l_+} &\rightarrow - \left( \frac{1}{i \omega} \cD r \cD \mathbf{h} + (\boldsymbol{1} \cdot) 4  \cD \cD \boldsymbol{\kappa} \right)^T \mathbf{b}_{0} \mathbf{Y}_{0} , \\
h^{(s=0)}_{m_+ m_+} &\rightarrow 
\left( \frac{a}{\omega} \mathbf{h}^T \, \mathbf{S}_{m_+m_+}^{(s=0) h} - 4 \, (\boldsymbol{1} \cdot \boldsymbol{\kappa}) \, \mathbf{S}_{m_+m_+}^{(s=0) \kappa} \right) \mathbf{Y}_{+2} , 
\end{align}
where
\begin{align}
\mathbf{S}_{m_+m_+}^{(s=0) h} &\equiv \mathbf{b}_{0} \left\{ \left( \hat{\boldsymbol{\Lambda}} - 2 a \omega \hat{\boldsymbol{\lambda}} \mathfrak{s}_{12} + a^2 \omega^2 (\mathfrak{s}^2)_{02} \right) \mathfrak{c}_{2} + \left( \hat{\boldsymbol{\lambda}} \mathfrak{s}_{12} - a \omega (\mathfrak{s}^2)_{02} \right) \right\} , \\
\mathbf{S}_{m_+m_+}^{(s=0) \kappa} &\equiv \mathbf{b}_{0} \left( \hat{\boldsymbol{\Lambda}} - 2 a \omega \hat{\boldsymbol{\lambda}} \mathfrak{s}_{12} + a^2 \omega^2 (\mathfrak{s}^2)_{02} \right) .
\end{align}
\end{subequations}
Here we are interpreting $\boldsymbol{\kappa}$ as a matrix whose entries $\kappa_{\ell \ell'}(r)$ are the functions introduced in Eq.~(\ref{eq:kappa-lj-def}), and $(\boldsymbol{1} \cdot)$ as a row vector that performs the sum over $\ell$.

 \subsection{Implementation details}

 \subsubsection{Determining the jumps \label{subsec:jumps1}}
Before computing the metric perturbation itself, it is necessary to compute the jumps in the radial functions. The jumps $\{ J_{0}{(P_{-2})} , J_{1}{(P_{-2})}, J_{0}{(h)}=0, J_{1}{(h)} \}$ are known in closed form from the Teukolsky and trace equations, and the jumps $\{ J_0(P_{-1}) , J_1(P_{-1}) , J_0(\kappa) , J_1(\kappa) \}$ are to be determined from the equations below. This step is purely algebraic: it does \emph{not} require the solutions of the ODEs.

For $|m| \ge 2$, we set up the system of equations:
\begin{subequations}
\begin{align}
\left[ \hrm_{l_+ l_+}^{\ell m}(r) \right] &= 0, \\
\left[ \hrm_{m_+ m_+}^{\ell m}(r) \right] &= 0, \label{eq:hmpmp-eqn} \\
\left[ \partial_{r} \hrm_{l_+ l_+}^{\ell m}(r) \right] &= -\frac{16 \pi \mu}{u^t \Delta_0} (u \cdot l_+)^2 Y_{0 \ell m}(\tfrac{\pi}{2})  , \\
\left[ \partial_{r} \hrm_{m_+ m_+}^{\ell m}(r) \right] &= -\frac{16 \pi \mu}{u^t \Delta_0} (u \cdot m_+)^2 Y_{2 \ell m}(\tfrac{\pi}{2}) ,  \label{eq:dhmpmp-eqn}
\end{align}
\label{eq:system}
\end{subequations}
for $\ell \in \{|m|, |m|+1, \ldots , \ell_{\text{max}} \}$, where $\left[ \mathcal{X} \right] \equiv \lim_{\epsilon \rightarrow 0^+} \{ \mathcal{X}(r_0 + \epsilon) - \mathcal{X}(r_0 - \epsilon) \}$ Here, each projected component involves the set of radial functions $\{ P_{\pm2 \ell}(r), P_{\pm1 \ell}(r), h_\ell(r), \kappa_{\ell \ell}(r) \}$ and their derivatives. To proceed, we took the following steps: (i) use the vacuum Teukolsky-Starobinskii identities to replace $P_{+2 \ell}(r)$ and $P_{+1 \ell}(r)$ and derivatives with $P_{-2 \ell}(r)$ and $P_{-1 \ell}(r)$ and derivatives; (ii) use the vacuum Teukolsky and trace equations to replace the second and higher derivatives of $P_{-2 \ell}(r)$, $P_{-1 \ell}(r)$, $h_{\ell}(r)$ and $\kappa_{\ell \ell}(r)$ with the zeroth and first derivatives; (iii) replace the zeroth and first derivatives with the known jumps (for $P_{-2}$ and $h$) and the unknowns $\{J_{0}{(P_{-1})} , J_{1}{(P_{-1})} , J_{0}{(\kappa)}, J_{1}{(\kappa)}\}$ (for $P_{-1}$ and $\kappa_{\ell \ell}$); (iv) set up as a linear system in the form $\mathbf{A} \mathbf{x} = \mathbf{b}$, and solve the linear system using a standard numerical method to obtain numerical values for the unknowns for each $\ell \in |m|, \ldots, \ell_{\text{max}}$. Finally, (vi) substitute the numerical solutions into the jump conditions for other projected components to check the consistency of the results.

For $m= \pm 1$, Eqs.~(\ref{eq:hmpmp-eqn}) and (\ref{eq:dhmpmp-eqn}) are trivial for $\ell = 1$, and so we supplement the equation set with the $\ell = 1$ components of $\rho \hrm_{l_+ m_+}$ (see Appendix \ref{appendix:projections}).

\emph{Validation:} Equations (\ref{eq:system}) represent four equations selected from a set of 20, per $(\ell, \ell+1)$. We use the remaining 16 equations to numerically test the legitimacy of the found solution. We find that all equations are satisfied at the expected level (typically to one part in $10^9$ -- $10^{12}$ for a machine precision calculation, or at higher accuracy with use of extended precision variables). The numerical violations become larger as $\ell$ approaches $\ell_{\text{max}}$. This is expected, as we have truncated a formally infinite set of equations at some finite order $\ell_\text{max}$, and despite increasing diagonal-dominance of the equation set at large-$\ell$, the truncation introduces some error. To maintain accuracy we do not use $\ell$-modes near the upper bound (typically, we discard modes satisfying $|\ell_{\text{max}} - \ell| \le 6$). Having an over-determined equation set serves as a useful test of the correctness of the projections in Appendix \ref{appendix:projections}, and it was used to eliminate bugs in the implementation.

Table \ref{tbl:jumps-a06} lists sample data for the jumps in modes up to $\ell =10$, for spin parameter $a=0.6M$ and a particle on a circular orbit at $r_0 = 6M$. Further validation data is given in Appendix \ref{sec:validation}. The data was calculated by solving the linear system with $\ell_{\text{max}} = 20$, and it was verified that the digits displayed do not change for $\ell_{\text{max}} = 25$.

\begin{table}
$$
\begin{array}{|r|r|r|r|}
 \hline
 \ell &  J_{0}{(P_{-2})} & J_{1}{(P_{-2})} & J_{1}{(h)} \\
 \hline
 2 & -0.43042261-22.341128 i & -15.174806-8.2706053 i & -0.58400168 \\
 3 & 1.0526256+32.554953 i & -10.378383+11.671792 i & 0 \\
 4 & -0.62261856+26.764723 i & 68.869415+10.924759 i & 0.50580439 \\
 5 & -0.64434894-62.33966 i & 18.755688-23.575077 i & 0 \\
 6 & 0.77655791-27.268381 i & -154.34993-11.257357 i & -0.49245178 \\
 7 & 0.46717261+89.511934 i & -26.534064+34.25991 i & 0 \\
 8 & -0.82977561+27.431825 i & 270.98654+11.368203 i & 0.48769095 \\
 9 & -0.36702867-115.83169 i & 34.126348-44.545681 i & 0 \\
 10 & 0.85460983-27.506157 i & -418.74335-11.419178 i & -0.48544304 \\
 \hline
\end{array}
$$
$$
 \begin{array}{|r|r|r|r|r|}
 \hline
 \ell& J_0(P_{-1}) & J_1(P_{-1})  & J_0(\kappa) & J_1(\kappa) \\
 \hline
 2 & -312.07481+93.561129 i & 66.594648-27.213203 i & -27.672578 & 35.062091 \\
 3 & 33.123174-89.559972 i & 20.702729-123.7698 i & 0 & 0 \\
 4 & 462.32145-43.380211 i & -131.86635+58.86907 i & 23.967245 & -99.57325 \\
 5 & -34.81874+93.114665 i & -21.403331+306.30278 i & 0 & 0 \\
 6 & -642.74608+29.016451 i & 194.24183-88.019575 i & -23.334539 & 202.83435 \\
 7 & 35.31824-94.138647 i & 21.601631-569.03607 i & 0 & 0 \\
 8 & 828.76827-21.912165 i & -255.8653+116.56298 i & 23.10895 & -343.87415 \\
 9 & -35.532485+94.574579 i & -21.685544+912.40633 i & 0 & 0 \\
 10 & -1016.9056+17.632607 i & 317.17617-144.85795 i & -23.002434 & 522.58898 \\
 \hline
\end{array}
$$
\caption{Jumps in radial functions at $r=r_0=6M$ for $a=0.6M$, $m=2$. Upper: Determined from Weyl scalars and the trace. Lower: determined from regularity on the sphere. Here $M=1$.}
\label{tbl:jumps-a06}
\end{table}

 \subsubsection{Radial functions}

To construct the metric perturbation itself, we require IN and UP solutions for a set of functions across the domain in $r$. The in/up functions for $P_{-2 \ell}(r)$, $P_{-1 \ell}(r)$ and $h_{\ell}(r)$, which satisfy homogeneous Teukolsky equations of spin $-2$, $-1$ and $0$ respectively, were calculated using {\tt TeukolskyRadial} function of the {\tt Teukolsky} package \cite{barry_wardell_2020_7037857} 
in the {\tt Black Hole Perturbation Toolkit} \cite{BHPToolkit} for Mathematica. The functions $\kappa_{\ell j}(r)$ for $\ell \neq j$ are proportional to $h_\ell(r)$: see Eq.~(\ref{eq:kappa-lj}). Finally, the function $\kappa_{\ell \ell}(r)$ satisfies an ODE with an extended source, Eq.~(\ref{eq:kappa-ll}). Two methods were used for calculating $\kappa_{\ell \ell}(r)$ numerically, which are detailed in Appendix \ref{appendix:kappa}.

\subsection{Results: $\ell m$-modes of the metric perturbation\label{subsec:results-lm}}

In this section we present a sample of numerical results for the metric perturbation, and we compare with other data sets. The  $\ell m$ radial modes for the 10 metric projections listed in Eq.~(\ref{eq:h-lm-modes}) are extracted using the projections described in Sec.~\ref{sec:construction} and Appendix~\ref{appendix:projections}.

\subsubsection{Schwarzschild $\ell$ modes\label{subsec:schw}}

In the non-spinning case ($a=0$), the $\ell m$ radial modes obtained with our new method can be directly compared with validation data from well-established methods for computing the metric perturbation in Lorenz gauge on Schwarzschild spacetime \cite{Barack:2005nr,Barack:2007tm,Barack:2009ux,Warburton:2011fk,Akcay:2010dx,Akcay:2013wfa}. 

\begin{figure}
\begin{center}
 \includegraphics[width=18cm]{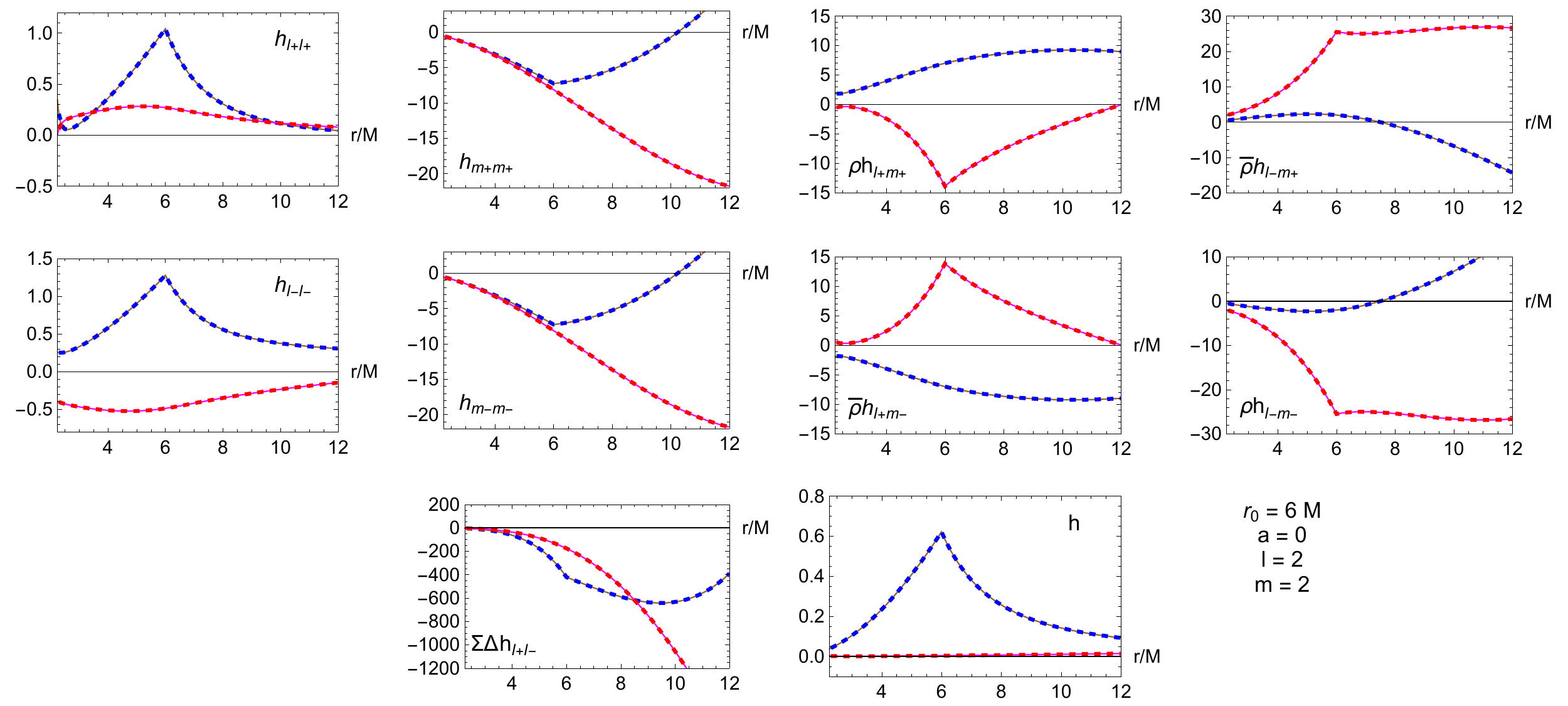} \\
 \includegraphics[width=10cm]{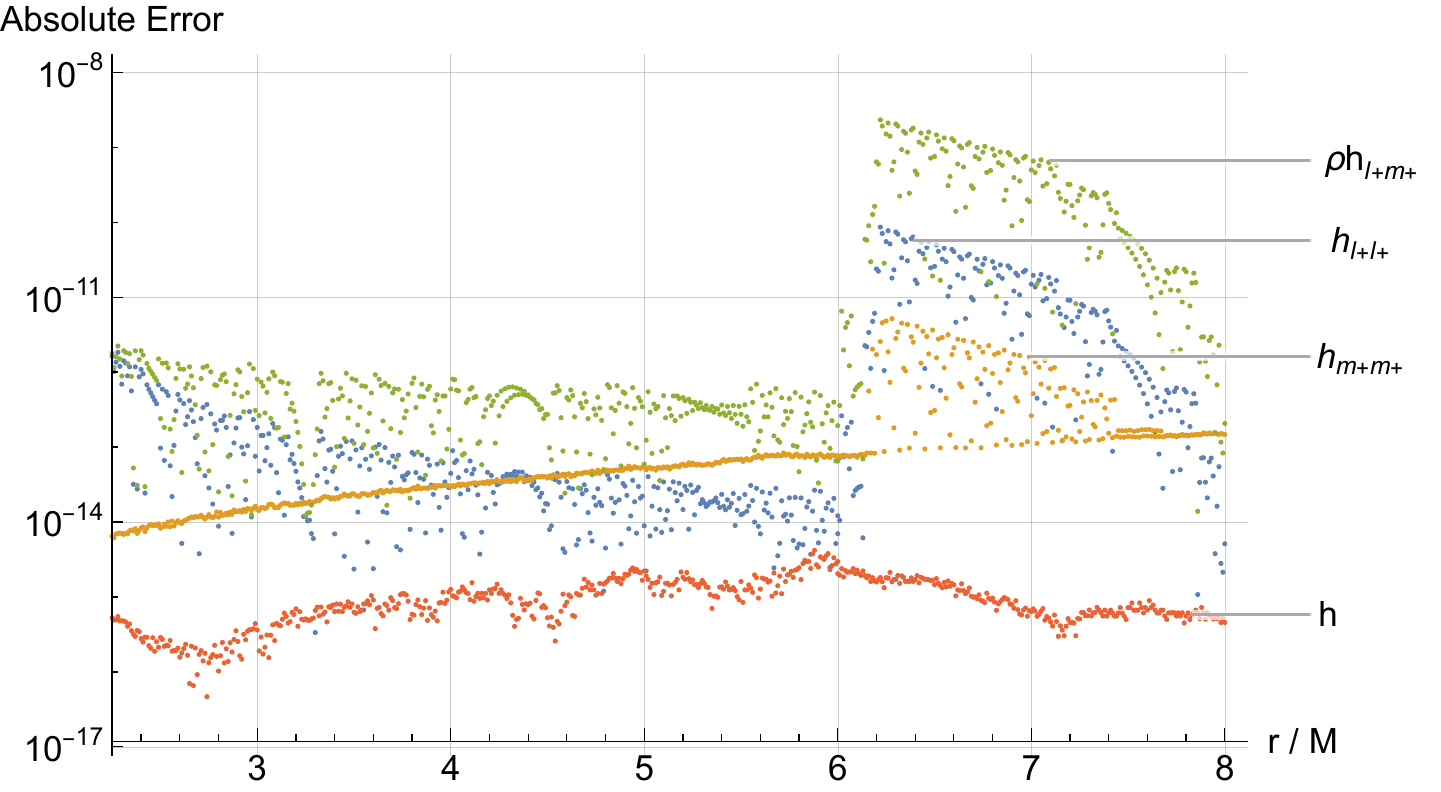}
\end{center}
 \caption{\emph{Upper:} A radial $\ell m$ mode of the metric projections of the Schwarzschild black hole, for $r_0 = 6M$ and $\ell = m = 2$. The mode extracted from the new method [dashed] is compared with validation data from an existing Schwarzschild code [solid]. The real [blue/brown] and imaginary [red/magenta] parts are shown for the 10 projections in Eq.~(\ref{eq:h-projections}).
 \emph{Lower:} The absolute value of the difference between the new data and the validation data, for four selected components of the metric perturbation.
 }
 % Plots generated by SchwarzschildComparison_New.nb
 \label{fig:schw-m2a}
\end{figure}

Figure \ref{fig:schw-m2a} shows a direct comparison for the mode $m=2$, $l=2$. Good agreement is found between the new results [dashed] and the validation data [solid] for all ten metric projections. Several other $\ell m$ modes were checked in this way (including $\ell = 4, 8, 12$ for $m=2$, and $m=4, 5$ for $\ell = 5$). The validation data was provided in the form of Barack-Lousto variables \cite{Barack:2005nr}, $\hb{i}(r)$ (with $i \in \{1\ldots10\}$), as defined with the conventions in Eq.~(16)--(18) of Ref.~\cite{Barack:2007tm}. The 10 metric projections (solid lines) shown in Fig.~\ref{fig:schw-m2a} were calculated from linear combinations of the Barack-Lousto variables, using Eq.~(\ref{eq:hBL}) of Appendix \ref{sec:BL}. 

The lower plot in Fig.~\ref{fig:schw-m2a} shows that the difference between the new data and the validation data is acceptably small across the radial domain. This absolute difference can be reduced further by changing internal parameters of the code (at the expense of run time). This suggests that the error is in the new data rather than the validation data (as expected). Moreover, the plot shows that the error in the UP solution ($r > r_0$) is typically greater than the error in the IN solution ($r < r_0$). 

In the Schwarzschild case, the projections of the metric perturbation in (\ref{eq:h-projections}) can easily be recast into Barack-Lousto variables, using Eq.~(\ref{eq:hBL-inverse}). An advantage of the BL representation is that it makes clear that for  $\ell + m$ even, the Lorenz-gauge metric is described by seven even-parity degrees of freedom ($i = 1 \ldots 7$), and for $\ell + m$ odd, in terms of three odd-parity degrees of freedom ($i = 8 \ldots 10$). Example radial profiles of the BL variables are shown in Appendix \ref{sec:BL}. In fact, the split by parity extends to the Kerr case, as we describe below.

  \subsubsection{Symmetries and parity}
The $\ell m$ modes of the metric perturbation exhibit certain symmetries after projection onto the spin-weighted spherical basis). For $\ell + m$ even,
\begin{subequations}\label{eq:symmetries}
\begin{align} 
\hrmlm_{m_+ m_+} = \hrmlm_{m_- m_-} , \quad \rho \hrmlm_{l_+ m_+} = - \rhoc \hrmlm_{l_+ m_-}, \quad \rhoc \hrmlm_{l_- m_+} = - \rho \hrmlm_{l_- m_-}.
\end{align}
For $\ell + m$ odd,
\begin{align}
\hrmlm_{l_+ l_+} = \hrmlm_{l_- l_-} = \hrmlm_{l_+ l_-} = \hrmlm_{m_+ m_-} = \hrmlm = 0, \quad \nn 
\hrmlm_{m_+ m_+} = - \hrmlm_{m_- m_-} , \\ \rho \hrmlm_{l_+ m_+} = \rhoc \hrmlm_{l_+ m_-}, \quad \rhoc \hrmlm_{l_- m_+} = \rho \hrmlm_{l_- m_-}.
\end{align}
\end{subequations}
These symmetries hold in the Kerr case, as well as the Schwarzschild case, and they are readily apparent in Figs.~\ref{fig:schw-m2a}, \ref{fig:comparison-m1}, \ref{fig:comparison-m2}, \ref{fig:comparison-m2-l34}, \ref{fig:comparison-m1-all} and \ref{fig:comparison-m2-all}.
Hence there are seven effective degrees of freedom for even-parity $\ell m$ modes, and three effective degrees of freedom for odd-parity $\ell m$ modes. In principle, one could use this property to construct a Kerr version of the Barack-Lousto variables, by extending Eqs.~\ref{eq:hBL-inverse}; this approach is not pursued further here. 

  \subsubsection{Kerr $\ell$ modes}

In the spinning case ($a \neq 0$), the results of the new method can be compared with the results of an existing 2+1D time-domain code that uses the effective source method at 4th order \cite{Dolan:2012jg,Isoyama:2014mja,Dolan-Barack-Wardell}. The time-domain code was run with a grid spacing of $\Delta r_\ast = M/12$ for $t_{\text{max}} = 250M$. The $\ell m$ profiles were extracted by projecting data in the $(r,\theta)$ domain onto spin-weighted spherical harmonics numerically. We made a comparison at $a=0.6M$, which is a parameter for which both methods are expected to perform robustly. Through this comparison, confidence increases that both methods are correctly implemented and complementary. 

\begin{figure}
\begin{center}
 \includegraphics[width=8cm]{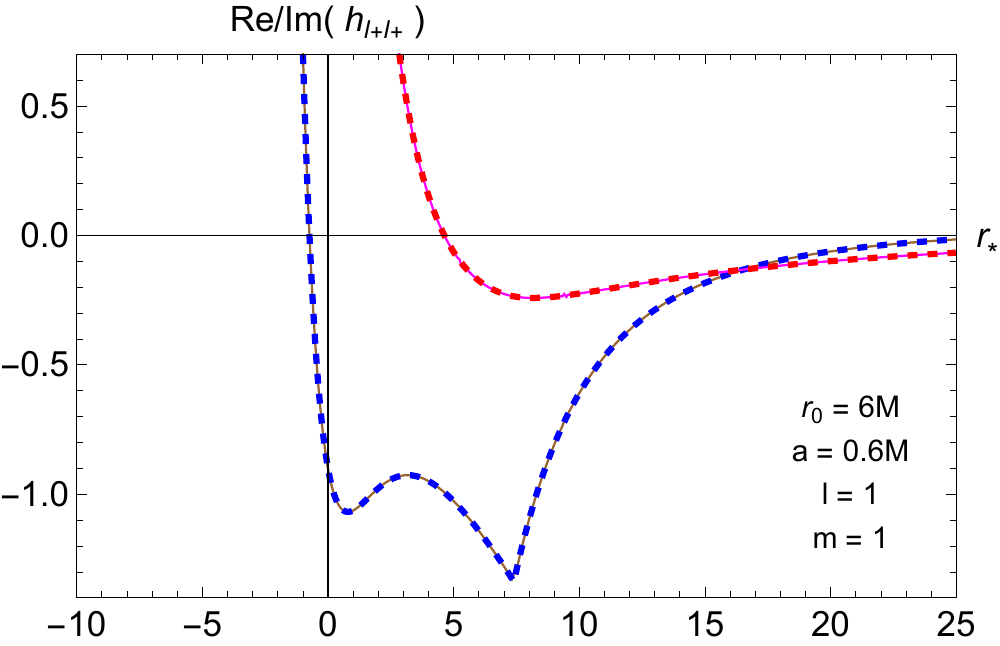} 
 \includegraphics[width=8cm]{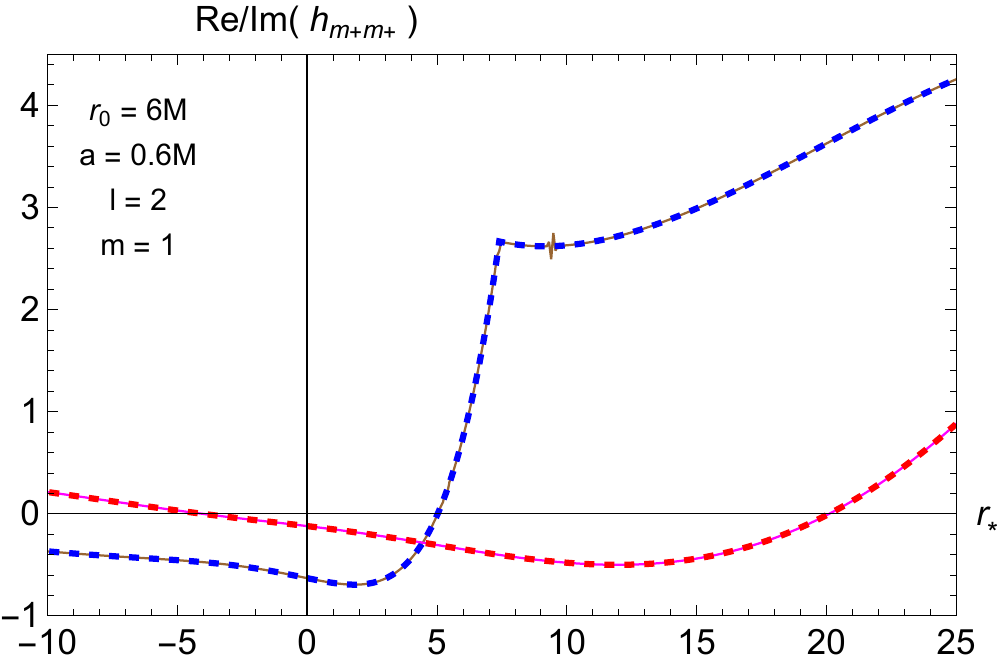}  \\
 \includegraphics[width=8cm]{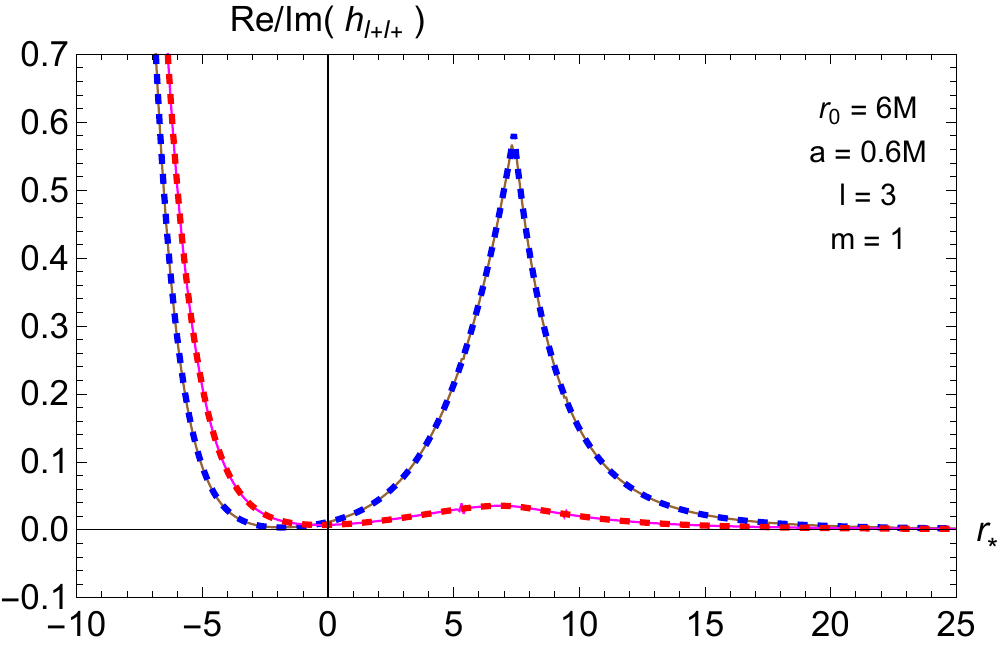}  
 \includegraphics[width=8cm]{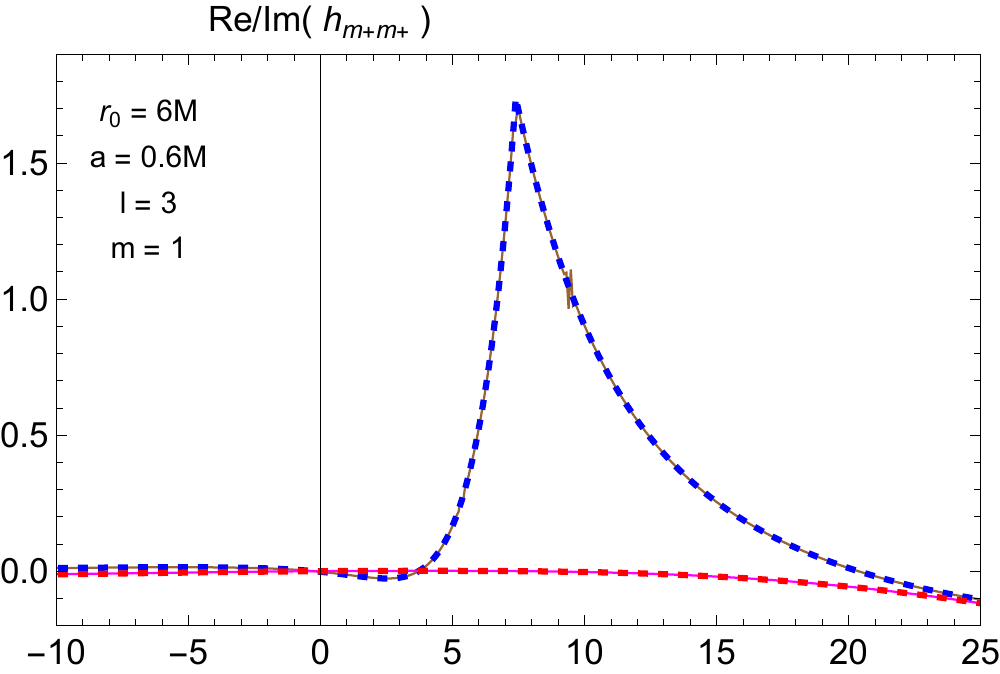} 
\end{center}
 \caption{A comparison between $lm$ modes extracted from the 2+1D time domain code [solid] and from the new separation of variables method [dashed], for $m=1$, $a=0.6M$ and $r_0 = 6M$ (see Fig.~\ref{fig:comparison-m2} for $m=2$). 
 Top left: $\ell = 1$, $h_{l_+l_+}$. Top right: $\ell = 2$, $h_{m_+m_+}$. Bottom: $\ell = 3$.
 N.B.~The glitch visible in the 2+1D data in $h_{m_+m_+}$ is associated with the boundary of the worldtube. 
 }
 % m-mode-wrangler-n12.nb
 \label{fig:comparison-m1}
\end{figure}

\begin{figure}
\begin{center}
 \includegraphics[width=8cm]{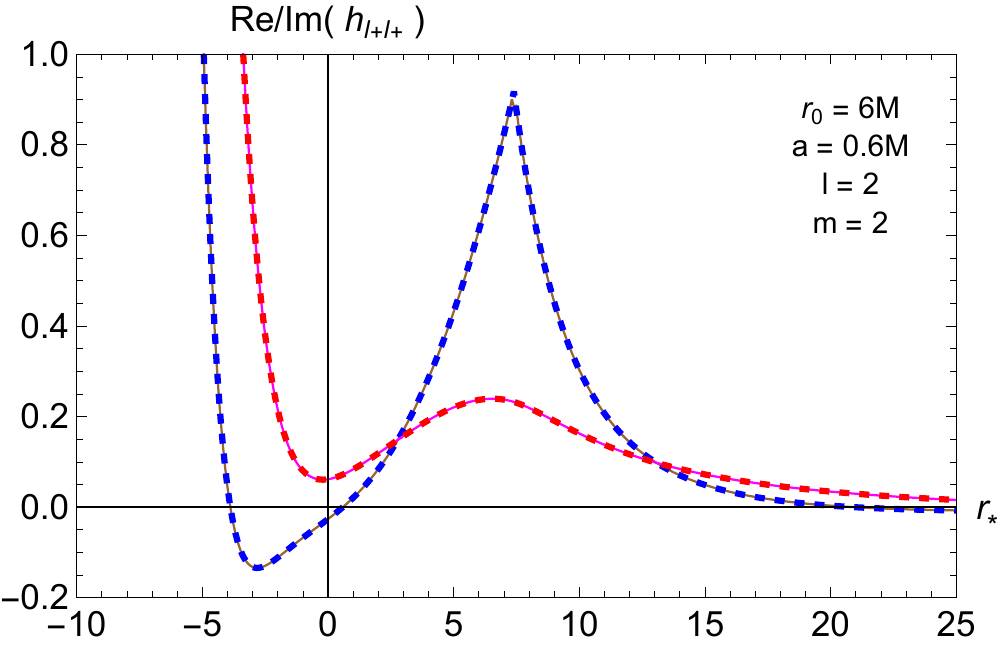} 
 \includegraphics[width=8cm]{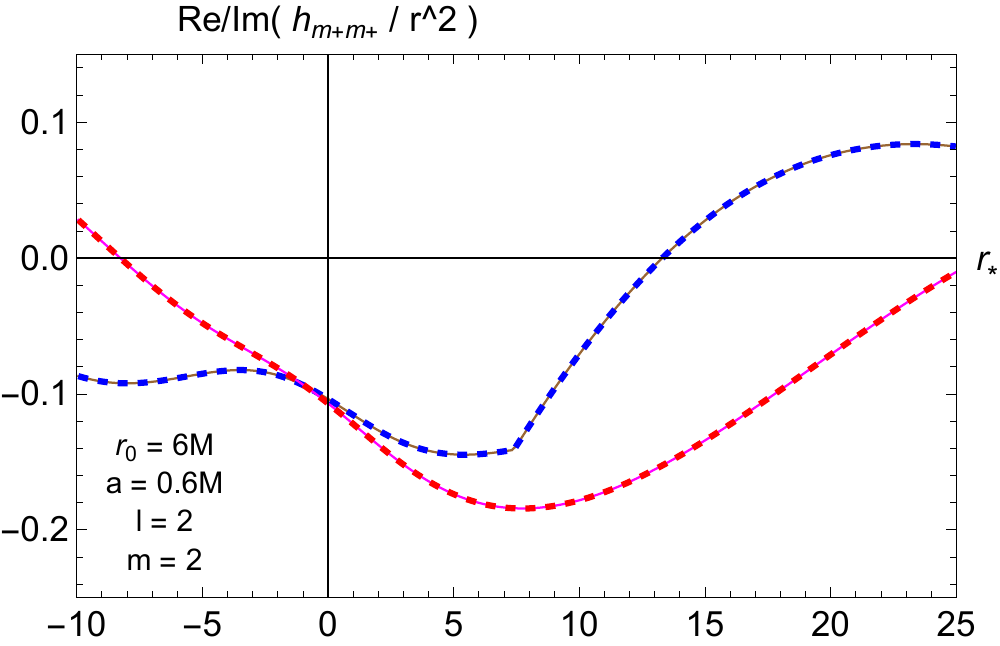}  \\
 \includegraphics[width=8cm]{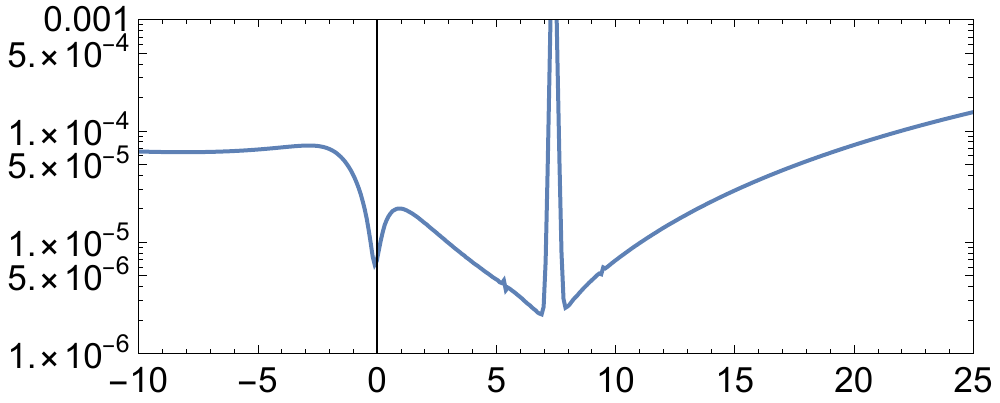}
 \includegraphics[width=8cm]{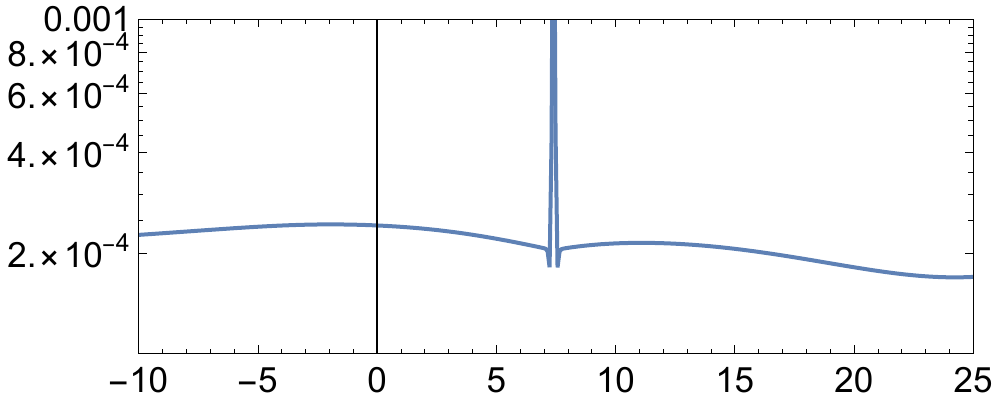} 
\end{center}
 \caption{A comparison between $lm$ modes extracted from the 2+1D time domain code [solid] and from the new separation of variables method [dashed], for $m=2$, $\ell - 2$, $a=0.6M$ and $r_0 = 6M$ (see Fig.~\ref{fig:comparison-m2} for $m=1$).
Left: $h_{l_+l_+}$. Right: $h_{m_+m_+}$.
The lower plots show the relative error (i.e.~the fractional difference between the two data sets). 
}
 \label{fig:comparison-m2}
\end{figure}

\begin{figure}
\begin{center}
 \includegraphics[width=8cm]{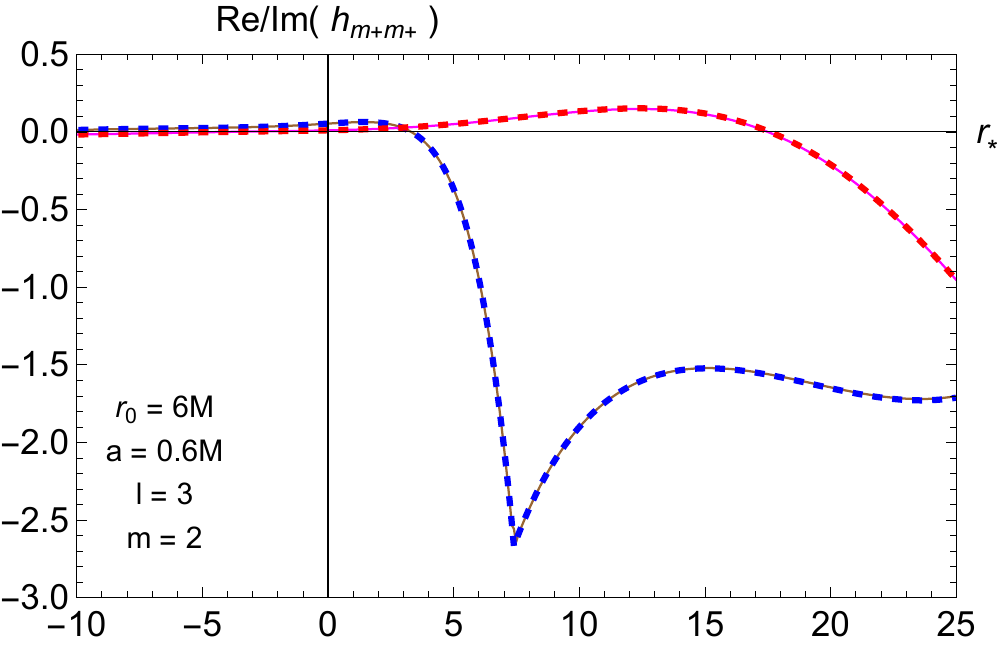}  \\
 \includegraphics[width=8cm]{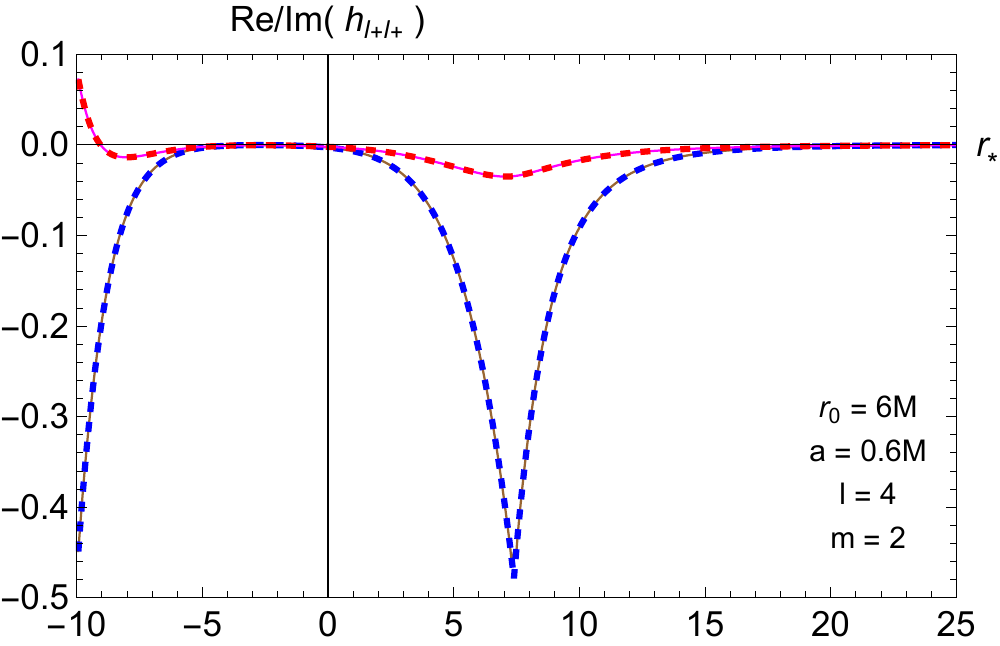} 
 \includegraphics[width=8cm]{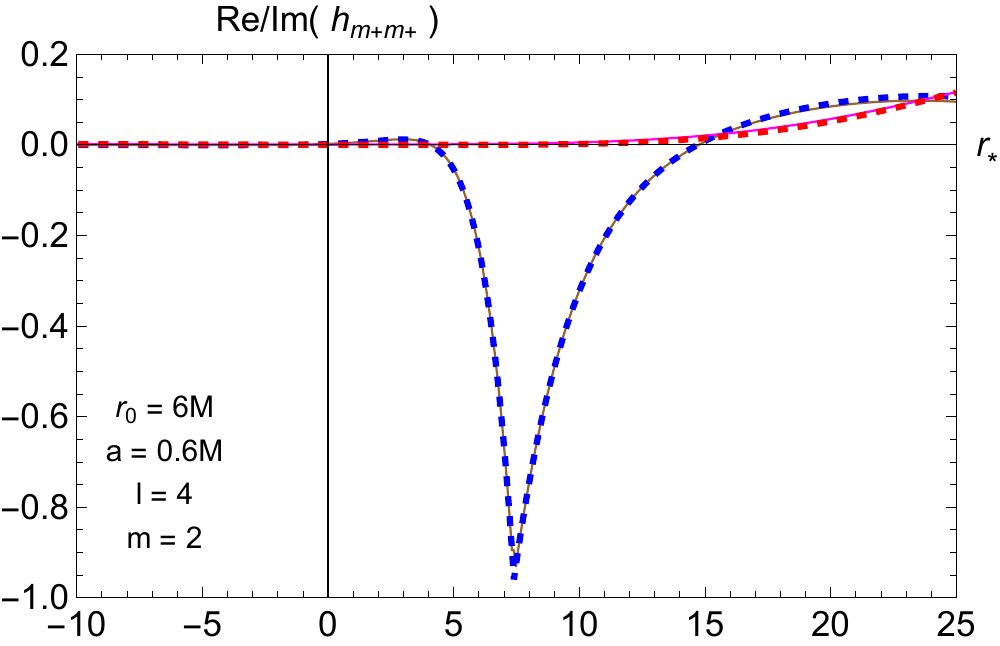}  
\end{center}
 \caption{As in Fig.~\ref{fig:comparison-m2}, but comparing the $\ell = 3$ (top) and $\ell = 4$ (bottom) modes.}
 \label{fig:comparison-m2-l34}
\end{figure}

Figure \ref{fig:comparison-m1} shows the components $\hrmlm_{l_+ l_+}$ and $\hrmlm_{m_+ m_+}$ for $m=1$ and $\ell = 1,2,3$. Figures \ref{fig:comparison-m2} and \ref{fig:comparison-m2-l34} show the same components for $m =2$ and $\ell = 2,3,4$. Robust agreement is found in every component, in every case checked. The profile of the relative error (i.e.~the absolute value of the fractional difference of the two data sets) is shown in the lower plots of Fig.~\ref{fig:comparison-m2}. The spike near $r=r_0$ is due to a limitation of the 2+1D data set: the puncture diverges as $r \rightarrow r_0$, $\theta \rightarrow \pi/2$ and is ill-defined in this limit. There is also a glitch associated with the boundary of the world tube in the 2+1D data. Over the rest of the domain, it appears that the residual difference is due to discretization error in the 2+1D (old) data, and the projecting the 2+1D data onto the spin-weighted spherical basis (particularly at higher $\ell$). We have verified that the level of agreement improves when using higher-resolution 2+1D data for which the discretization error in the finite-difference method is lessened.

\begin{figure}
\begin{center}
 \includegraphics[width=12cm]{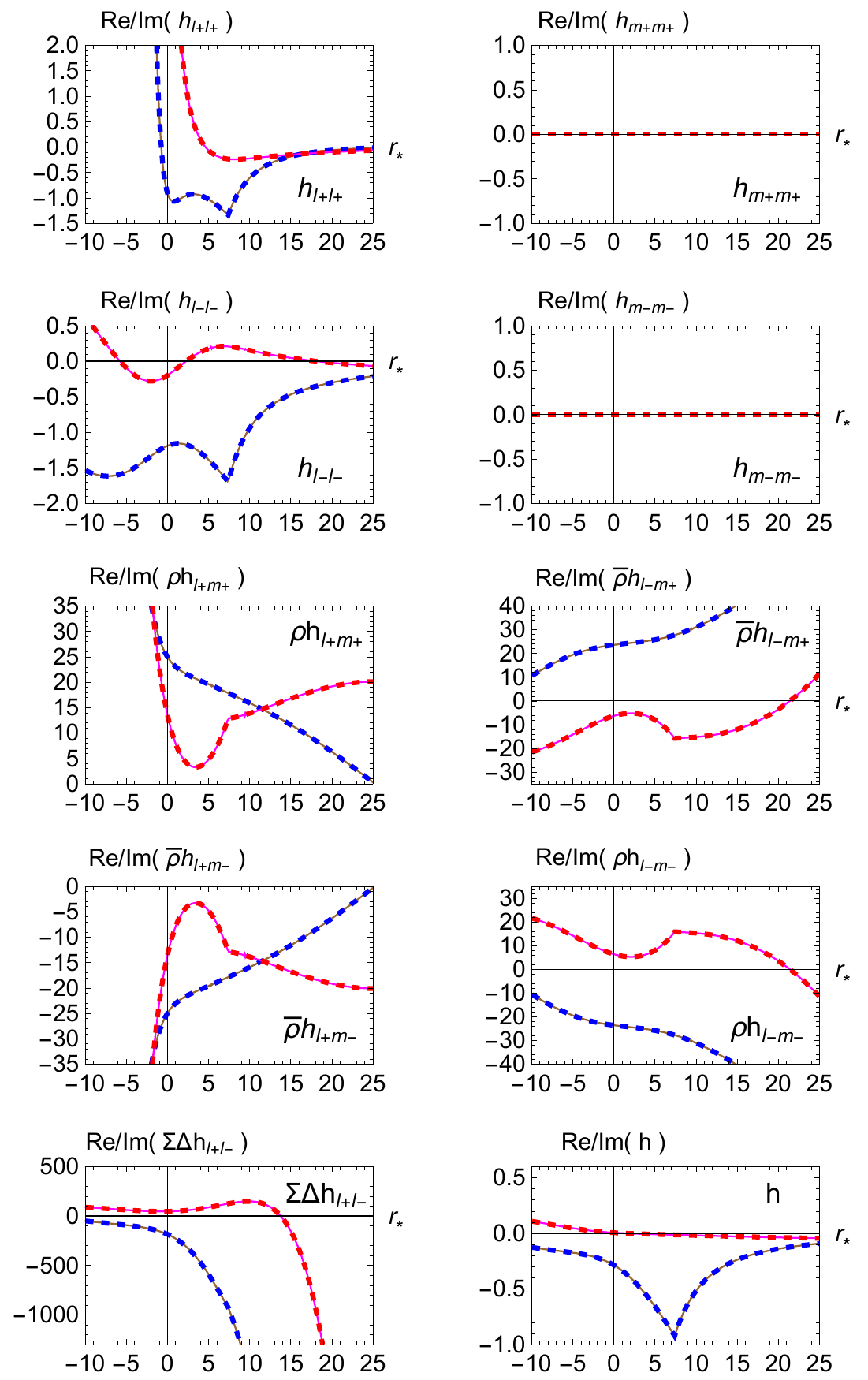} 
\end{center}
 \caption{A comparison between $lm$ modes extracted from the 2+1D time domain code [solid] and from the new separation of variables method [dashed], for $l = m=1$, $a=0.6M$ and $r_0 = 6M$ for all components.}
 \label{fig:comparison-m1-all}
\end{figure}

\begin{figure}
\begin{center}
 \includegraphics[width=12cm]{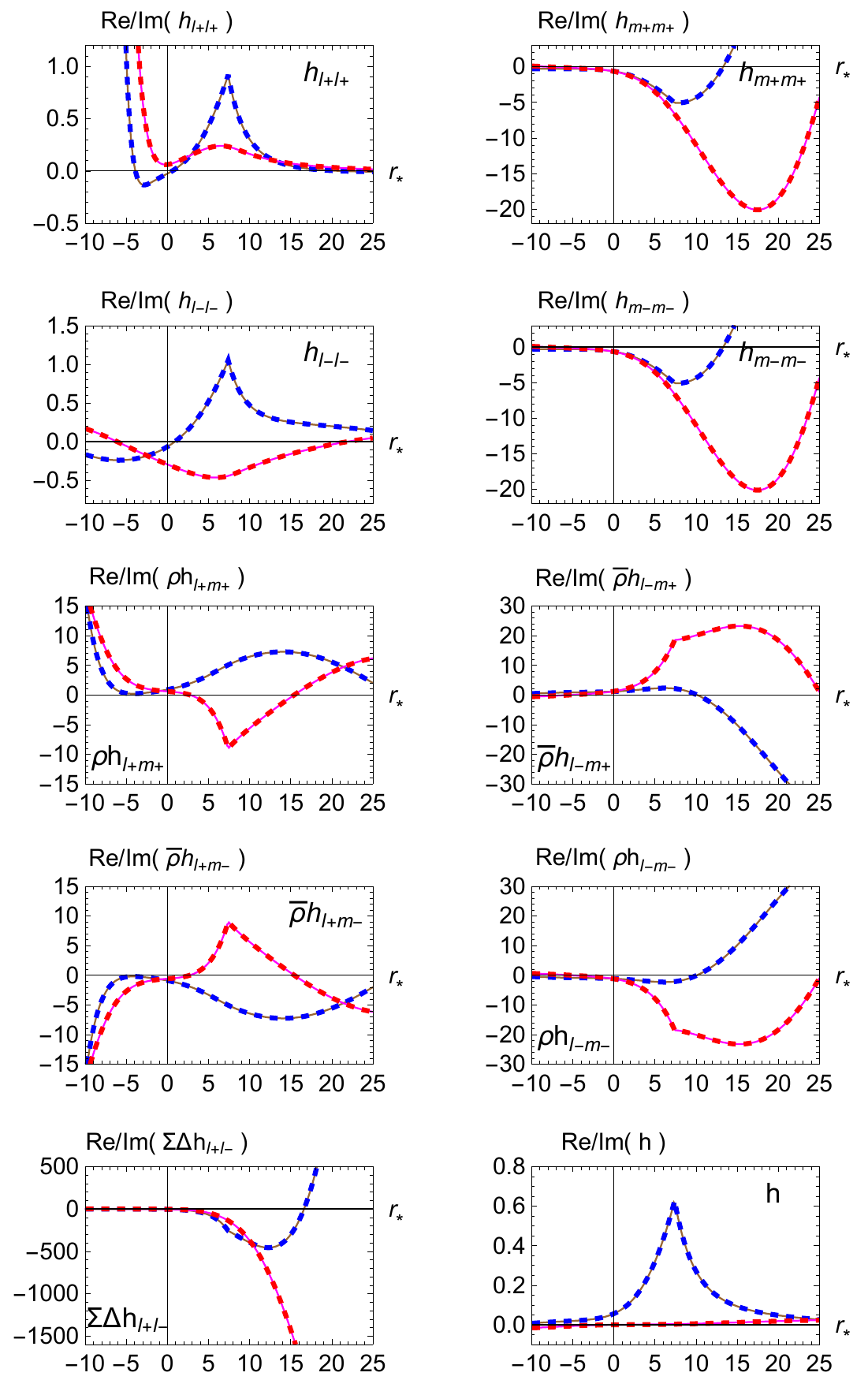} 
\end{center}
 \caption{A comparison between $lm$ modes extracted from the 2+1D time domain code [solid] and from the new separation of variables method [dashed], for $l = m = 2$, $a=0.6M$ and $r_0 = 6M$ for all components.}
 \label{fig:comparison-m2-all}
\end{figure}

Figures \ref{fig:comparison-m1-all} and \ref{fig:comparison-m2-all} show a comparison of \emph{all} 10 components listed in Eq.~(\ref{eq:h-lm-modes}), for the cases $\ell = m =1$ and $\ell = m = 2$, respectively. Good agreement is found in every component, in every case examined. In addition, the metric projections exhibit the even/odd parity symmetries described in the previous section. 

The agreement in the $m=1$ sector shown in Fig.~\ref{fig:comparison-m1-all} is particularly notable because a linear-in-$t$ gauge mode instability has been eliminated from the time-domain data by the application of a frequency filter, as described in Sec.~VIB in Ref.~\cite{Dolan:2012jg}. This sector is known to include a non-radiative $\ell = m = 1$ piece associated with the moving centre-of-mass of the system. Neither issue appears to stand in the way of agreement with the comparison data.

%Figure \ref{fig:m2-all-a099} shows $\ell m$ modes of the metric perturbation for a rapidly-spinning black hole with $a=0.99M$. The plots compare the contribution to the metric perturbation of the spin-2 piece derived from the Teukolsky scalars [dashed line], with the total metric perturbation formed from the sum of the $s=0,1,2$ pieces. The plots highlight once more (see also Fig.~\ref{fig:spins}) that it is necessary to include \emph{all} the pieces to form a Lorenz-gauge metric perturbation with modes that are $C^0$ at $r=r_0$. 
%
%\begin{figure}
%\begin{center}
% \includegraphics[width=16cm]{figs/hcomp_l2_m2_a0.pdf} \\
% \includegraphics[width=16cm]{figs/hcomp_l2_m2_a099.pdf}
%\end{center}
% \caption{Radial profiles of the metric perturbation components after projection onto the $l=m=2$ spherical basis. Upper set: $a = 0$. Lower set: $a = 0.99M$. The thin dashed lines show the real (blue) and imaginary (red) parts of the $s=2$ part of the metric only. The thick solid lines show the real and imaginary parts of the full metric perturbation (i.e.~the sum of $s=0$, $1$ and $2$ parts). }
%  \label{fig:m2-all-a099}
%\end{figure}

\subsubsection{Spin 0, 1 and 2 contributions}

\begin{figure}
\begin{center}
 \includegraphics[width=14cm]{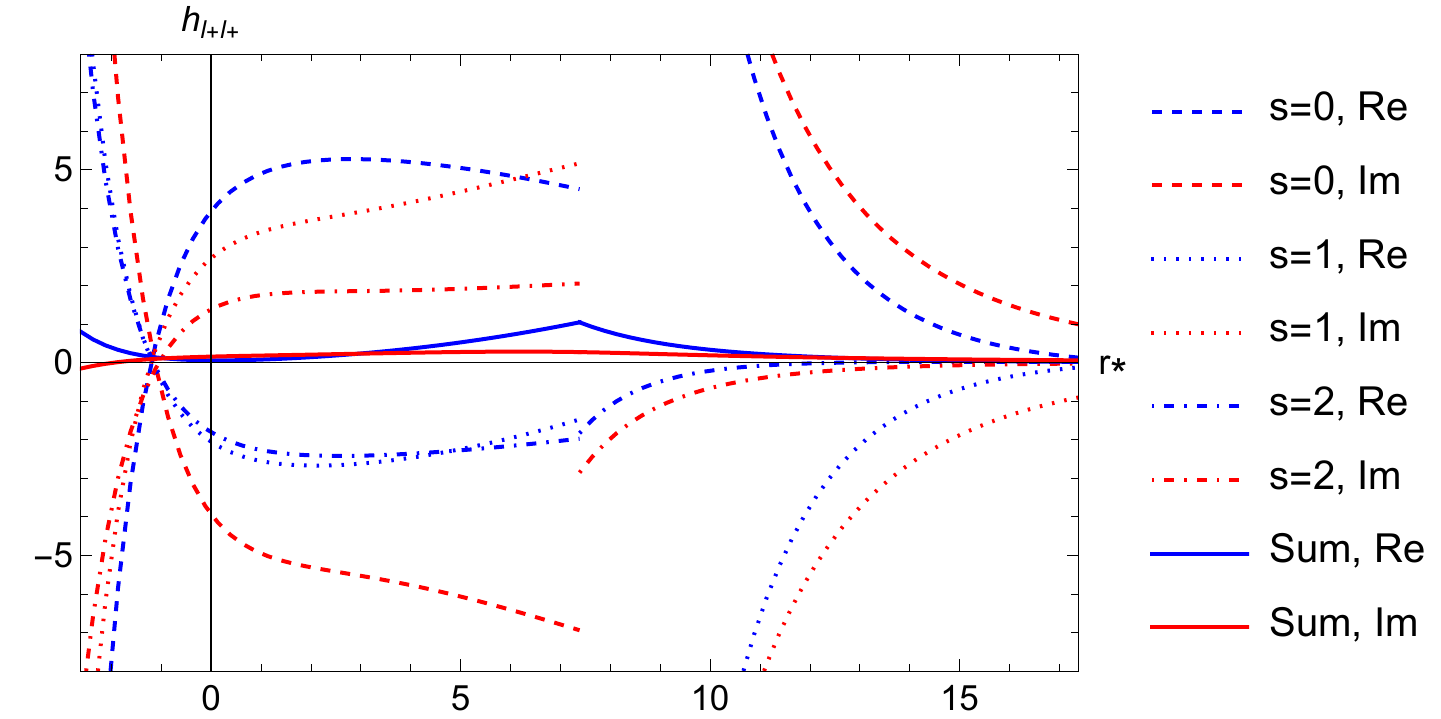} 
\end{center}
 \caption{Radial profiles of the spin $0$, $1$ and $2$ parts [dashed, dotted, dot-dashed] of the metric perturbation, and their sum [solid], for the component $h_{l_+l_+}$ and parameters $a = 0$, $r_0 = 6M$, $\ell = m = 2$.}
 \label{fig:spins}
\end{figure}

As described previously, the Lorenz-gauge metric perturbation is constructed from the sum of $s=0$, $s=1$ and $s=2$ contributions, each made from Teukolsky functions of the appropriate spin weight. The $\ell m$ modes of the individual $s=0$, $s=1$ and $s=2$ contributions are not expected to be continuous at $r=r_0$. This is illustrated in Fig.~\ref{fig:spins}, which shows the separate spin-$s$ contributions (in our classification) to the $\hrmlm_{l_+ l_+}$ component projected onto the $\ell = m = 2$ spherical harmonic. Only the sum itself is found to be continuous at $r=r_0$. Moreover, the sum is typically smaller in magnitude than the individual $s=0,1,2$ subcomponents. 

\subsubsection{Convergence}

\begin{figure}
\begin{center}
 \includegraphics[width=12cm]{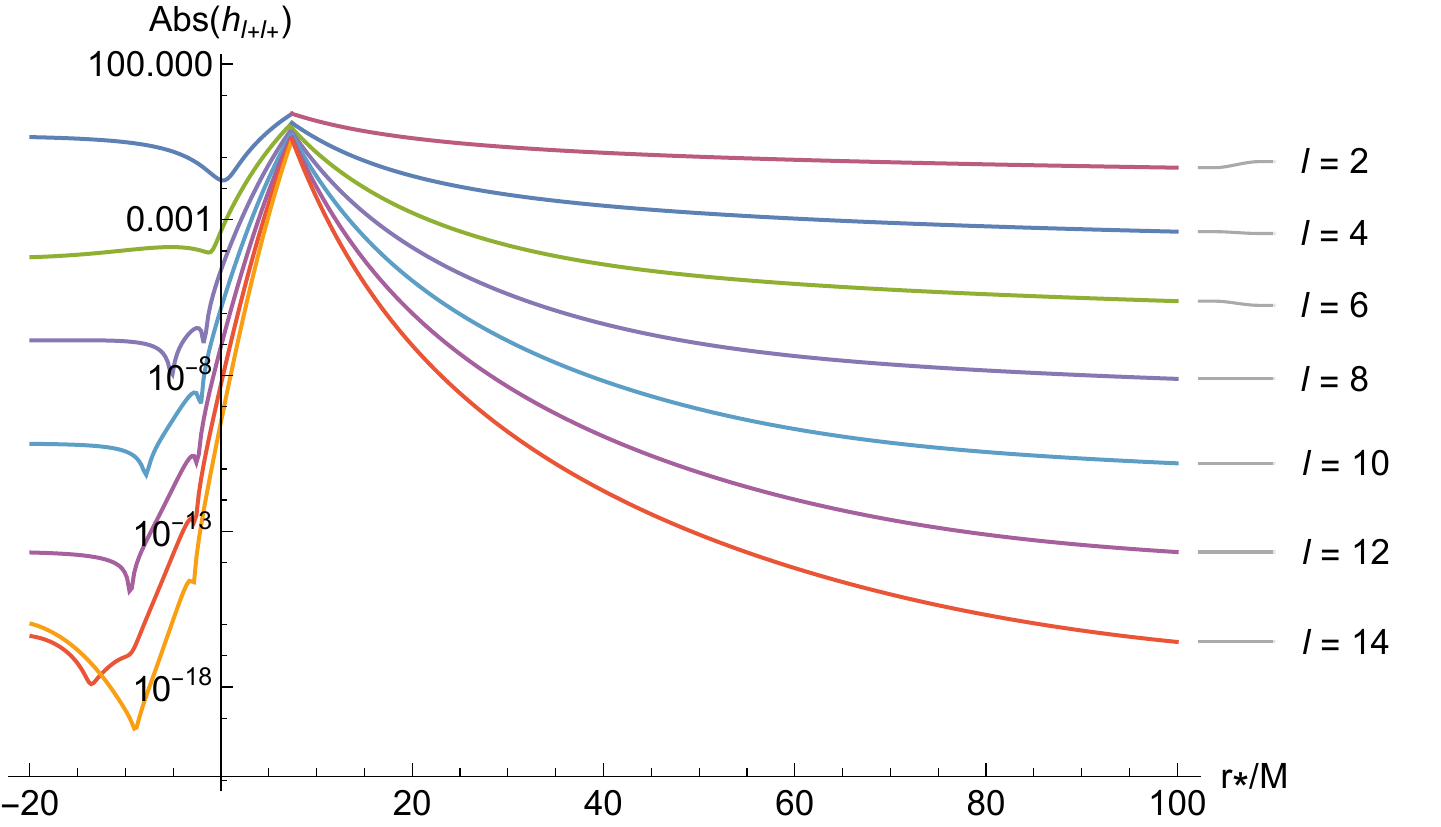} 
\end{center}
 \caption{
The $\ell$-modes of $h_{l_+ l_+}$ as a function of radius, for $a = 0.99M$, $r_0 = 6M$. The plot demonstrates exponential convergence, apart from at $r=r_0$. The convergence becomes more rapid as $|r - r_0|$ increases. 
 }
 \label{fig:lmodes-by-radius}
\end{figure}

We now consider the convergence of the $\ell$-mode sum in more detail. 
Figure \ref{fig:lmodes-by-radius} shows the $\ell$ modes of a particular component of the metric ($\hrmlm_{l_+ l_+}$) as a function of radius. The plot illustrates that the sum over $\ell$ is exponentially convergent away from $r=r_0$, and that the speed of convergence increases with $|r - r_0|$. In the far-field, only a few modes are needed for an accurate calculation. By contrast, near to the worldline at $r=r_0$ the situation is more interesting.

\begin{figure}
\begin{center}
 \includegraphics[width=12cm]{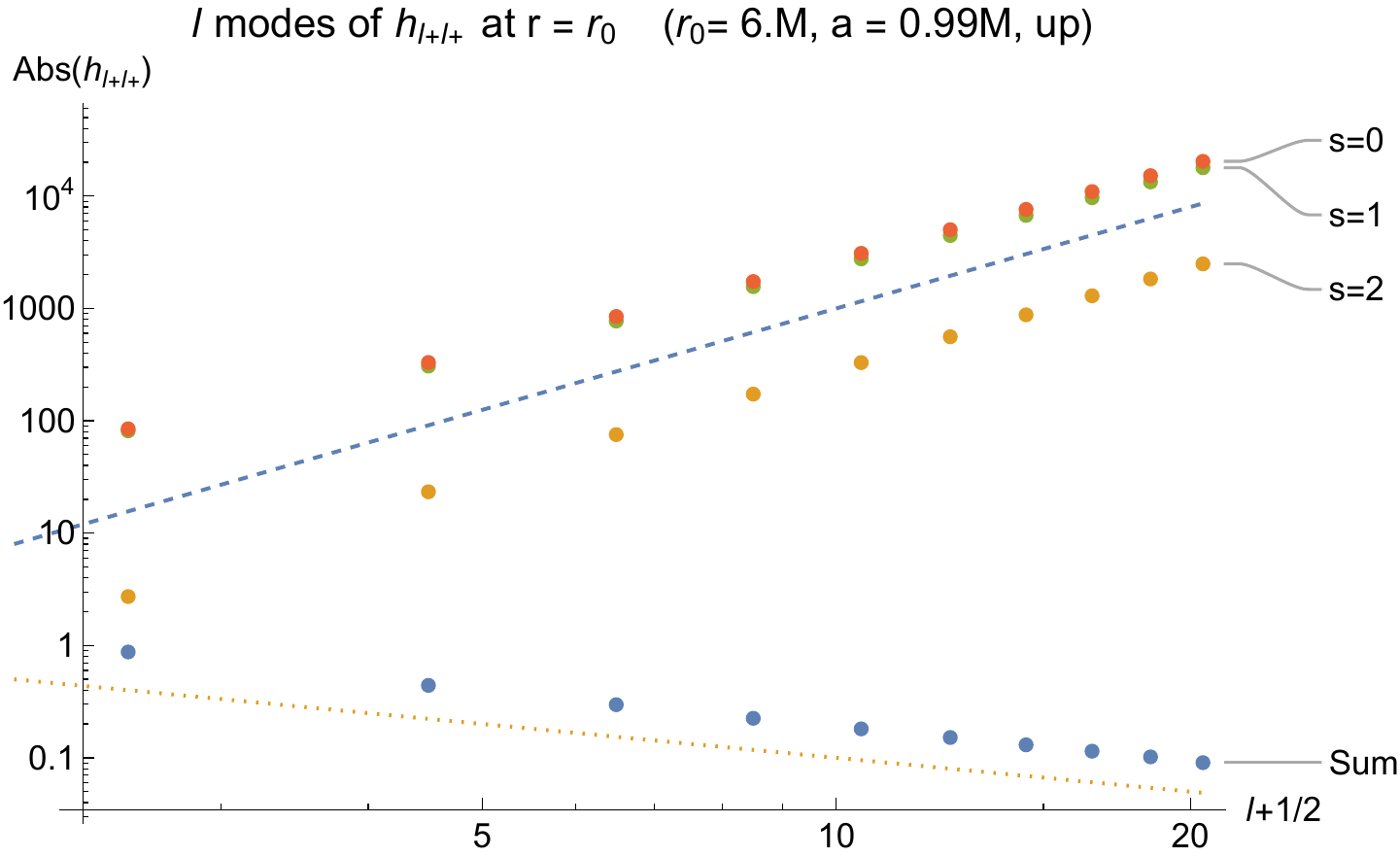} \\
 \includegraphics[width=8cm]{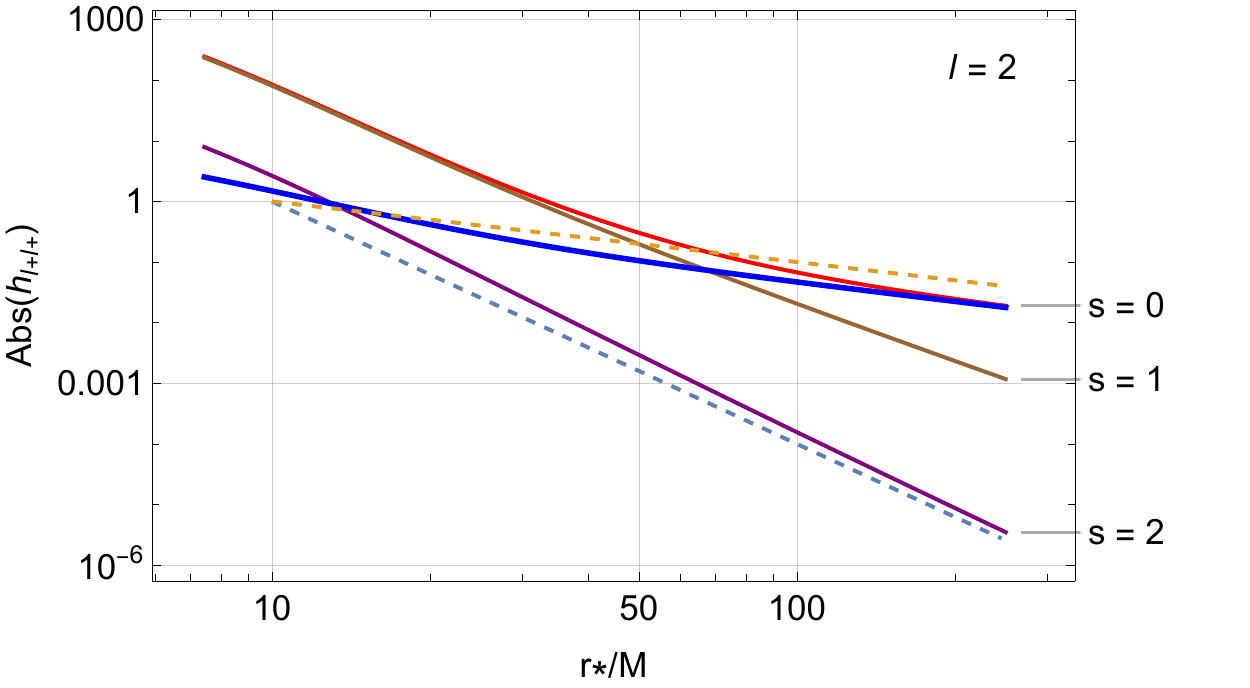}
 \includegraphics[width=8cm]{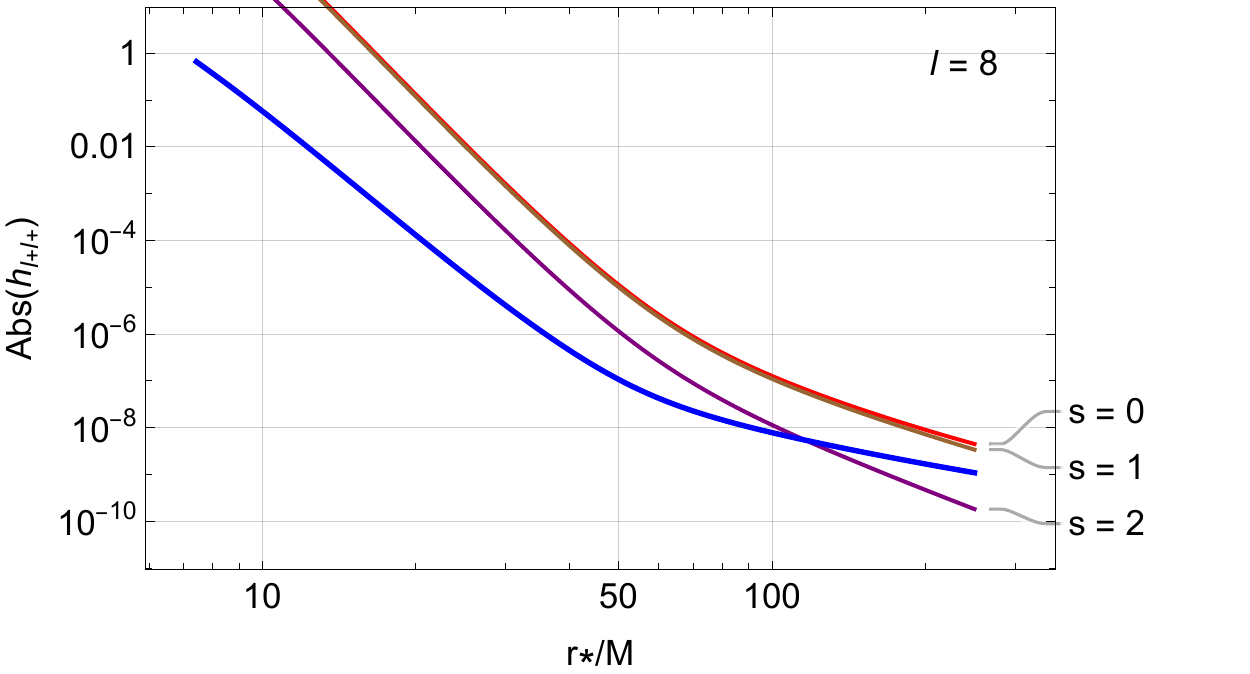} 
\end{center}
 \caption{The contribution of the spin $0$, $1$ and $2$ parts to $h_{l_+ l_+}$ for $a = 0.99M$, $r_0 = 6M$.
  \emph{Upper:} The contributions of parts of spin $0$, $1$ and $2$ at $r=r_0$ (in the right-hand side limit) as a function of $\ell + 1/2$. The dashed and dotted guidelines are $(\ell + 1/2)^3$ and $()^{-1}$, respectively. \emph{Lower:} Plots showing the relative contributions of the spin parts to the total (thick blue line) as a function of radius. The dashed guidelines are proportional to $r^{-4}$ and $r^{-1}$. 
 }
 \label{fig:lmodes-on-worldline}
\end{figure}

Figure \ref{fig:lmodes-on-worldline} shows the (absolute value of the) $\ell$ modes of $h_{l_+ l_+}$ evaluated on the circular orbit at $r = r_0$. These $\ell$ modes (blue dots) decay in power-law fashion, in proportion to $(\ell + 1/2)^{-1}$. Hence, and as expected, the mode sum is not convergent at $r=r_0$, which is not unexpected since the metric perturbation itself diverges on the worldline (at $r=r_0$, $\theta=\pi/2$); this issue typically addressed by employing a puncture scheme. The figure also shows the behaviour of the $s=0$, $s=1$ and $s=2$ parts with $\ell$. It is evident that these parts individually grow, in proportion to $(\ell+1/2)^3$, whereas their sum decays, in proportion to $(\ell + 1/2)^{-1}$. Hence there is significant cancellation between the individual spin parts at high-$\ell$; nevertheless, this can be handled using extended-precision arithmetic. The observed behaviour with $\ell$ is indicative of the non-smoothness of the individual parts as functions on the sphere at $r=r_0$. 

The lower plots of Fig.~\ref{fig:lmodes-on-worldline} show how the relative contributions from the $s=0$, $s=1$ and $s=2$ parts of the metric perturbation vary across the radial domain. In the far field ($r \gg M$) the $s=2$ and $s=1$ parts fall off more rapidly than the $s=0$ part. The latter falls off in proportion to $1/ r$, in the same way as the sum itself.

\subsection{Results: $m$-modes in $(r,\theta)$ space\label{subsec:results-m}}

In the previous section, we reviewed the projection of the metric perturbation onto \emph{spherical} $\ell m$ modes. The projection onto $\ell m$ modes is not necessary (other than to determine the jumps), however, if one wishes to compute the metric perturbation in the $(r,\theta)$ space. In this section, we examine data in the $(r,\theta)$ plane formed from a partial sum over the spin-weighted \emph{spheroidal} modes up to some $\ell_{\text{max}}$. For brevity, we focus on one component $h_{l_+ l_+}$ for $r_0 =6M$, $m=2$ and $a = 0.6M$. 

\begin{figure}
\begin{center}
 \includegraphics[width=8cm]{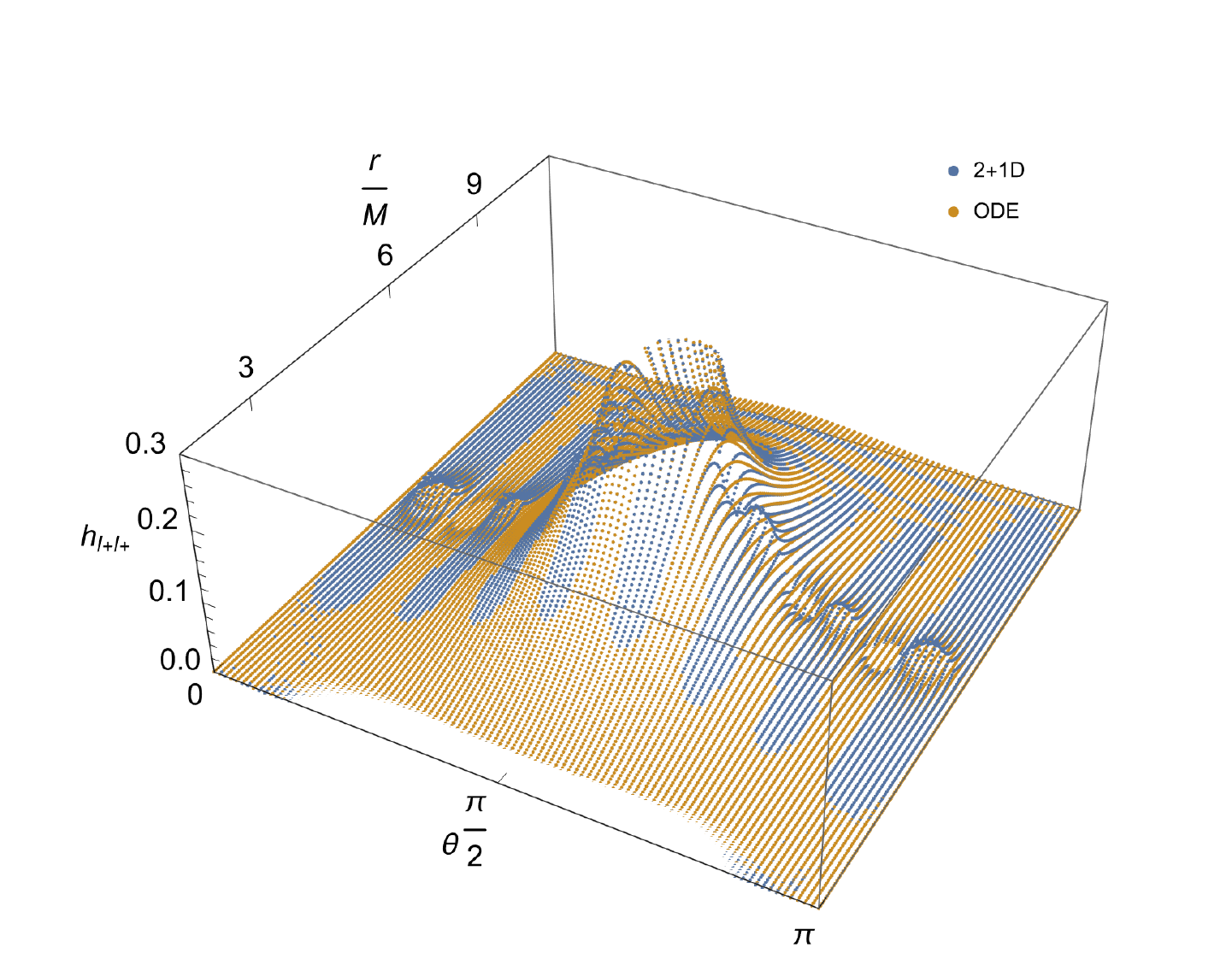} 
 \includegraphics[width=8cm]{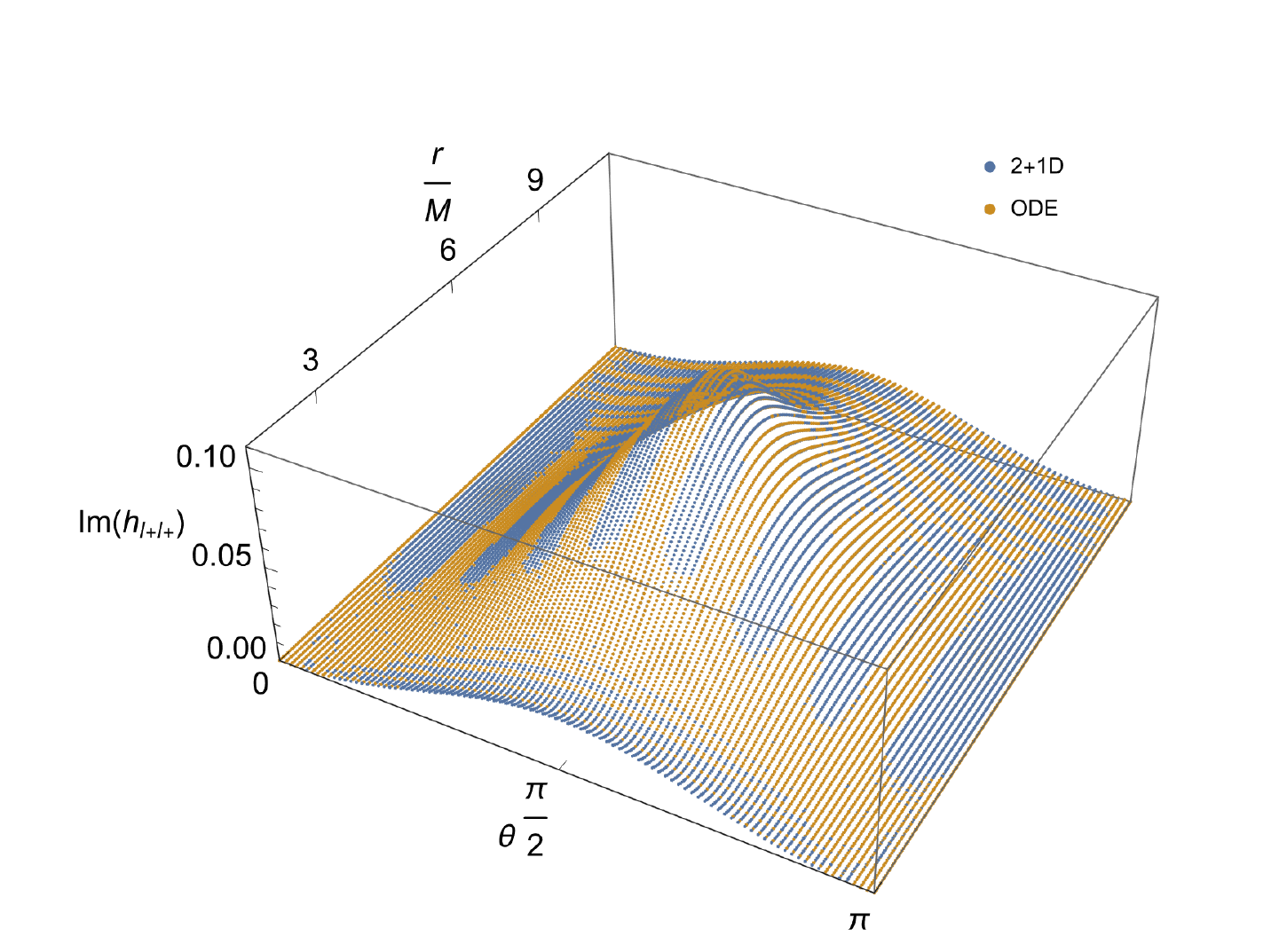}  \\
 \includegraphics[width=8cm]{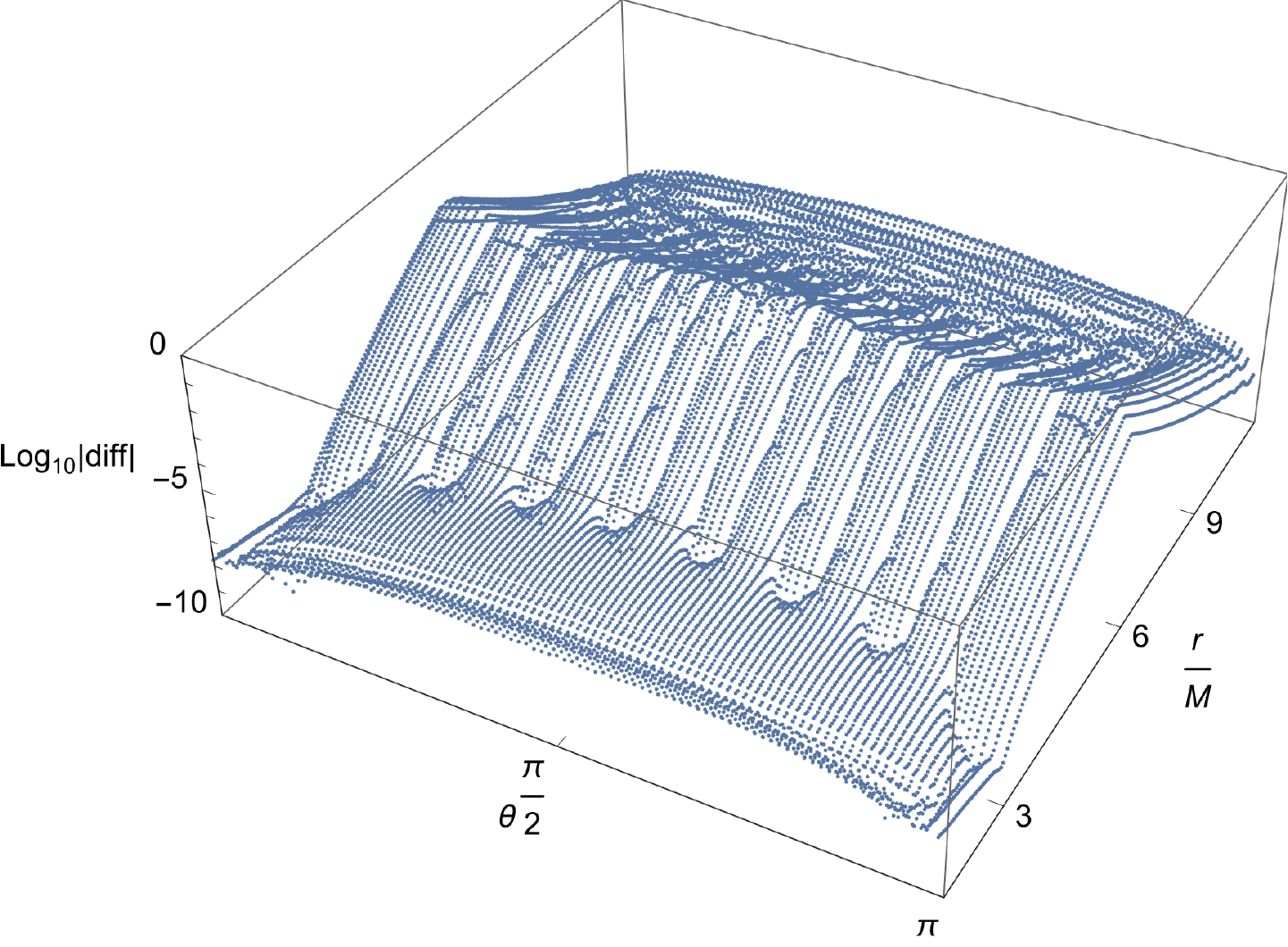} 
\end{center}
 \caption{The profile of $h_{l_+l_+}$ in the $(r,\theta)$ plane for $r_0 = 6M$, $m=2$ and $a=0.6 M$. The upper plots show the real [left] and imaginary [right] parts of $h_{l_+l_+}$ computed with the new method, summed to $\ell_{\text{max}} = 14$ [blue points]; and the results of the 2+1D time domain code [brown points]. The lower plot shows the absolute value of the difference between the two data sets, on a logarithmic scale.}
  \label{fig:m2-3D}
\end{figure}

Figure \ref{fig:m2-3D} confronts the results of the new method with the comparison data set from the 2+1D time-domain code in the $(r,\theta)$ domain. The partial sum of spheroidal $\ell$-modes is truncated at $\ell_{\text{max}} = 14$. The plots show that, whereas there is good agreement for the imaginary part across the $(r,\theta)$ domain, in the real part there is substantial disagreement near the particle radius $r_0$ and for all values of $\theta$. This is a consequence of the truncation of the mode sum at $\ell_{\text{max}} = 14$. The lower plot in Fig.~\ref{fig:m2-3D}, which shows the difference between the two data sets, makes it clear the truncation error falls off with $|r - r_0|$ in an approximately exponential manner. The floor on the error at $\sim 10^{-6}$ is likely to be due to discretization error in the (old) 2+1D data set. We have verified that changing the grid resolution for the 2+1D code changes this floor, as expected.

\begin{figure}
\begin{center}
 \includegraphics[width=12cm]{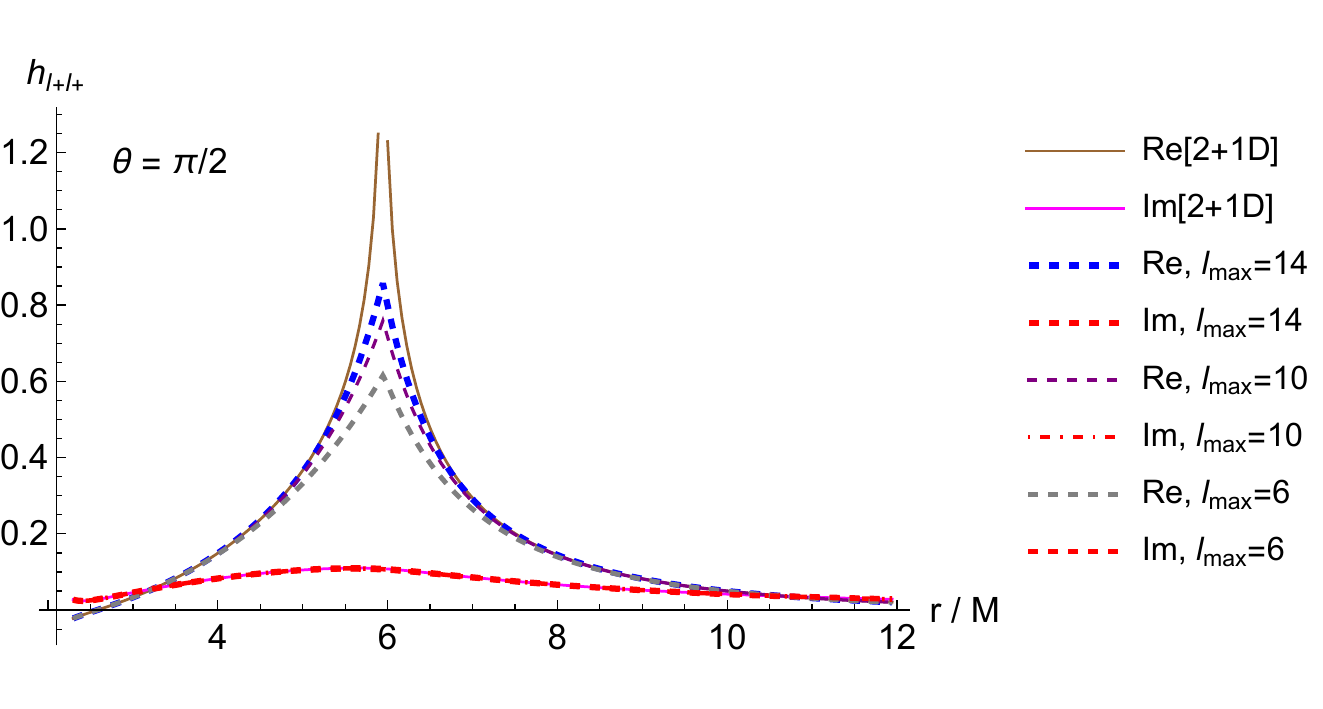} 
\end{center}
 \caption{The radial profile of $h_{l_+l_+}$ in the equatorial plane ($\theta=\pi/2$) for $r_0 = 6M$, $m=2$ and $a=0.6 M$. The solid lines show the real and imaginary components of the comparison data set (i.e.~data from the 2+1D time domain code). The dashed lines show results from the new method, with partial sums up to $\ell_{\text{max}} = 6$, $10$ and $14$.}
  \label{fig:m2-radial}
\end{figure}

\begin{figure}
\begin{center}
 \includegraphics[width=16cm]{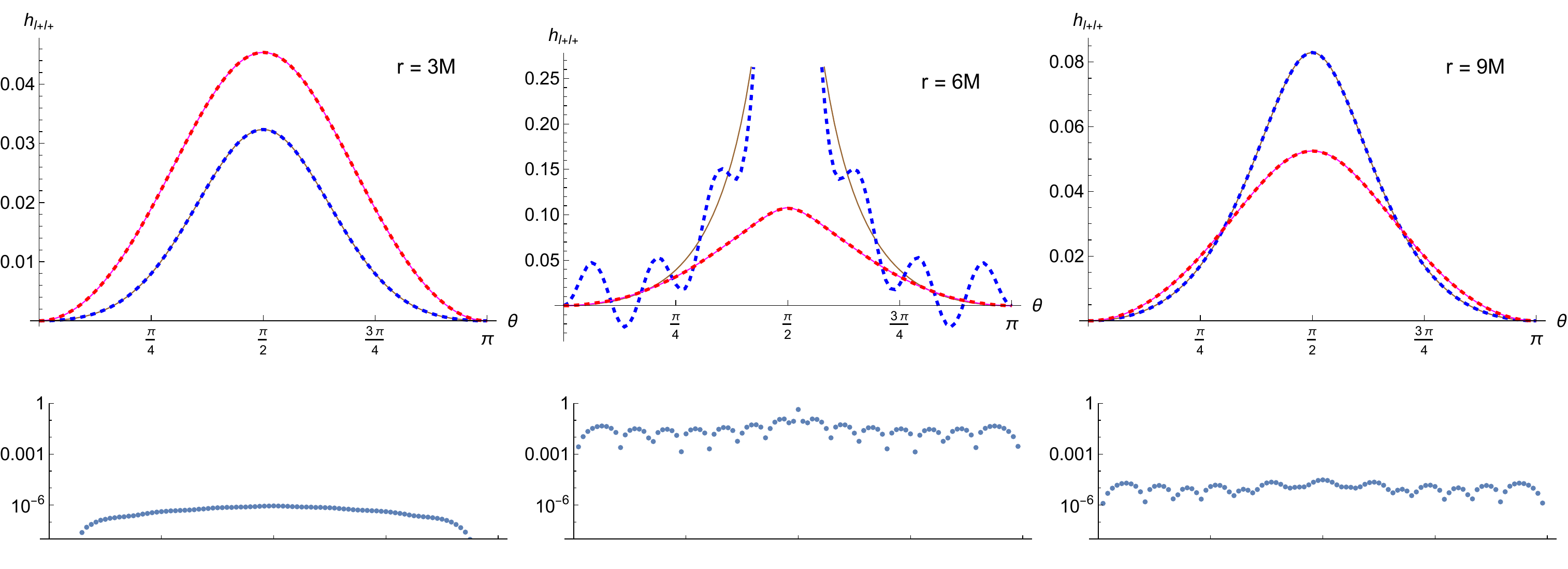} 
\end{center}
 \caption{The angular profile of $h_{l_+l_+}$ at $r = 3M$ (left), $6M$ (middle) and $9M$ (right) for $r_0 = 6M$, $m=2$ and $a=0.6 M$. In the upper plots, the results of the new method summed to $l_{\text{max}} = 14$ [dashed] are compared with the results of the 2+1D time-domain code [solid]. The lower plots show the absolute value of the difference between the data sets on a logarithmic scale.}
  \label{fig:m2-angular}
\end{figure}

Figures \ref{fig:m2-radial} and \ref{fig:m2-angular} show the radial and angular profiles of the data sets in Fig.~\ref{fig:m2-3D}. The radial profile in the equatorial plane shows the limitations of the $\ell$-sum in capturing the divergence of the metric perturbation as $r \rightarrow r_0$. While increasing the maximum value of $\ell$ in the sum does improve the fit, it does so slowly (logarithmically) in the vicinity of $r=r_0$. The angular profiles in Fig.~\ref{fig:m2-angular} highlight that, whereas at $r = 3M$ or $r=9M$ the truncated sum provides a good approximation, for $r=r_0 = 6M$ the angular profile of the truncated sum is in poor agreement with the 2+1D data. Moreover, the central plot ($r = 6M$) shows the expected angular oscillations associated with the $\ell = 14$ harmonic, indicating that this is an issue caused by the truncation of an infinite-sum representation of a divergent function. 

This issue of mode-sum-truncation error is entirely expected. We find that, qualitatively, it is no worse in the Kerr case than it is in the Schwarzschild case. Similar truncated $\ell$-mode sums have been widely used in a range of self-force calculations (including at second order) without causing any issues, suggesting that it is not likely to be a concern for most applications of interest.

Taken together, the results shown in Figs.~\ref{fig:schw-m2a}--\ref{fig:m2-angular} are evidence that we have successfully computed the Lorenz-gauge metric perturbation from combining spin-$2$, spin-$1$ and spin-$0$ metric perturbations that are constructed from the solutions of \emph{ordinary} differential equations. In other words, the first successful example of metric reconstruction from scalar variables in Lorenz gauge on Kerr spacetime.

\section{Static modes and completion}\label{sec:static}

Here we consider the static ($\omega=0$) part of the metric perturbation. For a particle on a circular equatorial orbit, the static part is also axially-symmetric, as the mode frequencies are $\omega_m = m \Omega$ for circular orbits. This sector requires separate consideration for two reasons. First, the Lorenz-gauge vacuum modes in Sec.~\ref{sec:mp} have factors of $\omega$ in the denominator, indicating that they are ill-defined for $\omega=0$. Second, the axially-symmetric part of the metric perturbation contains physical \emph{non-radiative} degrees of freedom, corresponding to perturbations in mass and angular momentum, as well as non-radiative gauge modes. These are known as the \emph{completion pieces}.

The overall approach is, {\it mutatis mutandis}, similar to that of Sec.~\ref{sec:construction}. First, we write down a complete set of Lorenz gauge modes, consisting of a set of modes derived from scalar variables, now augmented by completion pieces derived separately. Next, we construct the metric perturbation from `glueing' the outer and inner solutions at $r=r_0$, demanding that (i) the correct trace and Weyl scalars are obtained and (ii) the metric perturbation is $C^0$ on the sphere at $r=r_0$, and (iii) the MP is as regular as possible at the horizon and at infinity, and (iv) the MP has the correct mass and angular momentum at infinity. In fact, the four conditions cannot be simultaneously achieved even in the Schwarzschild case. We will construct a solution satisfying (i)--(iii) but with the incorrect mass and angular momentum, that is, the Kerr analogue of the Berndtson solution \cite{Berndtson:2007gsc} in the Schwarzschild case. 

Several simplifications arise in the static sector. First, the vector ($s=1$) gauge modes are not required for circular orbits. Second, the spin-weighted spheroidal harmonics reduce to spherical harmonics of the same spin-weight (because $a \omega = 0$), and so the projection of the metric perturbation onto spin-weighted spherical harmonics is comparatively straightforward. 

Third, for $\omega = 0$ the spectrum of radial and angular functions is degenerate, in the following sense: the spin-weighted angular functions ($Y_{\pm2}$ and $Y_{\pm1}$) can be obtained by acting upon the spin-zero angular functions ($Y_0$) with ladder operators ($\cLh_n$ and $\cLh^\dagger_n$), and moreover, in vacuum regions the spin-weighted radial functions ($P_{\pm2}$ and $P_{\pm1}$) can be obtained by acting upon the spin-weight zero radial function ($P_0$) with radial ladder operators (see e.g.~Sec.~IVB in Ref.~\cite{Maggio:2018ivz}). The former property is described in Appendix \ref{sec:spherical-harmonics}. The latter property is summarised below for the case $\omega=m=0$. 
Fourth, the radial functions in vacuum have closed-form solutions in terms of elementary special functions.

In source-free regions, the Teukolsky equations for the static, axially-symmetric sector ($\omega=m=0$) reduce to
\begin{subequations}
\begin{align}
\left\{ \partial_r \Delta \partial_r  - \lambda \right\} P_0 &= 0 ,   \label{eq:staticP0} \\
\left\{ \Delta \partial_r \partial_r - \lambda \right\} P_{\pm1} &= 0 ,  \label{eq:staticP1} \\
\left\{ \Delta \partial_r \partial_r - \Delta^\prime \partial_r - (\lambda - 2) \right\} P_{\pm 2} &= 0,  \label{eq:staticP2}
\end{align}
\end{subequations}
with $\lambda = l (l+1)$. From the Teukolsky-Starobinskii identities, we infer that  $P_{+2} = P_{-2}$ and $P_{+1} = P_{-1}$. It is straightforward to show (up to a choice of normalisation) that the homogeneous radial functions are related by
\begin{subequations}
\begin{align}
P_{\pm 1} &= \Delta \partial_r P_0 , 
&
P_0 &= \frac{1}{\lambda} \partial_r P_{\pm1} ,
\\
P_{\pm 2} &= \left( \Delta \partial_r - \Delta^\prime \right) P_{\pm 1} , 
&
P_{\pm 1} &= \frac{1}{\lambda-2} \partial_r P_{\pm 2} , \\
 &= \Delta^2 \partial_r \partial_r P_0 = \Delta \left(\lambda P_0 - \Delta^\prime \partial_r P_0 \right) &
P_0 &= \frac{1}{\lambda(\lambda-2)} \partial_r \partial_r P_{\pm 2} .  \label{eq:P0P2}
\end{align}
\label{eq:degeneracy}
\end{subequations}

 \subsection{From radiation gauge to Lorenz gauge}
 The metric perturbations in radiation gauge, detailed in Sec.~\ref{sec:radiation-gauge}, remain valid in the $\omega = 0$ limit. Conversely, in this same limit, the transformation to Lorenz gauge in Sec.~\ref{sec:hLorenz} appears to break down, as does the construction of the L(-) solution in Eq.~(\ref{eq:mp-s2}), due to factors of $\omega$ appearing in the denominator. For this reason, we must address anew the question of the appropriate transformation to Lorenz gauge.

  \subsubsection{Spin two\label{subsec:static-s2}}
In this section, we reconsider the transformation of the IRG solution (\ref{h-irg}) to Lorenz gauge. In the case $\omega = 0$, the  functions appearing in the gauge vector in (\ref{eq:HUsolution}) and (\ref{eq:chi}) are invalid, due to the factors of $\omega$ in the denominator; however, the construction (\ref{eq:xi-ansatz}) remains valid. Hence we seek an alternative solution to Eq.~(\ref{eq:HUode}). For $\omega = 0$, one can show that Eq.~(\ref{eq:HUode}) is satisfied by
\beq
\rhoc H_U = \frac{1}{2 \lambda} \left( \rhoc \cD \cLh_2^\dagger - 2 \left( \cLh_2^\dagger + i a \sin \theta \cD \right) \right) \psi ,  \label{eq:rhocHU-static}
\eeq
(now omitting $-$ subscripts).
Here we have assumed a decomposition into $\ell m$ modes, so that 
\beq
\psi = \sum_{\ell} \psi_{\ell} , \quad \quad \psi_\ell \equiv P_{-2 \ell}(r) Y_{-2 \ell 0}(\theta),
\eeq
where $Y_{-2 \ell 0}(\theta)$ are spherical harmonics of spin-weight $-2$. 
To complement this, we need a scalar field $\chi(r,\theta)$ satisfying the static equivalent of Eq.~(\ref{eq:Boxchi-rhs}).
Decomposing into harmonics,
\beq
\chi(r,\theta) = \sum_{\ell = 2}^\infty \chi_{\ell}(r,\theta)
\eeq
and using a standard form for the d'Alembertian, 
\begin{align}
 \left( \cD \Delta \cD^\dagger + \cLh_1 \cLh^\dagger \right) \chi_\ell &=  - \frac{1}{\lambda} \left(\cLh_1^\dagger - i a \sin \theta \cD \right) \left( \rhoc \cD \cLh_2^\dagger - 2 \left(\cLh_2^\dagger + i a \sin \theta \cD \right) \right) \psi_\ell  , \nn \\
&=  - \frac{1}{\lambda} \left[ \left(\rhoc \cD - 2 \right) \cLh_1^\dagger \cLh_2^\dagger - \rhoc_{,\theta} \left( \rhoc \cLh_2^\dagger - 2 \rhoc_{,\theta} \right) \cD \cD \right] \psi_\ell  \equiv S_\ell. \label{eq:chi-static-eq}
\end{align}
Despite some {\it ad hoc} effort, a neat closed-form solution to Eq.~(\ref{eq:chi-static-eq}) for $\chi_\ell$ in terms of $\psi_\ell$ has not been found (contrasting with Eq.~(\ref{eq:chi}) in the non-static case). From a practical perspective, it is sufficient that Eq.~(\ref{eq:chi-static-eq}) admits a solution that can be written as the sum of separable terms, as we now describe.

In the $m=0$ case, the Hertz potential $\psi$ is real, and we can split the source term $S_\ell$, and the function $\chi_\ell(r,\theta)$, into real and imaginary parts ($S_\ell = \Sre + i \Sim$ and $\chi_\ell =  \chire + i \chiim$). For $m=0$,  $\cLh_n = \cLh^\dagger_n = \partial_\theta + n \cot \theta$ and $\cD = \cD^\dagger = \partial_r$. This yields a pair of equations,
\begin{subequations}
\begin{align}
 \left( \partial_r \Delta \partial_r + \partial_\theta^2 + \cot \theta \partial_\theta \right) \chire &= S_\ell^{\text{re}} , \\
 \left( \partial_r \Delta \partial_r + \partial_\theta^2 + \cot \theta \partial_\theta \right) \chiim &= S_\ell^{\text{im}},
 \end{align}
with 
\begin{align}
\Sre &\equiv
 - \frac{1}{\lambda} \left[ (r \partial_r - 2) \cLh_1^\dagger \cLh_2^\dagger - a^2 \left( \sin \theta \cos \theta \cLh_2^\dagger + 2 \sin^2 \theta \right) \partial_r \partial_r \right] \psi_\ell , \label{eq:static-source-re} \\
\Sim &\equiv \frac{a}{\lambda} \left( \cos \theta \cLh^\dagger_1 + \sin \theta \, r \, \partial_r \right) \cLh^\dagger_2 \partial_r \psi_\ell \label{eq:static-source-im} .
\end{align}
 \label{eq:static-source-S}
\end{subequations}

\emph{Real part}: 
For $\omega=0$, the (spin-weighted) spheroidal harmonics reduce to (spin-weighted) spherical harmonics, and the angular operators $\cLh_n$ and $\cLh^\dagger_n$ act as spin-lowering and spin-raising operators (see Appendix \ref{sec:spherical-harmonics}). Hence the source term $\Sre$ in Eq.~(\ref{eq:static-source-re}) can be written as
\begin{align}
\Sre(r,\theta) &= - \frac{\hat{\Lambda}}{\lambda} (r \partial_r - 2) P_{-2\ell}(r) Y_{0\ell}(\theta) + \frac{1}{\lambda} a^2 (\partial_r \partial_r P_{-2\ell}) 
\sum_{k=-1}^{k=+1} M^{\text{re}}_{\ell, \ell+2k} Y_{0\,\ell + 2k}(\theta)
 \end{align}
where $\hat{\Lambda}$ is defined in Eq.~(\ref{eq:lambdah-def}), and the matrix elements
\beq
M^{\text{re}}_{\ell j} =  \left( - \hat{\boldsymbol{\lambda}}_2  \mathfrak{s}_{-10} \mathfrak{c}_0 + 2 (\mathfrak{s}^2)_{-2 0} \right)_{\ell j}
\eeq
couple to nearest neighbours of the same parity, i.e., the $\{l-2, l, l+2\}$ modes. Hence we can express a mode of the scalar field $\chire$ in the form
\beq
\chire(r,\theta) = - \frac{\hat{\Lambda}}{\lambda} \hat{\chi}^{(e)}_\ell(r) Y_{0 \ell}(\theta)  + \frac{a^2}{\lambda}  
 \sum_{k=-1}^1 M^{\text{re}}_{\ell, \ell+2k} \chi_{(e) \ell}^{[k]}(r) Y_{0 \, \ell+2k}(\theta) 
 , 
\eeq
where
\begin{subequations}
\begin{align}
\{ \partial_r \Delta \partial_r - \lambda \} \hat{\chi}^{(e)}_\ell &= (r \partial_r - 2) P_{-2\ell} , \label{eq:chieven} \\
\{ \partial_r \Delta \partial_r - \lambda_k \} \hat{\chi}^{[k]}_{(e) \ell} &= \partial_r \partial_r P_{-2 \ell} = \lambda (\lambda - 2) P_{0 \ell}, \label{eq:chiek}
\end{align}
\label{eq:chihat}
\end{subequations}
with $\lambda_{k} \equiv (l+2k)(l+2k+1)$.
Here the term $\partial_r \partial_r P_{-2\ell}$ has been replaced using Eq.~(\ref{eq:P0P2}).
Since $P_0$ satisfies Eq.~(\ref{eq:staticP0}), it is straightforward to verify that 
\begin{align}
 \hat{\chi}^{[\pm1]}_{(e)\ell}(r) &= \frac{\lambda (\lambda - 2)}{(\lambda - \lambda_{\pm 1})} P_{0\ell}(r) .
\end{align}
The other two functions $\hat{\chi}^{(e)}_\ell(r)$ and $\hat{\chi}^{[0]}_{(e)\ell}(r)$ are obtained by directly solving the radial equations (\ref{eq:chihat}), as described in Sec.~\ref{subsec:implementation2}.

\emph{Imaginary part}: For the imaginary scalar $\chiim$, there is coupling to the nearest-neighbour modes of opposite parity $(\ell - 1, \ell + 1)$ only. The source becomes
\beq
\Sim = \frac{a}{\lambda} \sum_{j=\pm1} \left( \partial_r P_{-2\ell} (\hat{\boldsymbol{\Lambda}} \mathfrak{c}_0 )_{\ell, \ell+j} + r \partial_r \partial_r P_{-2\ell}  (- \hat{\boldsymbol{\lambda}}_2 \mathfrak{s}_{-10})_{\ell, \ell+j} \right) Y_{0 \, \ell + j}
\eeq
and hence the function $\chiim$ can be written
\beq
\chiim(r,\theta) = \frac{a}{\lambda} \sum_{j=\pm1} \left( \hat{\chi}^{(o)c}_{[j]\ell} (r) (\hat{\boldsymbol{\Lambda}} \mathfrak{c}_0 )_{\ell, \ell+j} + \hat{\chi}^{(o)s}_{[j]\ell}(r) (- \hat{\boldsymbol{\lambda}}_2 \mathfrak{s}_{-10})_{\ell, \ell+j} \right) Y_{0 \, \ell + j}(\theta)
\eeq
where
\begin{subequations}
\begin{align}
\{ \partial_r \Delta \partial_r - \lambda_j \} \hat{\chi}^{(o)c}_{[j]\ell} &= \partial_r P_{-2\ell} , \label{eq:chiodd} \\
\{ \partial_r \Delta \partial_r - \lambda_j \} \hat{\chi}^{(o)s}_{[j]\ell} &= r \partial_r \partial_r P_{-2\ell} , \label{eq:chioj}
\end{align}
\end{subequations}
and here $\lambda_j = (l+j)(l+j+1)$. These ODEs can be solved directly for $\hat{\chi}_{[j]\ell}^{(o)c}(r)$ and $\hat{\chi}_{[j]\ell}^{(o)s}(r)$, as described in Sec.~\ref{subsec:implementation2}.

  \subsubsection{Spin zero and the trace\label{subsec:static-s0}}
In this section we seek a gauge vector $\xi^\mu$ that generates a metric perturbation $h_{\mu \nu} = 2 \xi_{( \mu ; \nu )}$ in Lorenz gauge, with a trace $h \equiv g^{\mu \nu} h_{\mu \nu} = 2 \nabla_{\mu} \xi^{\mu}$ that satisfies the vacuum equation $\Box h = 0$. Our starting point is to write $\xi^\mu$ as the sum of gradient and divergence terms, that is, $\xi = d \kappa + \delta \mathfrak{B}$, where $\kappa$ is a scalar and $\mathfrak{B}$ is a two-form. The scalar part $\kappa$ generates the trace, 
\beq
h = 2\delta \xi = 2\delta d \kappa = 2 \Box \kappa.
\eeq 
The Lorenz condition $\Box \xi_\mu = 0$ can be rewritten using $\Box \xi = (d \delta - \delta d) \xi = d \delta d \kappa - \delta d \delta \mathfrak{B}$. In other words, we seek a two-form $\mathfrak{B}$ such that
\beq
\delta d \delta \mathfrak{B} = d \delta d \kappa .
\eeq
Taking a co-derivative of the above, we see that $\delta d \delta d \kappa = 0$, i.e., $\Box \Box \kappa = \tfrac{1}{2} \Box h = 0$, as expected in vacuum. Conversely, if $\Box h \neq 0$, then a pure-gauge solution alone is insufficient, as we also might also expect. Taking an exterior derivative of the above, $(\delta d) (\delta d ) \mathfrak{B} = 0$. 

For the axisymmetric static case ($m=0$, $\omega=0$), a solution for the two-form can be constructed as follows:
\beq
\mathfrak{B}^{\mu \nu} = \frac{B(r,\theta)}{\Sigma} 2 \delta_r^{[\mu} \delta_\theta^{\nu]} 
\eeq
where $B(r,\theta)$ is a scalar function; likewise, $\kappa = \kappa(r,\theta)$. Then the gauge vector has the components
\beq
\xi_\mu = [0, \; \partial_r \kappa + \frac{1}{\Delta} \left(\partial_\theta + \cot \theta \right) B , \; \partial_\theta \kappa - \partial_r B, \; 0] . \label{eq:xi-gauge-scalar-static}
\eeq
We now expand the scalar functions in $\ell$ modes,
\begin{align}
h(r,\theta) = \sum_{\ell} h_{\ell}(r, \theta), \quad 
\kappa(r,\theta) = \sum_{\ell} \kappa_{\ell}(r, \theta),  \quad 
B(r,\theta) = \sum_{\ell} B_{\ell}(r, \theta). \label{eq:kappaB-expansion}
\end{align}
The trace condition and the Lorenz condition imply that, in vacuum, the $\ell$-modes of $h$, $\kappa$ and $B$ satisfy
\begin{align}
 \left(\partial_r \Delta \partial_r + \partial_\theta \partial_\theta + \cot \theta \partial_\theta \right) h_{\ell} &= 0 , \\
\left(\partial_r \Delta \partial_r + \partial_\theta \partial_\theta + \cot \theta \partial_\theta \right) \kappa_{\ell} &= \tfrac{1}{2} (r^2 + a^2 \cos^2 \theta) h_{\ell}(r,\theta) , \\
 \left(\Delta \partial_r \partial_r + \partial_\theta \partial_\theta + \cot \theta \partial_\theta - \frac{1}{\sin^2 \theta} \right) B_{\ell}  &= \tfrac{1}{2} (r^2 + a^2 \cos^2 \theta) \frac{\Delta \partial_r \partial_\theta h_{\ell}(r,\theta)}{\ell (\ell+1)} .
\end{align}
The trace equation is straightforwardly separable, with $h_{\ell} (r, \theta) = h_{\ell}(r) Y_{\ell 0} (\theta)$.  
In the Kerr case due to the coupling factor $\Sigma = r^2 + a^2 \cos^2 \theta$, a single $\ell$-mode of the trace will give contributions to $\{ \ell-2, \ell, \ell+2 \}$ harmonics of $B_{\ell}$ and $\kappa_{\ell}$.
More precisely, let
\beq
\kappa_{\ell}(r,\theta) = \sum_{k=-1}^{+1} \hat{\kappa}_{\ell k}(r) Y_{\ell+2k}(\theta) ,  \quad \quad
B_{\ell}(r,\theta) = \sum_{k=-1}^{+1} \hat{B}_{\ell k}(r) \partial_\theta Y_{\ell+2k}(\theta) , \label{eq:kappaB-lmodes}
\eeq
where the radial functions $\hat{\kappa}_{\ell k}(r)$ and $\hat{B}_{\ell k}(r)$ satisfy 
\begin{align}
\left\{ \partial_r \Delta \partial_r - \lambda_k \right\} \hat{\kappa}_{\ell k} &= \tfrac{1}{2} \left( \delta_{k0} \, r^2 h_{\ell}(r) + a^2 \left( \mathfrak{c}^2 \right)^{00}_{\ell,\ell+2k} h_{\ell}(r) \right) , \\
\left\{ \Delta \partial_r \partial_r - \lambda_k  \right\} \hat{B}_{\ell k} &= \frac{1}{2}  \Delta \left(\frac{1}{\ell (\ell+1)}  \delta_{k0} \, r^2 \partial_r h_{\ell}(r) + a^2 \left( \mathfrak{c}^2 \right)^{11}_{\ell,\ell+2k} \partial_r h_{\ell}(r) \right) ,  \label{eq:Bradial1}
\end{align}
with $\lambda_k \equiv  (\ell+2k)(\ell+2k+1)$ and the mixing coefficients as defined in Eq.~(\ref{eq:mixing-coeffs}).
Now decompose as follows,
\begin{subequations}
\begin{align}
\hat{\kappa}_{lk}(r) &= \tfrac{1}{2} \left( \delta_{k0} X_\ell(r) + a^2 \left( \mathfrak{c}^2 \right)^{00}_{\ell,\ell+2k}  Z_\ell^{[k]}(r) \right) , \\
\hat{B}_{lk}(r) &= \frac{1}{2} \Delta \left(  \frac{\delta_{k0}}{l(l+1)} (\partial_r X_\ell(r) + \delta B_\ell(r) )  + a^2 \left( \mathfrak{c}^2 \right)^{11}_{\ell,\ell+2k}  \partial_r Z^{[k]}_\ell \right) ,
\end{align}
\label{eq:kappaB-lk}
\end{subequations}
where the radial functions $X_\ell(r)$, $\delta B_\ell$ and $Z^{[k]}_\ell(r)$ with $k \in \{-1,0,1\}$ satisfy
\begin{align}
\left\{ \partial_r \Delta \partial_r - \lambda \right\}  X_\ell &= r^2 h_l , \label{eq:Xdef} \\
\left\{ \partial_r \Delta \partial_r - \lambda_k \right\} Z^{[k]}_\ell &= h_l ,  \label{eq:Zdef} \\
\left\{ \Delta \partial_r \partial_r - \lambda \right\} ( \Delta \delta B_{\ell} ) &= - 2 r \Delta h_{\ell} .  \label{eq:Bradial2}
\end{align}
Since $h_l(r)$ satisfies $\left\{ \partial_r \Delta \partial_r - \lambda \right\} h_l = 0$, the solutions for $Z_\ell^{[+1]}(r)$ and $Z_\ell^{[-1]}(r)$ are straightforward: 
\beq
Z^{[\pm 1]}_\ell(r) = h_{l}(r) / (\lambda - \lambda_{\pm 1}).
\eeq

In summary, the $s=0$ static vacuum Lorenz-gauge metric perturbation with trace $h$ is constructed from the gauge vector (\ref{eq:xi-gauge-scalar-static}) with the functions $\kappa(r,\theta)$ and $B(r,\theta)$ expanded into $\ell$ modes via Eqs.~(\ref{eq:kappaB-expansion}) and (\ref{eq:kappaB-lmodes}), and constructed from radial functions $\{ X_{\ell}(r), Z_\ell^{[k]}(r), \delta B_{\ell}(r) \}$ in Eq.~(\ref{eq:kappaB-lk}) satisfying Eqs.~(\ref{eq:Xdef}), (\ref{eq:Zdef}) and (\ref{eq:Bradial2}).
It turns out that $X_\ell(r)$ is not required, as it cancels out in the gauge vector (\ref{eq:xi-gauge-scalar-static}).

\subsection{Completion pieces\label{subsec:completion}}
The completion pieces are parts of the metric perturbation associated with the non-radiative, low-multipole ($m=0$, $\ell = 0$ and $\ell =1$) sector, that cannot be determined from Teukolsky scalars or the trace alone. They comprise physical mass and angular momentum perturbations, as well as gauge degrees of freedom. These modes are known in Lorenz gauge for the Schwarzschild case ($a=0$) in e.g.~Sec.~V of Ref.~\cite{Dolan:2012jg}. The Kerr case was partially considered in Ref.~\cite{Dolan:2021ijg}.

In Table \ref{tbl:completion}, we present a complete basis of Lorenz-gauge completion modes for the Kerr spacetime. Individually, each mode satisfies the vacuum field equations and the Lorenz-gauge condition, but not necessarily the physical boundary conditions at horizon or at infinity. The labelling is chosen to correspond to Ref.~\cite{Dolan:2012jg} as far as possible. Four of these modes are pure gauge: (B) and (C) are scalar gauge modes ($\xi_{[\mu ; \nu]} = 0)$, (D) is a vector gauge mode ($\xi_{[\mu ; \nu]} \neq 0$) and (F) is a dipole gauge mode. 

\begin{table}
\begin{center}
\begin{tabular}{l c l r r}
\hline \hline
Mode \quad & Metric pert.~or gauge vector \quad & Trace $h$ \quad & $Q_{(t)}$ & $Q_{(\phi)}$ \quad \\
\hline 
(A) & $g_{\mu \nu}$  & $h = 4$ & $M/2$ & $-aM$ \\
(B) & $\xi_\mu =  \left[ 0, \frac{r(r^2 + a^2)}{\Delta}, a^2 \sin \theta \cos \theta, 0 \right] - 2Mr_+^2 \xi_\mu^{(C)}$ & $h=6$ & $0$ & $0$ \\
(C) & $\xi_\mu =  \left[ 0, 1/\Delta, 0, 0 \right]$ & $h = 0$ & $0$ & $0$ \\
(D) & $\xi^\mu =  [t, \frac{M a^2 \cos^2\theta}{\Sigma}, 0, 0] + M \nabla^\mu y$ \quad \quad & $h = 2 + \frac{4M}{(r_+ - r_-)} \ln \left( \tfrac{r - r_+}{r - r_-} \right)$ \quad & $0$ & $0$ \\
(E) & $\frac{\partial}{\partial M} g_{\mu \nu} - 2 \nabla_\mu \nabla_\nu y$ & $h =  -\frac{4}{(r_+ - r_-)} \ln \left( \tfrac{r - r_+}{r - r_-} \right)$ & $1$ & $-a$ \\
\hline 
(F) & $\xi^\mu = \left[ 2 a t, - \frac{a M (r^2 - a^2) \cos^2 \theta}{\Sigma}, 0, t \right] + \nabla^\mu z$ & $h = 4 \, a$ & 0 & 0 \\
(G)& $\tfrac{\partial}{\partial a} g_{\mu \nu} - 2 \xi_{(\mu ; \nu)} , \quad \xi_\mu = [0, \frac{a (r \sin^2\theta + M\cos^2 \theta)}{\Delta}, a \sin \theta \cos \theta, 0 ]$ \; & $h=0$ & 0 & $-M$
 \\ \hline \hline
\end{tabular}
\end{center}
\caption{Completion pieces and gauge modes. All the above generate metric perturbations in Lorenz gauge in vacuum. The gauge modes are constructed as $h_{\mu \nu} = \xi_{\mu ; \nu} + \xi_{\nu ; \mu}$, and the trace is $h = g^{\mu \nu} h_{\mu \nu}$ ($h = 2 \nabla_\mu \xi^\mu$ for gauge modes). The scalar functions $y$ and $z$ are defined in Eq.~(\ref{eq:yz}). The labelling (A), (B), etc.~was chosen for consistency with the Schwarzschild completion modes listed in Ref.~\cite{Dolan:2012jg}.
}
\label{tbl:completion}
\end{table}

The mass and angular momentum content of each mode is determined by evaluating the conserved charges $Q_{(t)}$ and $Q_{(\phi)}$ associated with the Killing vectors of Kerr spacetime: see Sec.~IIE in Ref.~\cite{Dolan:2021ijg} for details. For the conformal (A), energy (E) and angular momentum (G) modes, these quantities are non-zero and are given in the last two columns of Table \ref{tbl:completion}. For all the pure-gauge modes, these quantities are trivially zero.

The set of modes is not fully linearly independent, as the conformal mode (A) can be constructed from a linear combination of the other modes (see Eq.~(37) in Ref.~\cite{Dolan:2012jg}). 

\subsubsection{Scalar equations}
The construction of the (E) and (F) modes requires the solution of equations for the scalars $y$ and $z$. These satisfy
\begin{subequations}
\begin{align}
\Box y &= \frac{2}{(r_+  - r_-)} \ln \left( \frac{r - r_+}{r - r_-} \right), \\
\Box z &= \frac{2 a M r \cos^2 \theta}{\Sigma} , 
\end{align}
\label{eq:yz}
\end{subequations}
Closed-form solutions can be found by applying separation of variables, after casting in the form
\begin{subequations}
\begin{align}
\left\{ \partial_r \Delta \partial_r + \partial_\theta \partial_\theta + \cot \theta \partial_\theta \right\} z &= \frac{2}{r_+ - r_-} \left( (r^2 + \tfrac{1}{3} a^2) + \tfrac{2}{3} a^2 P_2 (\cos \theta) \right) , \\
\left\{ \partial_r \Delta \partial_r + \partial_\theta \partial_\theta + \cot \theta \partial_\theta \right\} z &= 2 a M r \left( \tfrac{1}{3} + \tfrac{2}{3} P_2(\cos \theta) \right) .
\end{align}
\end{subequations}
In solving the equations, the complementary function is chosen to make $z$ as regular as possible at the outer horizon $r=r_+$. Then $z$ can be written in closed form as
\begin{align}
z &= \frac{1}{3} a M \left\{ r - r_+ + 2M \ln \left(\frac{r-r_-}{r_+ - r_-} \right)  
 -  \frac{(r-r_+)((r_- - 5r_+ ) r + r_+ ( r_+ + 3 r_-) )}{4(M^2-a^2)} P_2(\cos \theta) \right\} .
\end{align}
Similarly, $y$ can be determined in closed form, but the result is too long to quote here.

\subsection{Construction of the metric perturbation}

The axially-symmetric metric perturbation is the sum of three parts,
\beq
h^{(m=0)}_{\mu \nu} = \hcomp_{\mu \nu} + h^{(s=2)}_{\mu \nu} + h^{(s=0)}_{\mu \nu} . \label{eq:hm0}
\eeq
Here $\hcomp_{\mu \nu}$ is the completion piece (Sec.~\ref{subsec:completion}), $h^{(s=2)}_{\mu \nu}$ is derived from the Teukolsky scalars (Sec.~\ref{subsec:static-s2}), and  $h^{(s=0)}_{\mu \nu}$ generates the trace (Sec.~\ref{subsec:static-s0}). The physical solution is constructed by glueing IN and UP modes at the circular-orbit radius $r=r_0$,
\beq
h^{(m=0)}_{\mu \nu}(r,\theta) = \Theta(r-r_0) h^{\text{up}}_{\mu \nu}(r, \theta) +   \Theta(r_0-r) h^{\text{in}}_{\mu \nu}(r, \theta) , 
\eeq
with corresponding expressions for the three pieces individually. Here IN (UP) is a solution that satisfies physical boundary conditions at the outer horizon (at spatial infinity). The IN and UP homogeneous solutions also split into three parts, as in Eq.~(\ref{eq:hm0}). 

The completion piece is made from a sum of completion modes,
\beq
h^{(\text{comp.}) in/up}_{\mu \nu} =\sum_{\Lambda \in \{B,C,D,F,E,\text{G}\}} c^{\text{in/up}}_\Lambda h^{(\Lambda)}_{\mu \nu} 
\eeq
where $h^{(\Lambda)}_{\mu \nu} $ are listed in Table \ref{tbl:completion}
(N.B.~the conformal mode (A) is not required due to linear dependence). The jumps in the coefficients,
\beq
\Delta c_{\Lambda} = c_{\Lambda}^{\text{up}} - c_{\Lambda}^{\text{in}} , \label{eq:Delta-c-def}
\eeq 
are determined by the field equations, whereas to obtain individual values of $c_{\Lambda}^{\text{up}}$ and $c_{\Lambda}^{\text{in}}$ one must also impose boundary conditions. The mass and angular momentum conditions (i.e.~jumps in $Q_{(t)}$ and $Q_{(\phi)}$) give
\beq
\Delta c_{E} = E , \quad \quad \Delta c_{G} = (L - a E) / M ,  \label{eq:EG}
\eeq
where $E$ and $L$ are the particle's specific energy and angular momentum, respectively (here setting the particle mass $\mu$ to unity). The remaining jumps are determined numerically, as detailed below.

\subsubsection{Boundary conditions and the Berndtson solution\label{subsec:Berndtson}}
In the Schwarzschild case, it is well-known that the Lorenz-gauge metric perturbation with the correct mass and angular momentum at infinity, and which is regular at the horizon, is not asymptotically flat at spatial infinity: $h_{tt}$ approaches a non-zero constant value as $r\rightarrow\infty$. On the other hand, there exists a Lorenz-gauge solution which is asymptotically flat, and regular on the horizon, but which has the incorrect mass; this is the so-called Berndtson solution \cite{Berndtson:2007gsc}. The difference between the two solutions is a homogeneous completion mode.

Here we seek to construct the analogue of the Berndtson solution for Kerr spacetime: a metric perturbation which is regular on the boundaries, but which consequently has the incorrect mass \emph{and} incorrect angular momentum. This is the metric perturbation which emerges naturally from the 2+1D time domain code. If required, one can then add homogeneous completion modes to restore the correct mass and angular momentum, at the expense of regularity at spatial infinity and/or the horizon.  

To be asymptotically flat, the up coefficients of the B, D and F modes must be zero. At the horizon, the mode $B$ is regular, and one can form two further horizon-regular modes by taking a linear combination of the C, D, E and G modes, and a linear  combination the C, F and G modes. This is sufficient to fully determine the in and up coefficients, by using Eq.~(\ref{eq:Delta-c-def}) once the jumps are known. In summary, 
\begin{align}
c^{\text{in}}_C &= -2 M r_+^2 \Delta c_D - \frac{a}{6} \left( r_-^3 + r_-^2 r_+ + 5 r_- r_+^2 - 7 r_+^3 \right) \Delta c_F ,  & 
c^{\text{up}}_B &= 0 , \\
c^{\text{in}}_E &= -M \Delta c_D , & 
c^{\text{up}}_D &= 0 , \\
c^{\text{in}}_G &= -a \Delta c_D - \frac{r_+ (r_+ - r_-)^2}{r_+ + r_-} \Delta c_F , & 
c^{\text{up}}_F &= 0 . 
\end{align} 
where $r_\pm = M \pm \sqrt{M^2 - a^2}$. The remaining in/up coefficients are found by application of Eq.~(\ref{eq:Delta-c-def}). 

If instead one seeks a non-asymptotically-flat solution with the correct mass and angular momentum, one requires $ c^{\text{in}}_E =  c^{\text{in}}_G = 0$, and so to maintain regularity at the horizon, this implies that $c^{\text{in}}_C = c^{\text{in}}_D = c^{\text{in}}_F = 0$ too.

\subsection{Implementation details\label{subsec:implementation2}}

\subsubsection{Smoothing and jumps}
The full $m=0$ metric perturbation for a particle on a circular orbit at $r=r_0$ is constructed from: (i) the completion pieces; (ii) the radial functions $P_{-2 \ell}(r)$ and $h_\ell(r)$ that are determined uniquely by the Teukolsky and trace equations, respectively; and (iii) a stack of radial functions: $\{ \hat{\chi}^{(e)}_\ell, \hat{\chi}^{[k]}_{(e) \ell}, 
\hat{\chi}^{(o)c}_{[j]\ell}, \hat{\chi}^{(o)s}_{[j]\ell} \}$ arising in the $s=2$ sector from solving the $\chi$ equation, and $\{ Z_\ell^{[k]}(r), \delta B_{\ell}(r) \}$ arising in the $s=0$ sector. The radial functions in the stack satisfy ODEs with extended source terms determined from either $P_{-2 \ell}(r)$ or $h_\ell(r)$. For each radial function in the stack, a composite function can be made by glueing at $r=r_0$ an UP mode (regular at infinity and used for $r>r_0$) and an IN mode (regular at the horizon and used for $r<r_0$). The IN and UP functions are determined by the boundary conditions and the sourced equations, but only up to one complementary function of either side of $r=r_0$ (i.e.~an additive homogeneous degree of freedom). Naturally, the question arises of how to determine these homogeneous degrees of freedom.

We take the following approach. Where possible, we make a choice of complementary functions such that the composite radial functions in the stack are $C^1$ at $r=r_0$ (i.e.~continuous and differentiable). The jumps in the second and higher derivatives at $r=r_0$ then follow from the ODEs that the radial functions satisfy on either side of $r=r_0$. This smoothing is applied to $\{ \hat{\chi}^{(e)}_\ell, \hat{\chi}^{[0]}_{(e) \ell}, \hat{\chi}^{(o)c}_{[j]\ell}, \hat{\chi}^{(o)s}_{[j]\ell}, Z_\ell^{[0]}, \delta B_{\ell} \}$. On the other hand, the functions $\hat{\chi}^{[\pm1]}_{(e) \ell}$ and $Z_\ell^{[\pm1]}(r)$ are defined directly in terms of $h_\ell(r)$, and so these functions inherit their properties at $r=r_0$ directly from the in and up modes of the trace. With this \emph{choice} of $C^1$ radial functions, the resulting metric perturbation does \emph{not} satisfy the Lorenz-gauge field equations, but this is rectified with a final step. We now add a free-scalar mode, $h_{\mu \nu} = \nabla_\mu \nabla_\nu \kappa^{(0)}$, $\Box \kappa^{(0)} = 0$, $\kappa^{(0)} = \sum_\ell \kappa_\ell^{(0)}(r) Y_{0\ell}(\theta)$ in the scalar sector. Restoring this degree of freedom is necessary and sufficient to allow for a regular solution that satisfies the $m=0$ field equations. In short, the jumps at $r=r_0$ in the free-scalar modes are fixed by demanding continuity in the $l$-modes of any two components of the metric perturbation, and we then verify that the $l$-modes of all five independent components of the metric perturbation have the behaviours at $r=r_0$ (i.e.~continuity, with jumps in first derivatives that follow from the projections of the stress-energy $T_{\mu \nu}$). There is an element of choice in the above procedure; for example, we could equally well have removed the smoothing from $Z_\ell^{[0]}$ to restore two degree of freedom for each $\ell \ge 2$ (i.e.~the jumps in $\kappa_\ell^{(0)}$ and $\partial_r \kappa_\ell^{(0)}$ at $r=r_0$).

\subsubsection{Metric components}
For circular orbits, the $m=0$ static metric perturbation is described by five tetrad components:
\beq
\left[ h_{l_+ l_+} , h_{m_+ m_+}, \rho h_{l_+ m_+} , \Sigma \Delta h_{l_+ l_-}, h \right]. 
\eeq
The $\ell m$ spin-weighted spherical modes of these components are defined as in Sec.~\ref{sec:construction}. The $\ell m$ components $\hrm_{l_+ l_+}^{\ell 0}, \hrm_{m_+ m_+}^{\ell 0}, \Sigma \Delta \hrm_{l_+ l_-}^{\ell 0}, \hrm^{\ell 0}$ are real for even $\ell$, and are zero for odd $\ell$. The $\ell m$ component $\rho \hrm_{l_+ m_+}^{\ell 0}$ is real for even $\ell$, and imaginary for odd $\ell$. This property is consistent with Eqs.~(\ref{eq:symmetries}) after recalling that ${}_{-1}Y_{\ell 0}(\theta,\phi) = - {}_{+1}Y_{\ell 0}(\theta,\phi)$. The symmetries (\ref{eq:symmetries}) can be applied to determine the other five projections of the metric perturbation.

\subsubsection{Determining the jumps}
The method applied is similar to that described in Sec.~\ref{subsec:jumps1}. The jumps in the radial functions $P_{-2 \ell}$ and $h_\ell$ follow directly from the decoupled, separable Teukolsky and trace equations: see Eqs.~(\ref{eq:jumps}) and Eqs.~(\ref{eq:jumps-h}), after setting $m=\omega=0$ and using $J_k^{(P_{+2})} = J_k^{(P_{-2})}$  for $m=0$. The jumps in the free-scalar mode, and the completion pieces, are determined from the condition of regularity on the sphere, that is, the requirement that the $\ell m$ modes of the metric perturbation are $C^0$. 

To determine the jumps $\{\Delta c_B , \Delta c_C , \Delta c_D , \Delta c_F\}$ (with the jumps $\{\Delta c_E, \Delta c_G \}$ specified in Eq.~(\ref{eq:EG})), and $\{J_{0}^{(\kappa)}, J_{1}^{(\kappa)} \}$ for $\ell = 2, \ldots, \ell_{\text{max}}$, we used the condition of continuity in the components
\begin{align}
\hrm^{\ell=0}_{l_+ l_+} , \quad \Sigma \Delta \hrm^{\ell=0}_{l_+ l_-} , \quad \rho \hrm_{l_+ m_+}^{\ell=1}, \quad \rho \hrm_{l_+ m_+}^{\ell=2} , \nn \\
\hrm^{\ell}_{l_+ l_+},  \quad \hrm^{\ell}_{m_+ m_+}, \quad \ell = 2,4, \ldots, \ell_{\text{max}}.
\end{align}
A linear system of equations was set up and solved as described in Sec.~\ref{subsec:jumps1}. 
The components below were used as consistency checks:
\begin{align}
\rho \hrm^{\ell}_{l_+ m_+},  \quad \rho \hrm^{\ell+1}_{l_+ m_+}, \quad \Sigma \Delta \hrm^{\ell}_{l_+ l_-}, \quad \hrm^{\ell}, \quad  \ell = 2,4, \ldots, \ell_{\text{max}}.
\end{align}

Table \ref{tbl:jumps-m0-a06} lists sample results for the jumps in the radial functions and the completion coefficients in the $m=0$ sector, for a particle on a circular orbit at $r_0=6M$ about a spinning black hole with $a=0.6M$. 

\begin{table}
$$
 \begin{array}{|r|r|r|r||r|r|}
 \hline
 \ell & J_{0}^{(P_{2})} & J_{1}^{(P_{2})} & J_{1}^{(h)} & J_{0}^{(\kappa)} & J_{1}^{(\kappa)} \\ 
 \hline
 2 & 0.18205606 & -3.1459024 & 0.47686532 & -0.70225908 & 1.2944813 \\
 3 & 1.3498380 & 0.52293035 & 0 & 0 & 0 \\
 4 & -0.010511011 & 0.81732941 & -0.47983647 & 0.0073276691 & -0.48860610 \\
 5 & -0.31977902 & -0.12388313 & 0 & 0 & 0 \\
 6 & 0.0021929201 & -0.37893342 & 0.48057639 & -0.00083196519 & 0.23654968 \\
 7 & 0.12300649 & 0.047652996 & 0 & 0 & 0 \\
 8 & -0.00072391066 & 0.21890875 & -0.48086511 & 0.00018476421 & -0.13879233 \\
 9 & -0.059880516 & -0.023197849 & 0 & 0 & 0 \\
 10 & 0.00030569899 & -0.14262573 & 0.48100652 & -0.000059441396 & 0.091094585 \\
 \hline
\end{array}
$$

$$
\begin{array}{|r|r|}
 \hline
 \Delta c_B & -0.17354798 \\
 \Delta c_C  & -2.7383014 \\
 \Delta c_D  & 0.19491814 \\
 \Delta c_F  & 0.025072823 \\
 \Delta c_E & 0.92766353 \\
 \Delta c_G & 2.4250527 \\
 \hline
\end{array}
$$

 \caption{Jumps in radial functions and completion mode coefficients for $a = 0.6M$, $r_0 = 6M$.}
 \label{tbl:jumps-m0-a06}
\end{table}

\subsubsection{Radial functions}

To calculate the metric perturbation itself, it is necessary to obtain `in' and `up' solutions to the radial ODEs \eqref{eq:staticP0}, \eqref{eq:staticP2}, \eqref{eq:chieven}, \eqref{eq:chiek}, \eqref{eq:chiodd}, \eqref{eq:chioj}, \eqref{eq:Zdef} and \eqref{eq:Bradial2} for the stack of radial functions $\{ P_{-2\ell}(r), h_{\ell}(r), \hat{\chi}^{(e)}_\ell(r), \hat{\chi}^{[k]}_{(e) \ell}(r), \hat{\chi}^{(o)c}_{[j]\ell}(r), \hat{\chi}^{(o)s}_{[j]\ell}(r), Z_\ell^{[k]}(r), \delta B_{\ell}(r) \}$. 

Equation (\ref{eq:staticP0}) for $P_0(r)$ has solutions in terms of Legendre functions $P_\ell(x)$ and $Q_\ell(x)$ with $x \equiv \Delta_{,r} / (r_+ - r_-)$ (with branch cut on $x \in (-\infty,1)$). This gives the in and up solutions, respectively, for the trace $h_\ell (r)$. Moreover, applying differential operators as in Eq.~(\ref{eq:degeneracy}) yields in and up solutions of $P_{-2\ell}(r)$ satisfying Eq.~(\ref{eq:staticP2}). With these as inputs for the right-hand sides, we then used Mathematica to obtain `in' and `up' solutions in closed form for the full stack of radial functions. The `in' solutions consist of finite polynomials in $x$ composed with $\{1, \ln(1 + x)\}$, and the `up' solutions consist of finite polynomials in $x$ composed with $\{1, \ln(1 + x), \text{Li}_2((1-x)/2) \}$ (i.e.~the polylog function).

\subsection{Results: the $m=0$ metric perturbation\label{subsec:results-m0}}

\begin{figure}
 \begin{center}
  \includegraphics[width=16.5cm]{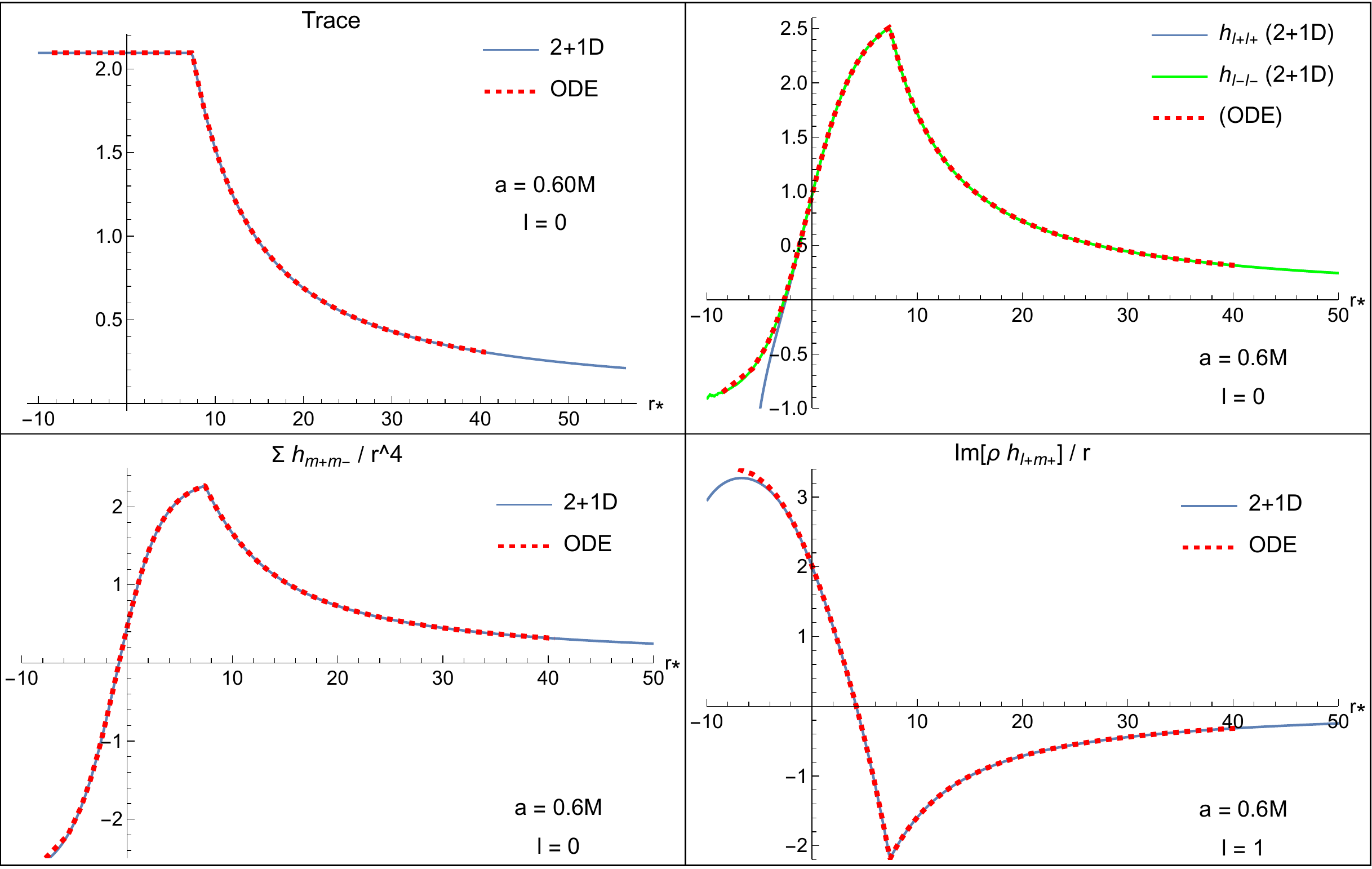} \\ 
    \caption{Static axisymmetric modes ($\ell=0$ and $\ell=1$). Comparing results of the new ODE method and the 2+1D code for $m=0$, $a=0.6M$ and $r_0=6M$. The first three plots show the (even-parity) components for $\ell = 0$, and the bottom-right plot shows (odd-parity) $\ell = 1$.
 \label{fig:m0l01}}
 \end{center}
\end{figure}

Figure \ref{fig:m0l01} shows the $\ell = 0,1$ modes of the metric perturbation, comparing the results of the approach described in this paper, with the results of the 2+1D time-domain code that was used in Ref.~\cite{Isoyama:2014mja}. The two data sets are in good agreement. 

The agreement shown in Fig.~\ref{fig:m0l01} is notable for two reasons. First, the construction of the solution in the 2+1D time-domain code is far from trivial, since it involves the use of a \emph{generalised} Lorenz gauge to tame a linear-in-$t$ gauge mode instability, specifically, a gauge in which $\nabla_{\mu} \bar{h}^{\mu \nu} \propto h_{tr}$. The solution shown in Fig.~\ref{fig:m0l01} [solid lines] is a linear combination of data from \emph{three} runs using three sets of initial data. A linear combination is taken to eliminate the $h_{tr}(r)$ component at two points in $r$; and this appears to have the effect of minimising $h_{tr}(r)$ across much of the domain, meaning that Lorenz gauge is (approximately) restored. However, this is a numerical approach, and there is associated numerical error and residual gauge violation. 

Second, it is notable that the time-domain solution, as constructed above, appears to approach the Kerr analogue of the Berndtson solution that we sought to construct explicitly in Sec.~\ref{subsec:Berndtson}. Without the completion pieces of Sec.~\ref{subsec:completion} to hand, this could not be demonstrated until now. We stress that the Kerr version of the Berndtson solution has the incorrect mass \emph{and} the incorrect angular momentum, in the sense that the additional mass and angular momentum associated with the metric perturbation are \emph{not} zero in the inner region ($r<r_0$), but that this can be simply remedied by adding a homogeneous completion piece at the expense of violating regularity at the boundaries; and in turn, regularity can be restored by applying a gauge transformation that moves out of Lorenz gauge.

The discrepancy for small-$r_\ast$ (i.e.~close to the horizon) visible in Fig.~\ref{fig:m0l01} is likely to be due to inaccuracy in the time-domain data due to the issues associated with the use of a generalised Lorenz gauge to tame the linear-in-$t$ gauge mode instability, discussed above. % This will be examined more closely in Ref.~\cite{Dolan-Barack-Wardell}.

\begin{figure}
 \begin{center}
  \includegraphics[width=16.5cm]{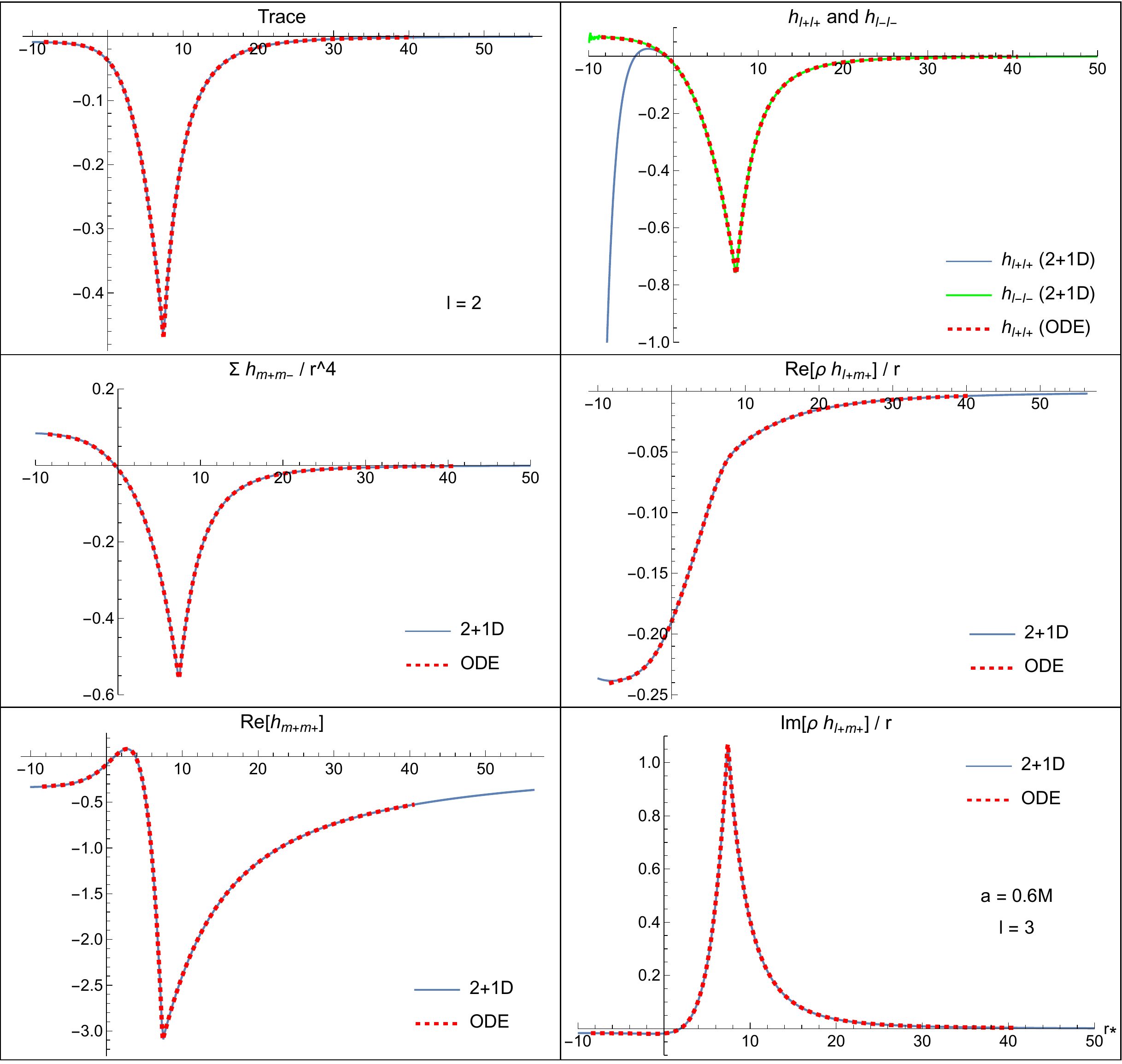} \\ 
    \caption{Static axisymmetric modes ($\ell=2$ and $\ell=3$). Comparing results of the new ODE method and the 2+1D code for $m=0$, $a=0.6M$ and $r_0=6M$. The first five plots show the (even-parity) components $\ell = 2$, and the bottom-right plot shows (odd-parity) $\ell = 3$.
 \label{fig:m0l23}}
 \end{center}
\end{figure}

\begin{figure}
 \begin{center}
  \includegraphics[width=16.5cm]{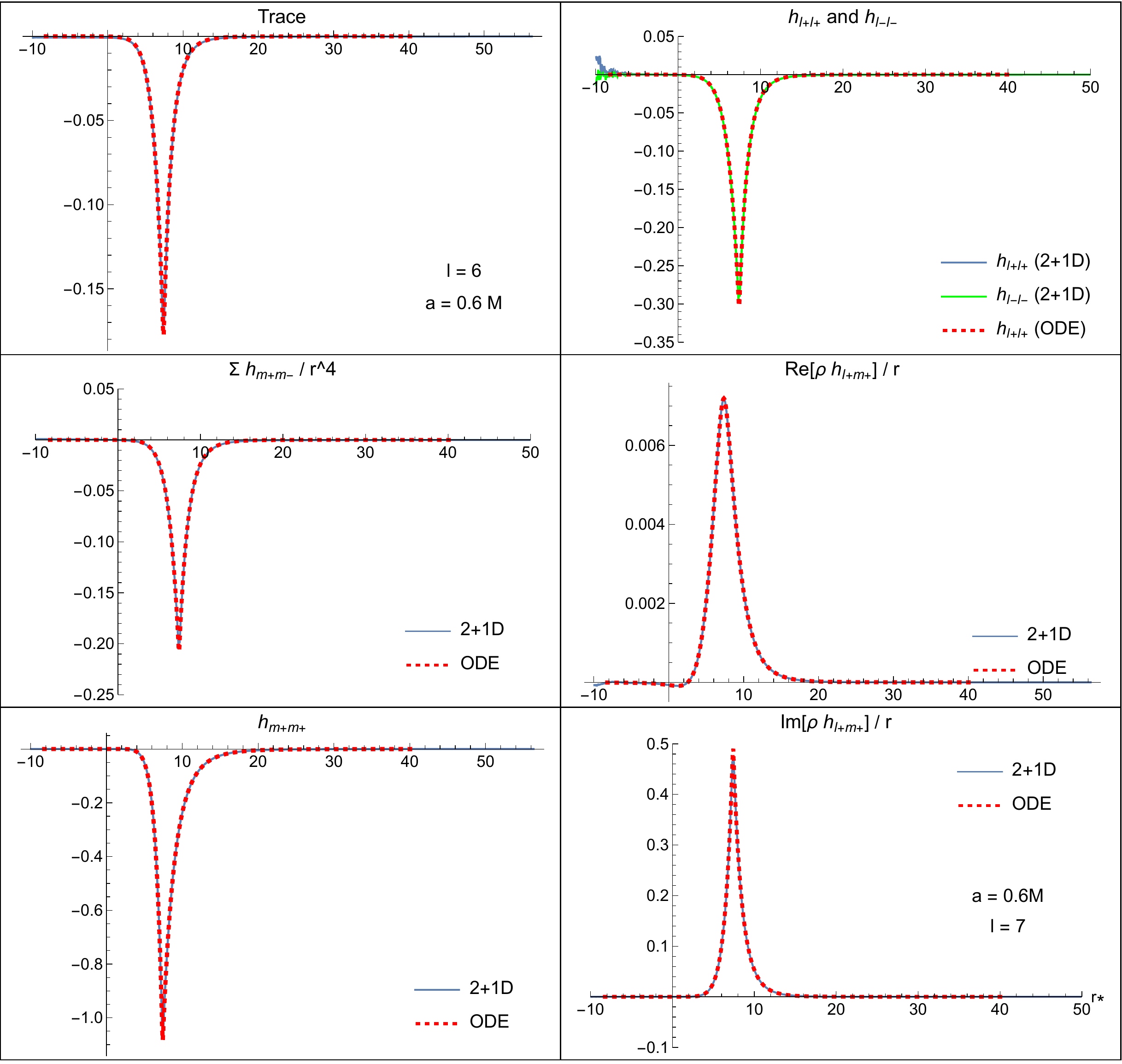} \\ 
    \caption{Static axisymmetric modes ($\ell=6$ and $\ell=7$). Comparing results of the new ODE method and the 2+1D code for $m=0$, $a=0.6 M$ and $r_0=6M$. The first five plots show the (even-parity) components $\ell = 6$, and the bottom-right plot shows (odd-parity) $\ell = 7$.
 \label{fig:m0l67}}
 \end{center}
\end{figure}

Figures \ref{fig:m0l23} and  \ref{fig:m0l67} show comparisons for the $\ell = 2, 3$ modes, and for the $\ell = 6, 7$ modes. There is good agreement between the data sets across most of the domain in $r$. In Fig.~\ref{fig:m0l23} there is a discrepancy close to the horizon, particularly in the $\hrmlm_{l_+ l_+}$ component, which is likely to be due to the challenges for the time-domain approach discussed above.

We take the agreement in all components shown in Figs.~\ref{fig:m0l01}, \ref{fig:m0l23} and \ref{fig:m0l67} to be persuasive evidence that the correct metric perturbation is recovered from the new method in the $m=0$ sector, and that the restoration of completion pieces has been correctly handled.

\section{Discussion and conclusions\label{sec:conclusions}}

In this paper we have successfully reconstructed the Lorenz-gauge metric perturbation on Kerr spacetime in the frequency domain, for the special case of a particle on a circular equatorial orbit. We have shown that the reconstructed metric perturbation is in good agreement with the numerical results of the 2+1D time-domain code \cite{Dolan:2012jg, Isoyama:2014mja, Dolan-Barack-Wardell}, in both static ($\omega  = 0$) and non-static ($\omega\neq0$) sectors (see Secs.~\ref{subsec:results-lm}, \ref{subsec:results-m} and \ref{subsec:results-m0}). Moreover, our results compare well with Schwarzschild data in the $a=0$ case (Figs.~\ref{fig:schw-m2a} and \ref{fig:schw-m2b}). Our aim here has been to demonstrate that this is an efficient and accurate method for calculating metric perturbations of Kerr spacetime, as it merely requires the solutions to ODEs (rather than PDEs) and certain linear systems of equations.

The method builds upon foundations set in the pioneering works of the 1970s by Teukolsky \cite{Press:1973zz,Teukolsky:1972my,Teukolsky:1973ha,Teukolsky:1974yv}, Chandrasekhar \cite{Chandrasekhar:1985kt}, Chrzanowski \cite{Chrzanowski:1975wv}, Wald \cite{Wald:1978vm} and others \cite{Cohen:1974cm}. Fundamentally, the metric perturbation is built from Teukolsky scalars of spin-weights $\pm2$, $\pm1$ and $0$, as well as certain auxiliary scalars. The re-use of the Teukolsky formalism comes with a practical advantage: accurate and well-tested methods for calculating Teukolsky functions are available in a community code, the Black Hole Perturbation Toolkit. Methods for calculating other elements, for example the functions $\kappa_{\ell\ell}(r)$, can feasibly be added to the codebase too.

A key motivation behind this work is the necessity of calculating first-order metric perturbations in sufficiently-regular gauges (such as Lorenz) at sufficient accuracy that they can serve as inputs for second-order self-force calculations; let us review the progress achieved towards this goal. Firstly, the mode functions $P_{s\ell m}(r)$ and $S_{s\ell m}(\theta)$ (and thus the modes of the metric perturbation) can be computed to very high accuracy. Likewise, the linear equations determining the jumps can be solved to high precision, and these equations do not present a significant source of error. In the current implementation, the leading source of numerical error is typically in the calculation of $\kappa_{\ell \ell}(r)$, which satisfies an ODE with an extended source, Eq.~(\ref{eq:kappa-ll}). This does not present a `hard' limit on accuracy because numerical error in $\kappa_{\ell \ell}(r)$ can be  reduced (at the expense of run time) by changing the internal parameters of the ODE solver and the IN/UP boundary conditions (see Appendix \ref{appendix:kappa-direct}). A more fundamental practical limitation on the accuracy of the metric reconstruction is the truncation of the infinite $\ell$-mode sum at some $\ell_{\text{max}}$. As shown in Figs.~\ref{fig:m2-3D}, \ref{fig:m2-radial}, and \ref{fig:m2-angular}, truncating the sums leads to significant (but regular) error in the vicinity of $r=r_0$ and across the domain in $\theta$. This is entirely expected, as it is a fundamental feature of a mode-sum representation of a function with a $1/r$-type divergence. It should be noted that the truncated sum will give a solution of superior accuracy across most of the domain in $(r,\theta)$ (i.e.~away from $r=r_0$) and in particular, in the asymptotic regimes, $r_\ast \rightarrow \pm \infty$. To mitigate the truncation error near the particle, we anticipate that, as well as using very high $\ell$-mode truncations, it will be helpful to use a puncture scheme, and work in this direction is underway. In addition, if necessary for $r \approx r_0$ other data sets can be used, such as $2+1$D time-domain codes \cite{Dolan:2012jg, Dolan-Barack-Wardell} or the elliptic-PDE approach under development in Ref.~\cite{Osburn:2022bby}. 

The axisymmetric, static sector ($m=0$, $\omega=0$) required special attention for two reason. First, the transformation to Lorenz gauge described in Ref.~\cite{Dolan:2021ijg} breaks down, and the difference between IRG and ORG metric perturbations becomes pure gauge. This caused us to seek out an alternative transformation to Lorenz gauge, which was somewhat less elegant but which nevertheless resulted in a practical scheme for calculations, due to the simplifications that are possible for $\omega=0$. Second, in this work we have elucidated the completion pieces in Lorenz gauge (Sec.~\ref{subsec:completion}). Putting these elements together, we were able to construct the $m=0$ Lorenz gauge solution in full. As expected from the Schwarzschild limit ($a=0$), the metric perturbation which is regular at the horizon and at infinity comes with the incorrect mass (as measured at infinity). In the Kerr case we found that the solution also has incorrect angular momentum. We confirmed (numerically) that the metric perturbation constructed in the frequency domain is in agreement with that from the time-domain code, which seeks out this (boundary-regular) solution naturally. Restoring the correct mass and angular momentum to the solution is simply a matter of adding the homogeneous solutions (E) and (G). This will lead to the violation of regularity at the boundaries, but this is an expected feature of the insistence on global Lorenz gauge, and not unexpected.

In the $m=0$ and $m=1$ sectors, the time-domain approach in Ref.~\cite{Dolan:2012jg} was impeded by linear-in-$t$ instabilities within Lorenz gauge. Our new method does not encounter this difficulty, since in the frequency domain such behaviour is naturally excluded. Importantly, by comparing with the results of the new method, we have verified that the steps taken in the time domain approach to eliminate the instability (i.e.~the use of a Generalised Lorenz Gauge and a linear combination for $m=0$, and the use of a frequency filter for $m=1$) are working as hoped,  with an associated numerical error which can now be quantified. In future, the new method will be employed to construct initial data for the time-domain code, to minimise the excitation of the instability. 

One aim of the work has been to clarify the relationship between Lorenz gauge, and the radiation gauges employed with success elsewhere in the literature \cite{Keidl:2006wk,Keidl:2010pm,Shah:2010bi,Shah:2012gu,vandeMeent:2015lxa,vandeMeent:2017bcc}. Building on Ref.~\cite{Dolan:2021ijg}, we have shown that, in vacuum, both the ingoing and outgoing radiation-gauge (IRG/ORG) metric perturbations can be transformed to Lorenz gauge. Next, we demonstrated that (for $\omega\neq0$ and in vacuum) by taking the \emph{difference} between the IRG and ORG solutions, one obtains a metric perturbation that can be written directly in terms of the Weyl scalars $\Psi_0$ and $\tilde{\Psi}_4$ that the metric perturbation will generate \cite{Aksteiner:2016pjt, Dolan:2021ijg}. This bypasses entirely the necessity of solving a fourth-order equation that relates the Hertz potential to the Weyl scalars, in the usual approach where \emph{either} the IRG or ORG solution is used on its own. Interestingly, this approach fails in both the static case ($\omega=0$) and the massless case ($M=0$), perhaps explaining why it has not emerged naturally from toy-model examples. 

A Lorenz-transformed solution derived from the Weyl scalars alone is not sufficient on its own, of course, since it is is traceless and since it is not regular at $r=r_0$. To recover the correct Lorenz-gauge solution it was necessary to add in the scalar-type trace modes, and vector-type traceless gauge modes, to restore regularity on the sphere $r=r_0$ (except on the particle worldline at $\theta=\pi/2$). The regularity of the metric perturbation in all 10 components convinces us that there are no missing pieces in this jigsaw. 

A valid criticism of the Lorenz-gauge calculation herein is that rather a lot of effort has been devoted to deriving modes which, in the vacuum regions $r \neq r_0$, can be entirely eliminated by means of a gauge transformation. But removing these parts of the metric perturbation with a gauge transformation comes at the cost of introducing distributions in the metric perturbation, as we argue below.

The Lorenz-gauge metric perturbation $h_{\mu \nu}^{L}$ satisfies a hyperbolic wave equation, Eq.~(\ref{eq:hLorenz-field-equation}), and thus it is expected to be free of distributions (i.e.~Dirac delta functions, etc), and this inference is consistent with the numerical results obtained here. The regularity of the Lorenz-gauge MP sets it apart from other metric perturbations that are reconstructed in the frequency domain. Reconstructed metric perturbations for point-like sources typically have distributions along radial strings (e.g.,~the ``full-string'' or ``half-string'' solutions \cite{Ori:2002uv}), on the sphere at $r=r_p(t)$ (for example, the ``no-string'' solution \cite{Pound:2013faa}), or confined to the worldline (e.g., the ``shadowless'' solution of Toomani \etal~\cite{Toomani:2021jlo,Green:2019nam}, or the AAB metric perturbation deriving from Eq.~(56) of Ref.~\cite{Aksteiner:2016pjt}). 
 Irregularity of the MP does not significantly impede {\it first-order} accurate self-force calculations (see e.g.~the calculations of Refs.~\cite{Keidl:2006wk,Keidl:2010pm,Shah:2010bi,Shah:2012gu,vandeMeent:2015lxa,vandeMeent:2017bcc} and the analysis of Refs.~\cite{Barack:2001ph,Ori:2002uv, Pound:2013faa, Merlin:2016boc}), but it does appear to pose a challenge at \emph{second-order}. This is because the source at second-order includes quadratic combinations of the first-order metric perturbations, and its derivatives, and generically these terms will include products of distributions; and such products are not themselves defined as distributions. Recently, Upton and Pound \cite{Upton:2021oxf,Upton:2022uhi} have introduced a class of highly-regular gauges designed to help with the second-order challenge. 

To see how distributions arise in the metric perturbation, consider a gauge transformation \emph{away} from Lorenz gauge, where in the now-familiar way, the gauge vector is made by glueing together IN and UP solutions at the particle's radius, $\xi_{\mu} = \xi_\mu^+ \Theta^+ + \xi_\mu^- \Theta^-$. For a particle on a circular orbit, the gauge-transformed MP is 
\begin{align}
h_{\mu \nu}^{L} - 2 \xi_{(\mu ; \nu)} 
 &= h_{\mu \nu}^{L} - \left(h_{\mu\nu}^+ \Theta^+ + h_{\mu\nu}^- \Theta^- \right) - 2 \delta(r - r_0) \left( \xi_{(\mu}^+ - \xi_{(\mu}^- \right) \delta_{\nu)}^r
\end{align}
where $h^{\pm}_{\mu \nu} = 2 \xi^\pm_{(\mu ; \nu)}$ are the homogeneous MPs associated with the IN and UP gauge vectors, and the last term is in proportion to delta-function on the sphere. This makes it clear that applying a regular gauge transformation to remove the $s=0$ and $s=1$ parts of the Lorenz-gauge metric perturbation, or the transformation from radiation-gauge to Lorenz gauge, comes at the expense of introducing distributions to the metric perturbation.

A natural extension of this work is to the challenge of \emph{generic orbits}, that is, the calculation of the Lorenz-gauge metric perturbations sourced by a particle on a non-circular, non-equatorial bound geodesic orbit on Kerr spacetime. Two approaches suggest themselves. First, in keeping with the approach in this paper, one could seek to construct a metric perturbation from a linear combination of \emph{vacuum} modes alone. As these modes have already been described and projected onto harmonics in this paper, much of the foundation is already in place. Again, the Weyl scalars and trace will be determined by sourced spin-2 and spin-0 Teukolsky equations (now with extended-but-compact sources), and the remaining ($s=0$ and $s=1$) coefficients will be determined by the conditions of regularity at $r=r_p(t)$ (and it seems most natural to choose either the periapsis or apoapsis). The metric perturbation at time $t$ will be reconstructed by glueing vacuum solutions at $r=r_p(t)$, relying on the validity of \emph{extended homogeneous solutions} for success (in essence, that the spacetime is vacuum everywhere except for on a compact worldline). It is likely that jumps must be determined separately for each $m, n, k$ numbers that accompany generic orbits in the frequency-domain representation.

A second, more general approach is founded on a key result in a theorem by Aksteiner, Andersson and B\"ackdahl \cite{Aksteiner:2016pjt}. Equation (56) in Ref.~\cite{Aksteiner:2016pjt} can be rearranged (in the frequency domain, and after omitting the unspecified gauge terms) into a reconstruction formula, giving a metric perturbation that is a full solution to the sourced Einstein equations in the presence of a conserved stress-energy tensor. The metric perturbation is derived from (differential operators acting on) the Weyl scalars $\Psi_0$ and $\tilde{\Psi}_4$, and an additional term that involves differential operators acting on the trace-free Ricci tensor (and thus, by the field equations, the stress-energy tensor itself). The first term has the form of the IRG-minus-ORG construction used in this work. For point-like sources, the second term is distributional on the worldline (and the first term is likely to be irregular across the sphere $r=r_p(t)$ too). We conceive that, by applying a regular gauge transformation and using techniques herein, this AAB solution can be ``Lorenzified'', rendering it regular away from the worldline, and entirely free from distributions. Work in this direction is in progress. Success in the endeavour would yield a method for calculating Lorenz-gauge solutions for the linearized Einstein equation with \emph{any} (conserved) stress-energy tensor. This would undoubtedly benefit second-order black hole perturbation theory, where the source in the linearised Einstein equation is extended across the entire domain in space.

At several stages in our point-particle calculation we have made use of the vacuum Teukolsky equations and Teukolsky-Starobinskii identities. In doing so, we have tacitly assumed that all distributions on the worldline that would appear through source terms ultimately cancel (with each other and with distributions that appear in the non-Lorenz gauge metric perturbation) by the time we reach Lorenz gauge. We defer an explicit demonstration that this is indeed the case to the aforementioned work in progress on the general appproach, where the cancellations are more explicitly manifest.

\begin{acknowledgments}
With thanks to Leor Barack, Adam Pound and Theo Torres for discussions. With special thanks to Niels Warburton for supplying Schwarzschild comparison data, and for help in calculating the boundary condition for the fourth-order equation at the horizon for the scalar gauge contribution. This work makes use of the Black Hole Perturbation Toolkit.
S.D.~acknowledges financial support from the Science and Technology Facilities Council (STFC) under Grant No.~ST/T001038/1 and Grant No.~ST/X000621/1, and from the European Union's Horizon 2020 research and innovation programme under the H2020-MSCA-RISE-2017 Grant No.~FunFiCO-777740. CK acknowledges support from Science Foundation Ireland under Grant number 21/PATH-S/9610. 
\end{acknowledgments}

\appendix

\section{Spin-weighted spherical harmonics and the ladder operators\label{sec:spherical-harmonics}}

In the $a \omega=0$ case, the spin-weighted \emph{spheroidal} harmonics ${}_s S_{lm}(\theta)$ reduce to spin-weighted \emph{spherical} harmonics ${}_s Y_{lm}(\theta)$, and the angular operators $\cL^\dagger_n$ and $\cL_n$ reduce to ladder operators $\cLh^\dagger_n$ and $\cLh_n$ that act on the spin-weighted spherical harmonics to raise and lower their spin-weight. Explicitly, 
\begin{subequations}
\begin{align}
\cLh ({}_0Y_{lm}) &= +\lambdah \, ({}_{-1} Y_{lm}) , &
\cLh^\dagger_1 \, ({}_{-1} Y_{lm}) &= - \lambdah \, ({}_0 Y_{lm}) , \\
\cLh^\dagger ({}_0Y_{lm}) &= - \lambdah \, ({}_{+1} Y_{lm}) , &
\cLh_1 \, ({}_{+1} Y_{lm}) &= + \lambdah \, ({}_0 Y_{lm}) , \\
\cLh_{-1} ({}_{-1} Y_{lm}) &= + \lambdah_2 \, ( {}_{-2}Y_{lm} ) &
\cLh_2^\dagger ({}_{-2} Y_{lm}) &= - \lambdah_2 \, ( {}_{-1} Y_{lm} ) ,\\
\cLh_{-1}^\dagger ({}_{+1} Y_{lm}) &= -\lambdah_2 \, ( {}_{+2}Y_{lm} )   &
\cLh_2 ({}_{+2} Y_{lm}) &= + \lambdah_2 \, ( {}_{+1} Y_{lm} ) ,
\end{align}
and hence
\begin{align}
\cLh_{-1} \cLh (Y_{lm}) &= \hat{\Lambda} \, ({}_{-2} Y_{lm}) , &
\cLh^\dagger_1  \cLh^\dagger_2 ({}_{-2} Y_{lm}) &= \hat{\Lambda} \, ({}_0 Y_{lm}) , \\
\cLh^\dagger_{-1} \cLh^\dagger (Y_{lm}) &= \hat{\Lambda} \, ({}_{+2} Y_{lm}) ,  &
\cLh_1 \cLh_2 ({}_{+2} Y_{lm}) &= \hat{\Lambda} \, ({}_0 Y_{lm}) , 
\end{align}
\end{subequations}
where 
\begin{subequations}
\begin{align}
\lambdah &\equiv \sqrt{l (l+1)} , \\
\lambdah_2 &\equiv \sqrt{(l-1) (l+2)} , \\
\hat{\Lambda} \equiv \lambdah \lambdah_2 &= \sqrt{(l-1)l(l+1)(l+2)} . 
\end{align}
\label{eq:lambdah-def}
\end{subequations}
(N.B.~This notation clashes with that used in Teukolsky section).

\section{Radiation gauge metric perturbations and Weyl scalars\label{appendix:weyl}}
The non-zero components of $h_{\mu \nu}^{\text{IRG}}$, defined in Eq.~(\ref{h-irg}), are
\begin{align}
\hirgup_{l_- l_-} &= - 2 \frac{\rhoc^2}{\Delta^2} \left( \cL_1^\dagger - \frac{2 \rhoc_{,\theta}}{\rhoc} \right) \cL_2^\dagger \psi_-  , \\
\hirgup_{m_- m_-} &= - 2 \rhoc^2 \left( \cD - \frac{2}{\rhoc} \right) \cD \psi_- , \\ 
\hirgup_{l_- m_-} &= \frac{2 \rhoc^2}{\Delta} \left\{   \cD \cL_2^\dagger  - \frac{1}{\rho} \left( \cL_2^\dagger + \rho_{,\theta} \cD \right) - \frac{1}{\rhoc} \left( \cL_2^\dagger + \rhoc_{,\theta} \cD \right)  \right\} \psi_-  .
\end{align} 
Similarly, the non-zero components of $h_{\mu \nu}^{\text{ORG}}$,  defined in Eq.~(\ref{h-org}), are
\begin{align}
\horgup_{l_+ l_+} &= - 2 \frac{\rhoc^2}{\Delta^2} \left( \cL_1 - \frac{2 \rhoc_{,\theta}}{\rhoc} \right) \cL_2 \psi_+  , \\
\horgup_{m_+ m_+} &= - 2 \rhoc^2 \left( \cD^\dagger - \frac{2}{\rhoc} \right) \cD^\dagger \psi_+ , \\ 
\horgup_{l_+ m_+} &= \frac{2 \rhoc^2}{\Delta} \left\{   \cD^\dagger \cL_2  - \frac{1}{\rho} \left( \cL_2 + \rho_{,\theta} \cD^\dagger \right) - \frac{1}{\rhoc} \left( \cL_2 + \rhoc_{,\theta} \cD^\dagger \right)  \right\} \psi_+  .
\end{align} 

The Weyl scalars of the metric perturbation are found by taking the expression for the perturbed Weyl tensor in terms of the perturbed metric, $C_{abcd}(g) \approx C_{abcd}(g+h) \approx C^0_{abcd}(g) + \delta C_{abcd}(g,h)$ and then projecting $\delta C_{abcd}(g,h)$ on to the background tetrad legs. The scalars of extremal spin weight are given in Eq.~(55) of Ref.~\cite{Pound:2021qin}, which in our notation translate as
\begin{align}
\Psi_0 &= \frac{1}{4 \rho} \left[ \cL_{-1}^\dagger \cL^\dagger (\rho^{-1} h_{l_+ l_+} ) + \cD \cD (\rho^{-1} h_{m_+ m_+}) \right] \nn \\
& \quad  - \frac{1}{4 \rho^2} \left\{ \left(\cD - \frac{1}{\rho}\right) \left(\cL_{-1}^\dagger + \frac{\rho_{,\theta}}{\rho} \right) 
 +   \left(\cL_{-1}^\dagger - \frac{\rho_{,\theta}}{\rho} \right) \left(\cD + \frac{1}{\rho}\right) \right\} h_{l_+ m_+}  ,  \label{eq:Psi0-weyl} \\
\tilde{\Psi}_4 &= 
 \frac{1}{4 \rho} \left[ \cL_{-1} \cL (\rho^{-1} h_{l- l-}) + \cD^\dagger \cD^\dagger (\rho^{-1} h_{m_- m_-} ) \right] \nn \\ 
 & \; \quad - \frac{1}{4 \rho^2} \left\{\left( \cD^\dagger - \frac{1}{\rho} \right) \left( \cL_{-1} + \frac{\rho_{,\theta}}{\rho} \right) + \left( \cL_{-1} - \frac{\rho_{,\theta}}{\rho} \right) \left( \cD^\dagger + \frac{1}{\rho} \right) \right\} h_{l_- m_-} .  \label{eq:Psi4-weyl}
\end{align}
Clearly, $\Psi_0 = 0$ for the IRG perturbation, and $\tilde{\Psi}_4 = 0$ for the ORG perturbation. Note, however that the ``primed'' versions are non-zero; a more detailed calculation that makes use of the vacuum Teukolsky equations yields the Weyl scalars summarised in Table \ref{tbl:weyl} (see e.g.~Ref.~\cite{Pound:2021qin} for further details).

 \section{Metric perturbation from a gauge vector\label{sec:spin-coefficients}}
Suppose we have a metric perturbation constructed from a gauge vector, $h_{\mu \nu}^{(\xi)} = - 2 \xi_{(\mu ; \nu)}$, with the gauge vector in the form
\beq
\xi^\mu = - \frac{1}{2\Sigma} \left[ A_+ \, \Delta l_+^\mu + A_- \, \Delta l_-^\mu  + B_+ \, m_+^\mu + B_- \, m_-^\mu \right] .
\eeq
The metric components are straightforward to calculate, using e.g.
\beq
{h}^{(\xi)}_{\mu \nu} l_+^\mu m_-^\nu = l_+^\mu ( m_{-}^{\nu} \xi_\nu  )_{,\mu} +  m_-^\mu ( l_{+}^{\nu} \xi_\nu  )_{,\mu} - (l_+^\mu \nabla_\mu m_-^\nu + m_-^\mu \nabla_\mu l_+^\nu ) \xi_\nu .
\eeq
Explicitly,
\begin{subequations}
\begin{align}
h_{l_+l_+} &= 2 \cD  A_-  , \\
h_{l_-l_-} &= 2 \cD^\dagger  A_+  , \\
h_{m_+m_+} &= 2 \cL_{-1}^\dagger B_- , \\
h_{m_-m_-} &= 2 \cL_{-1} B_+ , \\
h_{l_+l_-} &= \cD A_+ + \cD^\dagger A_- - \left( \frac{\Sigma_{,r}}{\Sigma} - \frac{\Delta_{,r}}{\Delta} \right) ( A_- + A_+ ) +  \frac{\Sigma_{,\theta}}{\Sigma \Delta}  (B_- + B_+), \\
h_{m_+m_-} &= \cL_1^\dagger B_+ + \cL_1 B_- + \frac{\Delta \Sigma_{,r}}{\Sigma} (A_- + A_+ ) - \frac{\Sigma_{,\theta}}{\Sigma} (B_- + B_+) , \\
h_{l_+m_+} &= \left(\cD - \frac{2}{\rho} \right) B_- + \left( \cL^\dagger - \frac{2 \rho_{,\theta}}{\rho} \right) A_- , \\
h_{l_+m_-} &= \left(\cD - \frac{2}{\rhoc} \right)  B_+ + \left( \cL - \frac{2 \rhoc_{,\theta}}{\rhoc} \right) A_- , \\
h_{l_-m_+} &= \left(\cD^\dagger - \frac{2}{\rhoc} \right) B_- + \left( \cL^\dagger - \frac{2 \rhoc_{,\theta}}{\rhoc} \right) A_+ , \\
h_{l_-m_-} &= \left(\cD^\dagger - \frac{2}{\rho} \right) B_+ + \left( \cL - \frac{2 \rho_{,\theta}}{\rho} \right) A_+  .
\end{align}
\label{h-from-xi}
\end{subequations}
Here we have made use of the following spin coefficients,
\begin{subequations}
\begin{align}
l_\pm^\nu \nabla_\nu l_\pm^\mu &= 0 , \\
m_\pm^\nu \nabla_\nu m_\pm^\mu &= \cot \theta \, m_\pm^\mu , \\
l_+^\nu \nabla_\nu m_+^\mu = m_+^\nu \nabla_\nu l_+^\mu &= \frac{\rho_{,\theta}}{\rho} l_+^\mu + \frac{\rho_{,r}}{\rho} m_+^\mu ,  \\
l_-^\nu \nabla_\nu m_+^\mu = m_+^\nu \nabla_\nu l_-^\mu &= \frac{\rhoc_{,\theta}}{\rhoc} l_-^\mu + \frac{\rhoc_{,r}}{\rhoc} m_+^\mu ,  \\
l_+^\nu \nabla_\nu l_-^\mu + l_-^\nu \nabla_\nu l_+^\mu &= \left( \frac{\Sigma_{,r}}{\Sigma} - \frac{\Delta_{,r}}{\Delta} \right) \left( l_+^\mu + l_-^\mu \right) - \frac{1}{\Delta} \frac{\Sigma_{,\theta}}{\Sigma} \left( m_+^\mu + m_-^\mu \right) , \\
m_+^\nu \nabla_\nu m_-^\mu + m_-^\nu \nabla_\nu m_+^\mu &= -\frac{\Delta \Sigma_{,r}}{\Sigma} \left( l_+^\mu + l_-^\mu \right) + \left(\frac{\Sigma_{,\theta}}{\Sigma} - \cot \theta \right) \left(m_+^\mu + m_-^\mu \right) .
\end{align}
\label{eq:spin-coefficients}
\end{subequations}

\section{Solving the sourced Teukolsky equation\label{appendix:sourced-teuk}}

The source terms in the Teukolsky equation (\ref{eq:sourced-teukolsky}) are (cf.~Eq.~(2.13) in Ref.~\cite{Press:1973zz})
\begin{align}
T_0 &= \left(\delta + \pi^\ast - \alpha^\ast - 3 \beta - 4 \tau \right) \left[ \left( D - 2 \epsilon - 2 \rho^\ast \right) T_{lm} - \left(\delta + \pi^\ast - 2 \alpha^\ast - 2 \beta \right) T_{ll} \right] \nn \\
 &+  \left( D - 3 \epsilon + \epsilon^\ast - 4 \rho - \rho^\ast \right) \left[ \left( \delta + 2 \pi^\ast - 2 \beta \right) T_{lm} - \left(D - 2\epsilon + 2\epsilon^\ast - \rho^\ast \right) T_{mm} \right] , \\  
T_4 &= \left( \Delta + 3 \gamma - \gamma^\ast + 4 \mu + \mu^\ast \right) \left[ \left(\delta^\ast - 2 \tau^\ast + 2 \alpha \right) T_{n\bar{m}} - \left(\Delta + 2 \gamma - 2 \gamma^\ast + \mu^\ast \right) T_{\bar{m}\bar{m}} \right] \nn \\
 &+ \left(\delta^\ast - \tau^\ast + \beta^\ast + 3 \alpha + 4 \pi\right) \left[ \left( \Delta + 2 \gamma + 2 \mu^\ast \right) T_{n\bar{m}} - \left( \delta^\ast - \tau^\ast + 2 \beta^\ast + 2 \alpha \right) T_{nn} \right]
\end{align}
so
\begin{align}
T_0 &= \frac{1}{2 \rho^2} \left\{ \left( \cL_{-1}^\dagger + \frac{4 \rhoc_{,\theta}}{\rhoc} - \frac{\rho_{,\theta}}{\rho} \right) \left[ \left( \cD + \frac{1}{\rho} \right) T_{l_+ m_+} - \left(\cL^\dagger - \frac{\rho_{,\theta}}{\rho} \right) T_{l_+ l_+} \right]
\right. \nn \\
& \quad \quad \quad + \left. \left( \cD + \frac{4}{\rhoc} - \frac{1}{\rho} \right) \left[\left( \cL_{-1}^\dagger + \frac{\rho_{,\theta}}{\rho} \right) T_{l_+ m_+} - \left( \cD - \frac{1}{\rho} \right) T_{m_+ m_+} \right] \right\}  , \\
\tilde{T}_4 \equiv \frac{4 \rhoc^4}{\Delta^2} T_4 &= \frac{1}{2 \rho^2} \left\{
\left(\cL_{-1} + \frac{4 \rhoc_{,\theta}}{\rhoc} - \frac{\rho_{,\theta}}{\rho} \right) \left[ \left(\cD^\dagger + \frac{1}{\rho} \right) T_{l_- m_-} - \left( \cL - \frac{\rho_{,\theta}}{\rho} \right) T_{l_- l_-} \right] 
 \right. \nn \\
& \quad \quad \quad + \left. 
 \left(\cD^\dagger + \frac{4}{\rhoc} - \frac{1}{\rho} \right) \left[ \left(\cL_{-1} + \frac{\rho_{,\theta}}{\rho} \right) T_{l_- m_-} - \left( \cD^\dagger - \frac{1}{\rho} \right) T_{m_- m_-} \right]
 \right\}
\end{align}

For a circular orbit, the stress-energy takes the form
\beq
T_{\mu \nu} = u_{\mu} u_{\nu} \mathcal{T} , \quad \quad u_{\mu} = [-E, 0, 0, L] ,
\eeq
where
\beq
\mathcal{T} = \frac{\mu}{r_0^2 u^t} \delta(r-r_0) \delta(\theta - \pi/2) \delta(\phi - \Omega t).
\eeq
and so its projections are
\begin{subequations}
\begin{align}
T_{l_+ l_+} = T_{l_- l_-} = -T_{l_- l_+} &= \mathcal{A}^2 \mathcal{T} , \\
T_{m_+ m_+} = T_{m_- m_-} = - T_{m_+ m_-} &= - \mathcal{B}^2 \mathcal{T}  , \\
T_{l_+ m_+} = T_{l_- m_-} = -T_{l_+ m_-} = -T_{l_- m_+} &= -i \mathcal{A} \mathcal{B} \mathcal{T} .
\end{align}
\end{subequations}
where
\begin{subequations}
\begin{align}
\mathcal{A} &= \frac{1}{\Delta_0} \left( E (r_0^2+a^2) - a L \right) \\ 
\mathcal{B} &= L - a E . 
\end{align}
\label{eq:AB}
\end{subequations}

We can take advantage of the fact that $\mathcal{T}$ is proportional to delta functions to evaluate several of the functions at the particle using $F(r,\theta)\delta(r-r_0) \delta(\theta - \pi/2) = F(r_0, \pi/2) \delta(r-r_0) \delta(\theta - \pi/2)$. Considering $\Sigma T_0$, we can use that
\begin{subequations}
\begin{align}
\frac{\rhoc}{\rho} \left( \cL_{-1}^\dagger + \frac{4 \rhoc_{,\theta}}{\rhoc} - \frac{\rho_{,\theta}}{\rho} \right) \left( \cD + \frac{1}{\rho} \right) T_{l_+ m_+} &= -i \mathcal{A} \mathcal{B} \left(\cL_{-1}^\dagger + \frac{3 i a}{r} \right) \left(\cD + \frac{1}{r_0} \right) \mathcal{T} , \\
\frac{\rhoc}{\rho} \left( \cL_{-1}^\dagger + \frac{4 \rhoc_{,\theta}}{\rhoc} - \frac{\rho_{,\theta}}{\rho} \right)  \left(\cL^\dagger - \frac{\rho_{,\theta}}{\rho} \right) T_{l_+ l_+} &= \mathcal{A}^2  \left( \cL_{-1}^\dagger + \frac{3 i a \sin \theta}{\rhoc} \right)  \left(\cL^\dagger - \frac{i a}{r_0} \right) \mathcal{T} \\
\frac{\rhoc}{\rho} \left( \cD + \frac{4}{\rhoc} - \frac{1}{\rho} \right) \left( \cL_{-1}^\dagger + \frac{\rho_{,\theta}}{\rho} \right) T_{l_+ m_+} &=  -i \mathcal{A} \mathcal{B} \left( \cD + \frac{3}{\rhoc} \right) \left( \cL_{-1}^\dagger - \frac{3 i a}{r_0} \right) \mathcal{T} \\
\frac{\rhoc}{\rho} \left( \cD + \frac{4}{\rhoc} - \frac{1}{\rho} \right) \left( \cD - \frac{1}{\rho} \right) T_{m_+ m_+} &= -\mathcal{B}^2 \left( \cD + \frac{3}{r} \right) \left( \cD - \frac{1}{r_0} \right) \mathcal{T} . 
\end{align}
\end{subequations}
Putting this together,
\begin{align}
2 \Sigma T_0 &= \left\{ - 2 i \mathcal{A} \mathcal{B} \left( \cL^\dagger \cD + \frac{2}{r_0} \cL^\dagger \right) - \mathcal{A}^2 \left(\cL_{-1}^\dagger \cL^\dagger + \frac{2 i a}{r_0} \cL^\dagger \right) + \mathcal{B}^2 \left( \cD \cD + \frac{2}{r_0} \cD \right) \right\} \mathcal{T} \\
 &= - 2 i \mathcal{A} \mathcal{B} \left[ \partial_r \partial_\theta - Q_0 \partial_r + \left(\frac{2}{r_0} - i W_0 \right)\partial_\theta - \left(\frac{2}{r_0} - i W_0 \right) Q_0 \right] \mathcal{T} \nn \\
 & \quad - \mathcal{A}^2 \left( \partial_\theta \partial_\theta + 2 \left(- Q_0 + \frac{i a}{r_0} \right) \partial_\theta + \left(Q_0^2 - \frac{2iaQ_0}{r_0} - 1 \right) \right) \mathcal{T} \nn \\
 & \quad + \mathcal{B}^2 \left( \partial_r \partial_r + 2 \left(- i W_0 + \frac{1}{r_0} \right)\partial_r + \left(i W_0^\prime - \frac{2 i W_0}{r_0} - W_0^2 \right) \right) \mathcal{T} . \label{eq:SigmaT0a}
\end{align}
where $Q = m (1 - a \Omega)$, $W_0 = m \left[ \Omega (r_0^2+a^2) - a \right] / \Delta_0$ and $W_0^\prime = \partial_{r_0} W_0$.

The spin-weighted spheroidal harmonics $S_{slm}(\theta)$ are orthonormal, with the standard normalisation
\beq
\int_0^\pi {}_sS_{l m}(\theta)  {}_sS_{l' m}(\theta) \sin \theta d\theta = \frac{1}{2 \pi} \delta_{l l'} .
\eeq
It is straightforward to establish that
\begin{align}
\delta(\theta - \pi/2) &= 2\pi \sum_{l=|m|}^{\infty} \left\{ {}_{s}S_{lm}(\tfrac{\pi}{2}) \right\} {}_{s}S_{lm}(\theta) , \\ 
\delta^\prime(\theta - \pi/2) &= 2\pi \sum_{l=|m|}^{\infty} \left\{ - {}_{s}S_{lm}^\prime(\tfrac{\pi}{2}) \right\} {}_{s}S_{lm}(\theta) , \\ 
 \quad \delta^{\prime\prime}(\theta - \pi/2) &= 2\pi \sum_{l=|m|}^{\infty} \left\{ 1 - \Lambda + Q_0^2 \right\} {}_{\pm 2}S_{lm}(\tfrac{\pi}{2}) {}_{\pm 2}S_{lm}(\theta) ,
\end{align}
and
\begin{align}
\delta(\phi - \Omega t) &= \frac{1}{2 \pi} \sum_{m=-\infty}^{\infty} e^{i m \phi} e^{- i \omega t} , 
\end{align}
where $\omega = m \Omega$.

Using these results in Eq.~(\ref{eq:SigmaT0a}),
\begin{align}
 -8 \pi \Sigma T_0 &=\mathfrak{D} \sum_{\ell m} \left\{ \mathfrak{C}_2 \delta''(r-r_0) + \mathfrak{C}_1 \delta'(r-r_0) + \mathfrak{C}_0 \delta(r-r_0) \right\} {}_{+2}S_{lm}(\theta) \chi_m \label{eq:SigmaT0b}
\end{align}
where $\chi_m = e^{im\phi} e^{-i \omega t}$ and
\begin{subequations}
\begin{align}
\mathfrak{D} &= -\frac{4\pi\mu}{r_0^2 u^t} \\
\mathfrak{C}_2 &= \mathcal{B}^2 S_0 \\
\mathfrak{C}_1 &= 2 \mathcal{B} \left[ i \mathcal{A} (S_0^\prime + Q_0 S_0) + \mathcal{B} \left(-iW_0 + \frac{1}{r_0}\right)S_0 \right] \\
\mathfrak{C}_0 &= 2 \mathcal{A}  \left[ i \mathcal{B} \left(\frac{2}{r_0} - i W_0\right) + \mathcal{A} \left(-Q_0 + \frac{ia}{r_0} \right) \right] \left(S_0^\prime + Q_0 S_0 \right) + \left[ \mathcal{A}^2 \Lambda + \mathcal{B}^2 \left(i W_0^\prime - \frac{2 i W_0}{r_0} - W_0^2 \right) \right] S_0
\end{align}
\end{subequations}
where $S_0 = S_{2lm}(\tfrac{\pi}{2})$, $S_0^\prime = \partial_\theta S_{2lm}(\tfrac{\pi}{2})$. Now we decompose the Weyl scalar in modes,
\beq
\Psi_0 = \sum_{lm} R_{+2}^{lm}(r) {}_{+2} S_{lm}(\theta) \chi_m 
\eeq
and insert into Eq.~(\ref{eq:sourced-psi0}) obtain the radial equation
\beq
\left\{ \Delta \cD_{1} \cD^\dagger_2 + 6 i \omega r - \Lambda \right\} R_{+2}^{lm} = \mathfrak{D} \sum_{\ell m} \left\{ \mathfrak{C}_2 \delta''(r-r_0) + \mathfrak{C}_1 \delta'(r-r_0) + \mathfrak{C}_0 \delta(r-r_0) \right\}.
\eeq
Now seek a solution in the form
\begin{align}
R_{+2}^{lm} = \frac{ \mathfrak{D}\mathfrak{C}_2 }{ \Delta_0 } \delta(r - r_0) + \alpha_+ R^{+}(r) \Theta(r - r_0) + \alpha_- R^{-}(r) \Theta(r_0 - r) ,   \label{eq:Rp2-ansatz}
\end{align}
where $R^{+}(r)$ and $R^{-}(r)$ are solutions to the homogeneous equation with appropriate boundary conditions, and $\alpha_+$ and $\alpha_-$ are constants to be determined. Expressing the operator on the l.h.s. in the form $\hat{D}_2 \equiv \left\{ \Delta \cD_{1} \cD^\dagger_2 + 6 i \omega r - \Lambda \right\}  = \Delta \partial_r \partial_r + 3 \Delta' \partial_r + V(r)$, 
it is quick to see that $\hat{D}_2 \delta = \Delta_0 \delta'' + \Delta_0' \delta' + (V_0 - 4) \delta$ and hence 
\begin{align}
\left( \Delta_0 J^{(0)}  + \frac{\mathfrak{D} \mathfrak{C}_2 \Delta_0^\prime}{\Delta_0} \right) \delta' + \left[ 2 \Delta_0' J^{(0)} + \Delta_0 J^{(1)} + \frac{\mathfrak{D}\mathfrak{C}_2}{\Delta_0} (V_0 - 4)  \right] \delta = \mathfrak{D} \mathfrak{C}_1 \delta' + \mathfrak{D} \mathfrak{C}_0 \delta
\end{align}
where $J^{(0)}$ and $J^{(1)}$ are the jumps in the radial function and its derivative, respectively, i.e.,
\begin{align}
J^{(0)} &\equiv \alpha_+ R^+(r_0)  - \alpha_- R^-(r_0) \\
J^{(1)} &\equiv (\alpha_+ \partial_r R^{+} - \alpha_-  \partial_r R^{-})_{r=r_0} .
\end{align}
Hence the jumps are determined by the linear system
\beq
\begin{pmatrix}
1 & 0 \\
\tfrac{2 \Delta_0^\prime}{\Delta_0} & 1
\end{pmatrix}
\begin{pmatrix} J^{(0)} \\ J^{(1)} \end{pmatrix} =
\frac{\mathfrak{D}}{\Delta_0} 
\begin{pmatrix}
\mathfrak{C}_1 - \tfrac{\Delta_0'}{\Delta_0} \mathfrak{C}_2 \\ 
\mathfrak{C}_0 - \tfrac{(V_0 - 4)}{\Delta_0} \mathfrak{C}_2 
\end{pmatrix} ,
\eeq
where
\beq
V_0 - 4 = \Delta_0 ( W_0^2 + i W_0') - i \Delta'_0 W_0 +  6 i m \Omega r - \Lambda .
\eeq
Jumps in the second and higher derivatives can be found using the homogeneous radial equation. For example,
\beq
J^{(2)} = -\frac{1}{\Delta_0} \left( 3 \Delta^\prime_0 J^{(1)} + V_0 J^{(0)} \right). 
\eeq
If we instead require the jumps $\tilde{J}^{(i)}$ in $P_{+2} = \Delta^2 R_{+2}$, then 
\beq
\tilde{J}^{(0)} = \Delta^2_0 J^{(0)}, \quad \tilde{J}^{(1)} = \Delta_0^2 \left( J^{(1)} + \tfrac{2 \Delta_0^\prime}{\Delta_0} J^{(0)} \right) , \quad \text{etc.}
\eeq
that is,
\beq
\begin{pmatrix} \tilde{J}^{(0)} \\ \tilde{J}^{(1)} \end{pmatrix}
= \Delta_0 \mathfrak{D} \begin{pmatrix}
\mathfrak{C}_1 - \tfrac{\Delta_0'}{\Delta_0} \mathfrak{C}_2 \\ 
\mathfrak{C}_0 - \tfrac{(V_0 - 4)}{\Delta_0} \mathfrak{C}_2 
\end{pmatrix} . \label{eq:jumps2}
\eeq

\section{Solving the sourced scalar equation\label{appendix:kappa}}

  \subsection{Calculating $\kappa$ with a direct method\label{appendix:kappa-direct}}
We seek IN and UP solutions to Eq.~(\ref{eq:kappa-ll}). 
First, the radial function is split into two parts, 
\beq
 \kappa_{\ell \ell}(r) = \kappa^{(2)}_{\ell \ell}(r) + a^2 \gamma_{\ell \ell} \, \kappa^{(0)}_{\ell \ell}(r), 
\eeq
where
\begin{align}
 \left(\cD \Delta \cD^\dagger - 2 i \omega r  - \lambda_{(\ell)} \right) \kappa^{(n)}_{\ell \ell} (r) = \tfrac{1}{2} r^n  h_{\ell}(r) . \label{eq:radial-kappa}
\end{align}
For large $r$, the UP solution, which is outgoing at null infinity, has an expansion of the form
\begin{align}
h_{\ell}(r) &= \frac{1}{r} \beta^{\text{up}}(r)  \sum_{k=0}^{\infty} c_{k}^{(h)} r^{-k} &
\kappa_{\ell \ell}^{(n)}(r) &= r^{n-2} \; \beta^{\text{up}}(r) \sum_{k=0}^{\infty} c_{k}^{(\kappa_{n})} r^{-k}    \label{eq:kappa-up-series}
\end{align}
where
\begin{align}
\beta^{\text{up}}(r) \equiv e^{i \omega r} \left( \frac{r}{2M} \right)^{2 i M \omega} .
\end{align}
The first few coefficients are
\begin{subequations}
\begin{align}
c_0^{(h)} &=  1 , &
c_0^{(\kappa_2)} &= - i / (4 \omega) , &
c_0^{(\kappa_0)} &= i / (4 \omega) , \\
c_0^{(h)} &= \frac{i (\lambda + 2am\omega - 8M^2 \omega^2)}{2 \omega} , &
c_0^{(\kappa_2)} &= 0 , &
c_0^{(\kappa_0)} &= \frac{1 - \lambda - 2 a m \omega + 8 M^2 \omega^2}{8 \omega^2} .
\end{align}
\end{subequations}
The UP solution for $\kappa$ is unique up to a multiplicative scaling, and the addition of a homogeneous UP solution. 

The IN solution, which is ingoing at the horizon, has an expansion of the form
\begin{align}
h_{\ell}(r) &= \beta^{\text{in}}(r) \sum_{k=0}^{\infty} d_{k}^{(h)} (r-r_+)^{k} &
\kappa_{\ell \ell}^{(n)}(r) &= \beta^{\text{in}}(r) \sum_{k=0}^{\infty} d_{k}^{(\kappa_{n})} (r-r_+)^{k+1} \label{eq:kappa-in-series}
\end{align}
where
\beq
\beta^{\text{in}}(r)  = \left(  r-r_+  \right)^{- \frac{2 i M r_+ \tilde{\omega}}{r_+ - r_-}}  \left(r_+ - r_-  \right)^{\frac{2 i M r_- \tilde{\omega}}{r_+ - r_-}} (2M)^{2iM \tilde{\omega}} e^{- i \tilde{\omega} r_+} . 
\eeq
with $\tilde{\omega} \equiv \omega - a m / (2 M r_+)$. 

The series expansions (\ref{eq:kappa-up-series}) and (\ref{eq:kappa-in-series}) can be taken to high orders; typically, we used $k_{\text{max}} = 5$. The series solutions serve as initial conditions in the far-field (UP) and near-horizon (IN) regions. The full UP and IN solutions for the domain $r\ge r_0$ and $r\le r_0$, respectively, are found by integrating the set of ODEs using a numerical method. More precisely, Eq.~(\ref{eq:radial-kappa}) for $n=0$ and $n=2$ with Eq.~(\ref{eq:radial-h}) is solved in Mathematica using {\tt NDSolve} with the {\tt StiffnessSwitching} option.

  \subsection{Calculating $\kappa$ with the method of partial annihilators}
 
 Equation (\ref{eq:kappa-ll}) can be written in the alternative form
 \beq
   \frac{r^3}{(r^2+a^2 \gamma_{\ell\ell})\Delta}\LNW \left(r \kappa_{\ell\ell}(r)\right)=\frac{1}{2}h_\ell(r)\label{EqNW-kappa}
 \eeq
  where
 \begin{align}
  \LNW &\equiv \frac{d^2}{dr_\ast^2} + V(r) , \\
  V(r) &\equiv \frac{((r^2+a^2)\omega-am)^2}{r^4}-\frac{\Delta}{r^2}\left(\lambda -2am\omega+a^2\omega^2+\frac{2(M r-a^2)}{r^2}\right) ,
 \end{align}
 defined in \cite{Warburton:2014bya}, with $dr_\ast/dr = r^2 / \Delta$. The trace function $h_\ell (r)$ satisfies Eq.~(\ref{eq:radial-h}), which can be written in the alternative form
 \beq
    \LNW \left(r h_{\ell}(r)\right) = - \frac{16 \pi \mu (r_0+a^2 \gamma_{\ell \ell})\Delta_0}{u^t r_0^3} S_{0l}(\tfrac{\pi}{2}) \delta(r-r_0) . 
    \label{EqNW-h}
 \eeq
 We can combine Eq.~(\ref{EqNW-kappa}) and (\ref{EqNW-h}) into a fourth-order equation
 \beq
 \LNW \left\{ \frac{r^4}{(r^2+a^2 \gamma_{\ell\ell})\Delta}\LNW \left(r \kappa_{\ell\ell}(r)\right) \right\} = - \frac{8 \pi \mu (r_0+a^2 \gamma_{\ell \ell}) \Delta_0}{u^t r_0^3} S_{0l}(\tfrac{\pi}{2}) \delta(r-r_0) . \label{eq:fourthorder}
 \eeq
 This equation has a compact, distributional source, and it can be solved using the method of variation of parameters. 
 We note that Eq.~(\ref{eq:fourthorder}) is of a comparable form to the equation
 \begin{equation}
\mathcal{L}^{RW}_0 \left(\frac{1}{f(r)} \mathcal{L}^{RW}_0 M_{2af}(r)\right) = S^{RW}_0,
\end{equation}
 for a certain scalar potential $M_{2af}$ in the construction of the Schwarzschild metric perturbation \cite{Berndtson:2007gsc, Durkan:2022fvm}.

The four homogeneous solutions to (\ref{eq:fourthorder}) are related to the asymptotic boundary conditions, two of which are at infinity and two at the horizon. One of the solutions at infinity and one at the horizon correspond to solutions to of the homogeneous version of the second-order ODE (\ref{EqNW-h}). The solutions of (\ref{EqNW-h}) are immediately also solutions to the homogeneous fourth-order equation, and can be found in Ref.~\cite{Warburton:2010eq} in the Schwarzschild case. The remaining solutions are solutions to the fourth-order equation only, and not the second-order equation. This ensures new, independent solutions have been found to give a basis of four independent solutions, as required by a fourth-order ODE. The homogeneous solutions can be expanded as follows:
\begin{align}
h_\ell^{+}(r) &= e^{i \omega r_{*} }\hspace{-0.1cm} \sum _{i=0}^{n_{\text{max}}} \frac{A^{+}_i}{(r)^{i}}\bigg\vert_{r = r_{\text{out}}},	&
\kappa_{\ell\ell}^{+}(r) &= e^{i \omega r_{*} }\hspace{-0.1cm} \sum _{i=0}^{n_{\text{max}}} \frac{B^{+}_i}{(r)^{i}}\bigg\vert_{r = r_{\text{out}}},	\\
h_\ell^{-}(r) &= e^{-i \gamma r_*} \hspace{-0.2cm} \sum _{i=0}^{n_{\text{max}}}A^{-}_i(r-r_+)^i\big\vert_{r = r_{\text{in}}}, 
	&
\kappa_{\ell\ell}^{-}(r) &= e^{-i \gamma r_*} \hspace{-0.2cm} \sum _{i=0}^{n_{\text{max}}}B^{-}_i(r-r_+)^i\big\vert_{r = r_{\text{in}}},
\end{align}
where 
\begin{equation}
\gamma=\frac{2M \omega r_+-ma}{r_+^2}.
\end{equation}
Numerically, $r_{\text{in}}$ and $r_{\text{out}}$ are chosen to be $2+10^{-5}M$ and $10^5M$ respectively. The value of $n_{\text{max}}$ is set to $n_{\text{max}}=20$, though can be set lower than this for accurate results. Due to the distributional source, we can write the inhomogeneous solution as
\begin{align}
\kappa_{\ell\ell}(r) = & \left[c^{h2,+} h_{\ell}^{+}(r) + c^{h4,+} \kappa_{\ell\ell}^{+}(r)\right]\Theta(r - r_0) \\
& +\left[c^{h2,-} h_{\ell}^{-}(r) + c^{h4,-} \kappa_{\ell\ell}^{-}(r)\right]\Theta(r_0 - r) , \label{kapparet}
\end{align}
where the constant coefficients (for a given $s,l,m,r_0$) are given by
\begin{align}
c^{i,+}\Theta(r - r_0) &=  \int_{-\infty}^{r_*} \hspace{-0.1cm}\frac{\mathcal{W}^{i,+}\left(r_*^{\prime}\right) \mathcal{S}\left(r_*^{\prime}\right)}{\mathcal{W}} \mathrm{d} r_*^{\prime},  \label{eq:ch4+}\\
c^{i,-}\Theta(r_0 - r) &=   \int_{r_*}^{\infty} \frac{\mathcal{W}^{i,-}\left(r_*^{\prime}\right) \mathcal{S}\left(r_*^{\prime}\right)}{\mathcal{W}} \mathrm{d} r_*^{\prime},\label{eq:ch4-}
\end{align}
with $i \in \{h2,h4\}$. Here we have defined
\begin{equation}
\mathcal{S} \equiv \frac{8 \pi \mu (r_0+a^2 \gamma_{\ell \ell}) \Delta_0}{u^t r_0^3} S_{0l}(\tfrac{\pi}{2}) \delta(r-r_0). \label{stsource}
\end{equation} The fourth-order Wronskian $\mathcal{W}$ is also constant by Abel's Theorem and has the standard definition,
\begin{equation}
\hspace{-0.2cm} \mathcal{W} = \det\left(\begin{array}{cccc}
{h_\ell^{-}}&{\kappa_{\ell\ell}^{ -}}&{h_\ell^{+}}&{\kappa_{\ell\ell}^{ +}}\\
{\partial_{r_*}h_\ell^{-}}&{\partial_{r_*}\kappa_{\ell\ell}^{-}}&{\partial_{r_*}h_\ell^{+}}&{\partial_{r_*}\kappa_{\ell\ell}^{+}}\\
{\partial^2_{r_*}h_\ell^{-}}&{\partial^2_{r_*}\kappa_{\ell\ell}^{-}}&{\partial^2_{r_*}h_\ell^{+}}&{\partial^2_{r_*}\kappa_{\ell\ell}^{+}}\\
{\partial^3_{r_*}h_\ell^{-}}&{\partial^3_{r_*}\kappa_{\ell\ell}^{-}}&{\partial^3_{r_*}h_\ell^{+}}&{\partial^3_{r_*}\kappa_{\ell\ell}^{+}}
\end{array}\right),\label{eq:Ws4}
\end{equation}
and $\mathcal{W}^{h2/4,\pm}$ are given by the determinant of the matrices obtained from deleting the final row and the column containing the corresponding field, $\kappa_{\ell \ell}^{\pm}$ or $ h_{\ell}^\pm$ and its derivatives from the matrix in Eq.~(\ref{eq:Ws4}) \cite{Hopper:2012ty,ODEs}.

\section{Projection of the metric perturbation onto a spherical basis\label{appendix:projections}}
The method for projecting the metric perturbation onto a spin-weighted spherical basis was outlined in Sec.~\ref{subsec:projections}. Below are the projections of all components in Eq.~(\ref{eq:h-projections}) for spins 0, 1 and 2. 

\subsection{Spin one}

A complete set of projections for the spin-one metric perturbation in Eq.~(\ref{eq:mp-s1}) are 
\begin{subequations}
\begin{align}
h^{(s=1)}_{l_+ l_+} &=  2 \Delta^{-1} \left( \pm \mathbf{\cD_{-1} P_{+1}}  \right)^T  \mathbf{S}^{(s=1)}_{0} \mathbf{Y}_0 , \\
h^{(s=1)}_{l_- l_-} &= 2 \Delta^{-1} \left( \mathbf{\cD^\dagger_{-1} P_{-1}} \right)^T  \mathbf{S}^{(s=1)}_{0} \mathbf{Y}_0 , \\
h^{(s=1)}_{m_+ m_+} &= \left[ \pm 2 \left(\cD P_{-1} \pm \cD^\dagger P_{+1} \right) \right]^T  \mathbf{b}_{+1} \left( - \mathbf{\hat{\boldsymbol{\lambda}}_2} + a \omega  (\mathfrak{s})_{12} \right) \mathbf{Y}_{+2} ,  \\
h^{(s=1)}_{m_- m_-} &= - 2 \left(\cD P_{-1} \pm \cD^\dagger P_{+1} \right)^T  \mathbf{b}_{-1} \left( + \mathbf{\hat{\boldsymbol{\lambda}}_2} - a \omega  (\mathfrak{s})_{-1,-2} \right) \mathbf{Y}_{-2} , \\
\rho \, h^{(s=1)}_{l_+ m_+} &= \left\{
\left[ \pm (r \cD - 2) \mathcal{P} \right]^T \, \mathbf{b}_{+1} + i a ( \pm \cD \mathcal{P})^T \, \mathbf{b}_{+1} \mathfrak{c}_{+1} \right. \nn \\
 & \quad \left. \left[ \pm \Delta^{-1} P_{+1} \right]^T \mathbf{S}_0^{(s=1)} \left(  
- r \hat{\boldsymbol{\lambda}} - i a \hat{\boldsymbol{\lambda}} \mathfrak{c}_{+1} + (a \omega r + 2 i a) \mathfrak{s}_{01} + i a^2 \omega \mathfrak{s}_{01} \mathfrak{c}_{+1}  \right) \right\} \mathbf{Y}_{+1}, \\
\rhoc \, h^{(s=1)}_{l_+ m_-} &= \left\{
-(r \cD - 2) \mathcal{P}^T \, \mathbf{b}_{-1} + i a ( \cD \mathcal{P})^T \, \mathbf{b}_{-1} \mathfrak{c}_{-1} \right. \nn \\
 & \left. \quad [ \pm \Delta^{-1} P_{+1}]^T \mathbf{S}_0^{(s=1)} \left(  
 r \hat{\boldsymbol{\lambda}} - i a \hat{\boldsymbol{\lambda}} \mathfrak{c}_{-1} - (a \omega r + 2 i a) \mathfrak{s}_{0-1} + i a^2 \omega \mathfrak{s}_{0-1} \mathfrak{c}_{-1}
 \right) \right\} \mathbf{Y}_{-1}, \\
 \rhoc \, h^{(s=1)}_{l_- m_+} &= \left\{ 
[\pm (r \cD^\dagger - 2) \mathcal{P}]^T \, \mathbf{b}_{+1} + [ \mp i a ( \cD^\dagger \mathcal{P})]^T \, \mathbf{b}_{+1} \mathfrak{c}_{+1} \right. \nn \\
 & \left. \quad + \Delta^{-1} P_{-1}^T \mathbf{S}_0^{(s=1)} \left(  
 - r \hat{\boldsymbol{\lambda}} + i a \hat{\boldsymbol{\lambda}} \mathfrak{c}_{+1} + (a \omega r - 2 i a) \mathfrak{s}_{01} - i a^2 \omega \mathfrak{s}_{01} \mathfrak{c}_{+1}
 \right) \right\} \mathbf{Y}_{+1}, \\
\rho \, h^{(s=1)}_{l_- m_-} &= \left\{
- (r \cD^\dagger - 2) \mathcal{P}^T \, \mathbf{b}_{-1} - i a ( \cD^\dagger \mathcal{P})^T \, \mathbf{b}_{-1} \mathfrak{c}_{-1} \right. \nn \\
 & \left. \quad + \Delta^{-1} P_{-1}^T \mathbf{S}_0^{(s=1)} \left(  
  r \hat{\boldsymbol{\lambda}} + i a \hat{\boldsymbol{\lambda}} \mathfrak{c}_{-1} - (a \omega r - 2 i a) \mathfrak{s}_{0-1} - i a^2 \omega \mathfrak{s}_{0-1} \mathfrak{c}_{-1} 
 \right) \right\} \mathbf{Y}_{-1} , \\
\Sigma \Delta h^{(s=1)}_{l_+ l_-} 
 &= \left\{ \calP^T \mathbf{S}_0^{(s=1)} \left( r^2 + a^2 (\mathfrak{c}^2)_0 \right) - 2 r (P_{-1} \pm P_{+1})^T \mathbf{S}_0^{(s=1)} + 2 a^2 \calP^T \left( \mathbf{b}_{-1} \mathfrak{s}_{-10} \mp \mathbf{b}_{+1} \mathfrak{s}_{10} \right) \mathfrak{c}_0 \right\} \mathbf{Y}_0 \\
  &= - \Sigma h_{m_+ m_-}^{(s=1)} , 
\end{align}
\end{subequations}
where
\begin{align}
\mathbf{S}^{(s=1)}_{0} &=  - ( \mathbf{b}_{-1} \pm \mathbf{b}_{+1} ) \boldsymbol{\hat{\lambda}} + a \omega \left(\mathbf{b}_{-1} \mathfrak{s}_{-10} \pm \mathbf{b}_{+1} \mathfrak{s}_{+10}  \right)  , 
\end{align}

\subsection{Spin two}
A complete set of projections for the spin-two metric perturbation in Eq.~(\ref{eq:mp-s2}) are
\begin{subequations}
\begin{align}
h_{l_+ l_+}^{(s=2)} &= - \frac{1}{6 \omega^2 \Delta^2} {P_{+2}}^T \mathbf{S}_{l_+ l_+}^{(s=2)} \mathbf{Y}_0 , \\
h_{l_- l_-}^{(s=2)} &= -\frac{1}{6 \omega^2 \Delta^2} {P_{-2}}^T \mathbf{S}_{l_- l_-}^{(s=2)} \mathbf{Y}_0 , \\
h_{m_+ m_+}^{(s=2)} &= \frac{-1}{6 \omega^2} \left( \pm \cD \cD P_{-2} + \cD^{\dagger} \cD^{\dagger} P_{+2} \right)^T  \mathbf{b}_{+2} \mathbf{Y}_{+2} , \\
h_{m_- m_-}^{(s=2)} &= \frac{-1}{6 \omega^2} \left( \pm \cD \cD P_{-2} + \cD^{\dagger} \cD^{\dagger} P_{+2} \right)^T \left( \pm \mathbf{b}_{-2} \right) \mathbf{Y}_{+2} , 
\end{align}
and (omitting the transpose)
\begin{align}
\pm \beta^{-1} \, \rho h_{l_+ m_+}^{(s=2)} &= + \frac{1}{12 \omega^2} \left( (\cD \cD P_{-2}) \mathbf{S}_{+} - (\cD^\dagger \cD^\dagger P_{+2}) \mathbf{S}_{-}  \right) (\hat{\boldsymbol{\lambda}} - a \omega \mathfrak{s}_{01}) \mathbf{Y}_{+1} \nn \\
& \quad - \frac{1}{12\omega^2} \left( (\cD \cD \cD P_{-2}) \mathbf{S}_{+} - (\cD \cD^\dagger \cD^\dagger P_{+2}) \mathbf{S}_{-} \right) \left((\hat{\boldsymbol{\lambda}} - a \omega \mathfrak{s}_{01}) (r \mathbf{I} + i a \mathfrak{c}_{+1}) - i a \mathfrak{s}_{01} \right)  \mathbf{Y}_{+1} \nn \\
& \quad - \frac{1}{3 i \omega} \left[ (\cD \cD \cD P_{-2}) \mathbf{M}_{+2} \left(r \mathbf{I} + i a \mathfrak{c}_{+1} \right)^2 - 2 (\cD \cD P_{-2} ) \mathbf{M}_{+2} \left(r \mathbf{I} + i a \mathfrak{c}_{+1} \right) + 2 (\cD P_{-2}) \mathbf{M}_{+2} \right] \mathbf{Y}_{+1} \nn \\
& \quad - \frac{(\cD^\dagger P_{+2})}{3 i \omega \Delta} \mathbf{M}_{-2} \left[
 \left( \hat{\boldsymbol{\lambda}}^2 - 2 a\omega \hat{\boldsymbol{\lambda}} \mathfrak{s}_{01} + a^2 \omega^2 (\mathfrak{s}^2)_{-1 +1} \right) \left(r \mathbf{I} + i a \mathfrak{c}_{+1} \right)^2
\right. \nn \\
& \quad \quad \quad \quad \quad \quad \quad \quad \quad \quad \left. 
- 2 i a \left(\hat{\boldsymbol{\lambda}} \mathfrak{s}_{01} - a \omega (\mathfrak{s}^2)_{-1 +1} \right) \left( r \mathbf{I} + i a \mathfrak{c}_{+1} \right) - 2 a^2 (\mathfrak{s}^2)_{-1+1} 
 \right]  \mathbf{Y}_{+1},  \\
\beta^{-1} \, \rhoc h_{l_+ m_-}^{(s=2)} &= -\frac{1}{12 \omega^2} \left( (\cD \cD P_{-2}) \mathbf{S}_{-} - (\cD^\dagger \cD^\dagger P_{+2}) \mathbf{S}_{+}  \right) (\hat{\boldsymbol{\lambda}} - a \omega \mathfrak{s}_{0-1}) \mathbf{Y}_{-1} \nn \\
& \quad +\frac{1}{12\omega^2} \left( (\cD \cD \cD P_{-2}) \mathbf{S}_{-} - (\cD \cD^\dagger \cD^\dagger P_{+2}) \mathbf{S}_{+} \right) \left((\hat{\boldsymbol{\lambda}} - a \omega \mathfrak{s}_{0-1}) (r \mathbf{I} - i a \mathfrak{c}_{-1}) - i a \mathfrak{s}_{0-1} \right)  \mathbf{Y}_{-1} \nn \\
& \quad + \frac{1}{3 i \omega} \left[ (\cD \cD \cD P_{-2}) \mathbf{M}_{-2} \left(r \mathbf{I} - i a \mathfrak{c}_{-1} \right)^2 - 2 (\cD \cD P_{-2} ) \mathbf{M}_{-2} \left(r \mathbf{I} - i a \mathfrak{c}_{-1} \right) + 2 (\cD P_{-2}) \mathbf{M}_{-2} \right] \mathbf{Y}_{-1} \nn \\
& \quad + \frac{(\cD^\dagger P_{+2})}{3 i \omega \Delta} \mathbf{M}_{+2} \left[
 \left( \hat{\boldsymbol{\lambda}}^2 - 2 a\omega \hat{\boldsymbol{\lambda}} \mathfrak{s}_{0-1} + a^2 \omega^2 (\mathfrak{s}^2)_{+1 -1} \right) \left(r \mathbf{I} - i a \mathfrak{c}_{-1} \right)^2
\right. \nn \\
& \quad \quad \quad \quad \quad \quad \quad \quad \quad \quad \left. 
- 2 i a \left(\hat{\boldsymbol{\lambda}} \mathfrak{s}_{0-1} - a \omega (\mathfrak{s}^2)_{+1 -1} \right) \left( r \mathbf{I} - i a \mathfrak{c}_{-1} \right) - 2 a^2 (\mathfrak{s}^2)_{+1-1} 
 \right]  \mathbf{Y}_{-1},  \\
 \beta^{-1} \, \rhoc h_{l_- m_+}^{(s=2)} &= \frac{1}{12 \omega^2} \left( (\cD \cD P_{-2}) \mathbf{S}_{-} - (\cD^\dagger \cD^\dagger P_{+2}) \mathbf{S}_{+} \right) (\hat{\boldsymbol{\lambda}} - a \omega \mathfrak{s}_{01}) \mathbf{Y}_{+1} \nn \\
& \quad - \frac{1}{12\omega^2} \left( (\cD^\dagger \cD \cD P_{-2}) \mathbf{S}_{-} - (\cD^\dagger \cD^\dagger \cD^\dagger P_{+2}) \mathbf{S}_{+} \right) \left((\hat{\boldsymbol{\lambda}} - a \omega \mathfrak{s}_{01}) (r \mathbf{I} - i a \mathfrak{c}_{1}) + i a \mathfrak{s}_{01} \right)  \mathbf{Y}_{1} \nn \\
& \quad - \frac{1}{3 i \omega} \left[ (\cD^\dagger \cD^\dagger \cD^\dagger P_{+2}) \mathbf{M}_{+2} \left(r \mathbf{I} - i a \mathfrak{c}_{1} \right)^2 - 2 (\cD^\dagger \cD^\dagger P_{+2} ) \mathbf{M}_{+2} \left(r \mathbf{I} - i a \mathfrak{c}_{+1} \right) + 2 (\cD^\dagger P_{+2}) \mathbf{M}_{+2} \right] \mathbf{Y}_{+1} \nn \\
& \quad - \frac{(\cD P_{-2})}{3 i \omega \Delta} \mathbf{M}_{-2} \left[
 \left( \hat{\boldsymbol{\lambda}}^2 - 2 a\omega \hat{\boldsymbol{\lambda}} \mathfrak{s}_{01} + a^2 \omega^2 (\mathfrak{s}^2)_{-1 +1} \right) \left(r \mathbf{I} - i a \mathfrak{c}_{+1} \right)^2
\right. \nn \\
& \quad \quad \quad \quad \quad \quad \quad \quad \quad \quad \left. 
+ 2 i a \left(\hat{\boldsymbol{\lambda}} \mathfrak{s}_{01} - a \omega (\mathfrak{s}^2)_{-1 +1} \right) \left( r \mathbf{I} - i a \mathfrak{c}_{+1} \right) - 2 a^2 (\mathfrak{s}^2)_{-1+1} 
 \right]  \mathbf{Y}_{+1},  \\
\pm \beta^{-1} \, \rho h_{l_- m_-}^{(s=2)} &= -\frac{1}{12 \omega^2} \left( (\cD \cD P_{-2}) \mathbf{S}_{+} - (\cD^\dagger \cD^\dagger P_{+2}) \mathbf{S}_{-}  \right) (\hat{\boldsymbol{\lambda}} - a \omega \mathfrak{s}_{0-1}) \mathbf{Y}_{-1} \nn \\
& \quad +\frac{1}{12\omega^2} \left( (\cD^\dagger \cD \cD P_{-2}) \mathbf{S}_{+} - (\cD^\dagger \cD^\dagger \cD^\dagger P_{+2}) \mathbf{S}_{-} \right) \left((\hat{\boldsymbol{\lambda}} - a \omega \mathfrak{s}_{0-1}) (r \mathbf{I} + i a \mathfrak{c}_{-1}) + i a \mathfrak{s}_{0-1} \right)  \mathbf{Y}_{-1} \nn \\
& \quad + \frac{1}{3 i \omega} \left[ (\cD^\dagger \cD^\dagger \cD^\dagger P_{+2}) \mathbf{M}_{-2} \left(r \mathbf{I} + i a \mathfrak{c}_{-1} \right)^2 - 2 (\cD^\dagger \cD^\dagger P_{+2} ) \mathbf{M}_{-2} \left(r \mathbf{I} + i a \mathfrak{c}_{-1} \right) + 2 (\cD^\dagger P_{+2}) \mathbf{M}_{-2} \right] \mathbf{Y}_{-1} \nn \\
& \quad + \frac{(\cD P_{-2})}{3 i \omega \Delta} \mathbf{M}_{+2} \left[
 \left( \hat{\boldsymbol{\lambda}}^2 - 2 a\omega \hat{\boldsymbol{\lambda}} \mathfrak{s}_{0-1} + a^2 \omega^2 (\mathfrak{s}^2)_{+1 -1} \right) \left(r \mathbf{I} + i a \mathfrak{c}_{-1} \right)^2
\right. \nn \\
& \quad \quad \quad \quad \quad \quad \quad \quad \quad \quad \left. 
+ 2 i a \left(\hat{\boldsymbol{\lambda}} \mathfrak{s}_{0-1} - a \omega (\mathfrak{s}^2)_{+1 -1} \right) \left( r \mathbf{I} + i a \mathfrak{c}_{-1} \right) - 2 a^2 (\mathfrak{s}^2)_{+1-1} 
 \right]  \mathbf{Y}_{-1},  
\end{align}
and
\begin{align}
\Sigma \Delta \beta^{-1} h_{l_+ l_-}^{(s=2)} &= \phantom{-} 
\frac{1}{24 \omega^2} \left\{
[(\cD \Delta \cD^\dagger + \cD^\dagger \Delta \cD)\cD \cD P_{-2}]^T \mathbf{S}_{-} (r^2+a^2 (\mathfrak{c}^2)_0)  - 4 r \Delta (\partial_r \cD \cD P_{-2})^T \mathbf{S}_{-}  \right. \nn \\
& \quad \quad \quad \quad \quad \left. -2 a^2 (\cD \cD P_{-2})^T \mathbf{S}_{-} \hat{\boldsymbol{\lambda}} (\mathfrak{s}_{-10} - \mathfrak{s}_{10}) \mathfrak{c}_0 \right\} \mathbf{Y}_0 \nn \\
& \phantom{=}
- \frac{1}{24 \omega^2} \left\{
[(\cD \Delta \cD^\dagger + \cD^\dagger \Delta \cD)\cD^\dagger \cD^\dagger P_{+2}]^T \mathbf{S}_{+} (r^2+a^2 (\mathfrak{c}^2)_0)  - 4 r \Delta (\partial_r \cD^\dagger \cD^\dagger P_{+2})^T \mathbf{S}_{+}  \right. \nn \\
& \quad \quad \quad \quad \quad \left. -2 a^2 (\cD^\dagger \cD^\dagger P_{+2})^T \mathbf{S}_{+} \hat{\boldsymbol{\lambda}} (\mathfrak{s}_{-10} - \mathfrak{s}_{10}) \mathfrak{c}_0 \right\} \mathbf{Y}_0 \nn \\
& \phantom{=} + \frac{1}{3 i \omega} \left\{ (\cD \cD P_{-2})^T \mathbf{S}_{-2} (r^2 + a^2 (\mathfrak{c}^2)_0)(r - i a \mathfrak{c}_0 ) - (\cD P_{-2})^T \mathbf{S}_{-2} (r - i a \mathfrak{c}_0)^2 \right. \nn \\
& \quad \quad \quad \quad \left. - (\cD \cD P_{-2})^T \mathbf{M}_{-2}  \mathfrak{s}_{-10} (2 a^2 r \mathfrak{c}_0 - i a (r^2 + 3 a^2 (\mathfrak{c}^2)_0)) - 2 i a (\cD P_{-2})^T \mathbf{M}_{-2}  \mathfrak{s}_{-10} (r + i a \mathfrak{c}_0)  \right\} \nn \\
& \phantom{=} + \frac{1}{3 i \omega} \left\{ (\cD^\dagger \cD^\dagger P_{+2})^T \mathbf{S}_{+2} (r^2 + a^2 (\mathfrak{c}^2)_0)(r - i a \mathfrak{c}_0 ) - (\cD^\dagger P_{+2})^T \mathbf{S}_{+2} (r - i a \mathfrak{c}_0)^2 \right. \nn \\
& \quad \quad \quad \quad \left. + (\cD^\dagger \cD^\dagger P_{+2})^T \mathbf{M}_{+2}  \mathfrak{s}_{10} (2 a^2 r \mathfrak{c}_0 - i a (r^2 + 3 a^2 (\mathfrak{c}^2)_0)) + 2 i a (\cD^\dagger P_{+2})^T \mathbf{M}_{+2}  \mathfrak{s}_{10} (r + i a \mathfrak{c}_0) \right\} ,
\end{align}
\end{subequations}
where 
\begin{subequations}
\begin{align}
\mathbf{M}_{+2} &\equiv \mathbf{b}_{+2} (\hat{\boldsymbol{\lambda}}_2 - a \omega \mathfrak{s}_{21}) , \\
\mathbf{M}_{-2} &\equiv \mathbf{b}_{-2} (\hat{\boldsymbol{\lambda}}_2 - a \omega \mathfrak{s}_{-2-1}) , \\
\mathbf{S}_+ &\equiv \mathbf{b}_{+2} \left( \hat{\boldsymbol{\Lambda}} - 2 a \omega \hat{\boldsymbol{\lambda}}_2 \mathfrak{s}_{10} + a^2 \omega^2 \mathfrak{s}_{20} \right) , \\
\mathbf{S}_- &\equiv \mathbf{b}_{-2} \left( \hat{\boldsymbol{\Lambda}} - 2 a \omega \hat{\boldsymbol{\lambda}}_2 \mathfrak{s}_{-10} + a^2 \omega^2 \mathfrak{s}_{-20} \right) , \\
\mathbf{S}_{l_+ l_+}^{(s=2)} &\equiv \mathbf{S}_+ \pm \mathbf{S}_- , \\
\mathbf{S}_{l_- l_-}^{(s=2)} &\equiv \pm \mathbf{S}_+ + \mathbf{S}_- . 
\end{align}
\end{subequations}

\subsection{Spin zero}
A complete set of projections for the spin-zero metric perturbation in Eq.~(\ref{eq:mp-s0}) are
\begin{subequations}
\begin{align}
h^{(s=0)}_{l_+ l_+} &= - \left( \frac{1}{i \omega} \cD r \cD \mathbf{h} + (\boldsymbol{1} \cdot) 4  \cD \cD \boldsymbol{\kappa} \right)^T \mathbf{b}_{0} \mathbf{Y}_{0} , \label{eq:s0-1} \\
h^{(s=0)}_{l_- l_-} &= - \left(- \frac{1}{i \omega} \cD^\dagger r \cD^\dagger \mathbf{h} + (\boldsymbol{1} \cdot) 4  \cD^\dagger \cD^\dagger \boldsymbol{\kappa} \right)^T \mathbf{b}_{0} \mathbf{Y}_{0} , \\
h^{(s=0)}_{m_+ m_+} &=
\left( \frac{a}{\omega} \mathbf{h}^T \, \mathbf{S}_{m_+m_+}^{(s=0) h} - 4 \, (\boldsymbol{1} \cdot \boldsymbol{\kappa}) \, \mathbf{S}_{m_+m_+}^{(s=0) \kappa} \right) \mathbf{Y}_{+2} , \\
h^{(s=0)}_{m_- m_-} &= 
\left( - \frac{a}{\omega} \mathbf{h}^T \, \mathbf{S}_{m_-m_-}^{(s=0) h} - 4 \, (\boldsymbol{1} \cdot \boldsymbol{\kappa}) \, \mathbf{S}_{m_-m_-}^{(s=0) \kappa} \right) \mathbf{Y}_{-2} .  \label{eq:s0-4}
\end{align}
and 
\begin{align}
\rho h^{(s=0)}_{l_+ m_+} &= \quad  \frac{1}{2i\omega} (\cD h) \mathbf{b}_{0} \left( \hat{\boldsymbol{\lambda}}  - a \omega \mathfrak{s}_{01} \right) \left(r^2 \mathbf{I} + a^2 (\mathfrak{c}^2)_{+1} \right) \mathbf{Y}_{+1} \nn \\
&\quad - \frac{a}{\omega} \left((r \cD h)^T \mathbf{b}_{0} \mathfrak{s}_{01} - h^T \mathbf{b}_{0} \left( \hat{\boldsymbol{\lambda}} - a \omega \mathfrak{s}_{01} \right)  \mathfrak{c}_{+1} \right) \mathbf{Y}_{+1} \nn \\
& \quad + 4 (\cD \kappa)^T \mathbf{b}_{0} \left( \hat{\boldsymbol{\lambda}}  - a \omega \mathfrak{s}_{01} \right) \left( r \mathbf{I} + i a \mathfrak{c}_{+1} \right)  \mathbf{Y}_{+1} \nn \\
& \quad - 4 \left( (\kappa)^T \mathbf{b}_{0} \left( \hat{\boldsymbol{\lambda}}  - a \omega \mathfrak{s}_{01} \right) +  i a (\cD \kappa)^T \mathbf{b}_{0} \mathfrak{s}_{01} \right)  \mathbf{Y}_{+1} ,  \label{eq:s0-5} \\
\rhoc h^{(s=0)}_{l_+ m_-} &= \quad - \frac{1}{2i\omega} (\cD h) \mathbf{b}_{0} \left( \hat{\boldsymbol{\lambda}}  - a \omega \mathfrak{s}_{0-1} \right) \left(r^2 \mathbf{I} + a^2 (\mathfrak{c}^2)_{-1} \right) \mathbf{Y}_{-1} \nn \\
&\quad + \frac{a}{\omega} \left((r \cD h)^T \mathbf{b}_{0} \mathfrak{s}_{0-1} + h^T \mathbf{b}_{0} \left( \hat{\boldsymbol{\lambda}} - a \omega \mathfrak{s}_{0-1} \right)  \mathfrak{c}_{-1} \right) \mathbf{Y}_{-1} \nn \\
& \quad - 4 (\cD \kappa)^T \mathbf{b}_{0} \left( \hat{\boldsymbol{\lambda}}  - a \omega \mathfrak{s}_{0-1} \right) \left( r \mathbf{I} - i a \mathfrak{c}_{-1} \right)  \mathbf{Y}_{-1} \nn \\
& \quad + 4 \left( (\kappa)^T \mathbf{b}_{0} \left( \hat{\boldsymbol{\lambda}}  - a \omega \mathfrak{s}_{0-1} \right) +  i a (\cD \kappa)^T \mathbf{b}_{0} \mathfrak{s}_{0-1} \right)  \mathbf{Y}_{-1} , \\
\rhoc h^{(s=0)}_{l_- m_+} &= \quad - \frac{1}{2i\omega} (\cD^\dagger h) \mathbf{b}_{0} \left( \hat{\boldsymbol{\lambda}}  - a \omega \mathfrak{s}_{01} \right) \left(r^2 \mathbf{I} + a^2 (\mathfrak{c}^2)_{+1} \right) \mathbf{Y}_{+1} \nn \\
&\quad - \frac{a}{\omega} \left((r \cD^\dagger h)^T \mathbf{b}_{0} \mathfrak{s}_{01} - h^T \mathbf{b}_{0} \left( \hat{\boldsymbol{\lambda}} - a \omega \mathfrak{s}_{01} \right)  \mathfrak{c}_{+1} \right) \mathbf{Y}_{+1} \nn \\
& \quad + 4 (\cD^\dagger \kappa)^T \mathbf{b}_{0} \left( \hat{\boldsymbol{\lambda}}  - a \omega \mathfrak{s}_{01} \right) \left( r \mathbf{I} - i a \mathfrak{c}_{+1} \right)  \mathbf{Y}_{+1} \nn \\
& \quad - 4 \left( (\kappa)^T \mathbf{b}_{0} \left( \hat{\boldsymbol{\lambda}}  - a \omega \mathfrak{s}_{01} \right) -  i a (\cD^\dagger \kappa)^T \mathbf{b}_{0} \mathfrak{s}_{01} \right)  \mathbf{Y}_{+1} , \\
\rho h^{(s=0)}_{l_- m_-} &= \quad  \frac{1}{2i\omega} (\cD^\dagger h) \mathbf{b}_{0} \left( \hat{\boldsymbol{\lambda}}  - a \omega \mathfrak{s}_{0-1} \right) \left(r^2 \mathbf{I} + a^2 (\mathfrak{c}^2)_{-1} \right) \mathbf{Y}_{-1} \nn \\
&\quad + \frac{a}{\omega} \left((r \cD^\dagger h)^T \mathbf{b}_{0} \mathfrak{s}_{0-1} + h^T \mathbf{b}_{0} \left( \hat{\boldsymbol{\lambda}} - a \omega \mathfrak{s}_{0-1} \right)  \mathfrak{c}_{-1} \right) \mathbf{Y}_{-1} \nn \\
& \quad - 4 (\cD^\dagger \kappa)^T \mathbf{b}_{0} \left( \hat{\boldsymbol{\lambda}}  - a \omega \mathfrak{s}_{0-1} \right) \left( r \mathbf{I} + i a \mathfrak{c}_{-1} \right)  \mathbf{Y}_{-1} \nn \\
& \quad + 4 \left( (\kappa)^T \mathbf{b}_{0} \left( \hat{\boldsymbol{\lambda}}  - a \omega \mathfrak{s}_{0-1} \right) - i a (\cD^\dagger \kappa)^T \mathbf{b}_{0} \mathfrak{s}_{0-1} \right)  \mathbf{Y}_{-1} , \\
\Sigma \Delta h_{l_+ l_-} 
 &= \left\{ - h^T \mathbf{b}_0 \left(\frac{K_r}{\omega} - 2 \left(r^2+a^2 (\mathfrak{c}^2)_0 \right) \right) \left( r^2+a^2 (\mathfrak{c}^2)_0 \right) \right. \nn \\
 & \quad\quad - 2 \left[ \left(\cD \Delta \cD^\dagger + \cD^\dagger \Delta \cD \right) \kappa \right]^T  \mathbf{b}_0 \left(r^2+a^2 (\mathfrak{c}^2)_0 \right) \nn \\
 & \left. \quad\quad + 8 r \Delta (\partial_r \kappa)^T \mathbf{b}_0 + 4 a^2 \kappa^T  \mathbf{b}_0 \hat{\boldsymbol{\lambda}} \left( \mathfrak{s}_{-10} - \mathfrak{s}_{10} \right) \mathfrak{c}_0 \right\} \mathbf{Y}_0.   \label{eq:s0-9}
\end{align}
\end{subequations}
where
\begin{align}
\mathbf{S}_{m_+m_+}^{(s=0) h} &= \mathbf{b}_{0} \left\{ \left( \hat{\boldsymbol{\Lambda}} - 2 a \omega \hat{\boldsymbol{\lambda}} \mathfrak{s}_{12} + a^2 \omega^2 (\mathfrak{s}^2)_{02} \right) \mathfrak{c}_{2} + \left( \hat{\boldsymbol{\lambda}} \mathfrak{s}_{12} - a \omega (\mathfrak{s}^2)_{02} \right) \right\} , \\
\mathbf{S}_{m_+m_+}^{(s=0) \kappa} &= \mathbf{b}_{0} \left( \hat{\boldsymbol{\Lambda}} - 2 a \omega \hat{\boldsymbol{\lambda}} \mathfrak{s}_{12} + a^2 \omega^2 (\mathfrak{s}^2)_{02} \right) , \\
\mathbf{S}_{m_-m_-}^{(s=0) h} &= \mathbf{b}_{0} \left\{ \left( \hat{\boldsymbol{\Lambda}} - 2 a \omega \hat{\boldsymbol{\lambda}} \mathfrak{s}_{-1,-2} + a^2 \omega^2 (\mathfrak{s}^2)_{0,-2} \right) \mathfrak{c}_{-2} - \left( \hat{\boldsymbol{\lambda}} \mathfrak{s}_{-1,-2} - a \omega (\mathfrak{s}^2)_{0,-2} \right) \right\} , \\
\mathbf{S}_{m_+m_+}^{(s=0) \kappa} &= \mathbf{b}_{0} \left( \hat{\boldsymbol{\Lambda}} - 2 a \omega \hat{\boldsymbol{\lambda}} \mathfrak{s}_{-1,-2} + a^2 \omega^2 (\mathfrak{s}^2)_{0,-2} \right) .
\end{align}
In Eqs.~(\ref{eq:s0-1})--(\ref{eq:s0-4}), $\boldsymbol{\kappa}$ is a matrix, and $(\boldsymbol{1} \cdot)$ as a row vector that performs the sum over $\ell$. In Eqs.~(\ref{eq:s0-5})--(\ref{eq:s0-9}), the $(\boldsymbol{1} \cdot)$ term has been omitted for brevity,

\section{Validation data\label{sec:validation}}
Tables \ref{tbl:jumps-a0} and \ref{tbl:jumps-a099} list numerical values of the jumps in modes up to $\ell =10$, for a particle in orbit at $r_0 = 6M$ about a Schwarzschild black hole (\ref{tbl:jumps-a0}) and for a rapidly-spinning black hole with $a=0.99M$ (Table \ref{tbl:jumps-a099}).

\begin{table}
$$
\begin{array}{|r|r|r||r|}
 \hline
 \ell &  J_{0}{(P_{-2})} & J_{1}{(P_{-2})} & J_{1}{(h)} \\
 \hline
 2 & -0.9341652-32.035193 i & -14.012478-10.106341 i & -0.57205702 \\
 3 & 2.2106383+43.319488 i & -19.895745+13.086095 i & 0 \\
 4 & -1.2533141+42.979722 i & 77.705477+15.094069 i & 0.49541591 \\
 5 & -1.3855909-88.243759 i & 37.410955-28.566089 i & 0 \\
 6 & 1.6004396-44.73709 i & -181.03797-15.892428 i & -0.48234195 \\
 7 & 1.0112635+128.80814 i & -53.596965+42.316776 i & 0 \\
 8 & -1.7225155+45.357304 i & 322.32572+16.173922 i & 0.47768153 \\
 9 & -0.79669766-167.82924 i & 69.312696-55.455203 i & 0 \\
 10 & 1.7798579-45.64908 i & -501.39643-16.306296 i & -0.47548123 \\
 \hline
\end{array}
$$
$$
 \begin{array}{|r|r|r|r|r|}
 \hline
 \ell& J_0(P_{-1}) & J_1(P_{-1})  & J_0(\kappa) & J_1(\kappa) \\
 \hline
 2 & -269.03958+109.83495 i & 67.259895-36.611649 i & -30.891079 & 32.178207 \\
 3 & 62.91587-102.74119 i & 31.457935-158.39266 i & 0 & 0 \\
 4 & 350.94586-52.09929 i & -143.56876+62.953309 i & 26.752459 & -82.486749 \\
 5 & -65.986722+107.75587 i & -32.993361+408.57432 i & 0 & 0 \\
 6 & -471.56933+35.003159 i & 214.3497-90.424827 i & -26.046465 & 163.87568 \\
 7 & 66.881148-109.21646 i & 33.440574-769.06589 i & 0 & 0 \\
 8 & 599.8828-26.475798 i & -283.72835+118.03793 i & 25.794803 & -275.14456 \\
 9 & -67.263179+109.84031 i & -33.631589+1240.2802 i & 0 & 0 \\
 10 & -731.16863+21.32131 i & 352.52773-145.69562 i & -25.675987 & 416.16495 \\
 \hline
\end{array}
$$
\caption{Jumps in radial functions at $r=r_0=6M$ for $a=0$, $m=2$. Upper: Determined from Weyl scalars and the trace. Lower: determined from regularity on the sphere.}
\label{tbl:jumps-a0}
\end{table}

\begin{table}
$$
\begin{array}{|r|r|r|r|}
 \hline
 \ell &  J_{0}{(P_{-2})} & J_{1}{(P_{-2})} & J_{1}{(h)} \\
 \hline
 2 & -0.21782443-16.92872 i & -15.55513-7.3009438 i & -0.58007105 \\
 3 & 0.54358853+25.612564 i & -6.2410468+10.794723 i & 0 \\
 4 & -0.329755+19.185444 i & 64.799463+8.9513826 i & 0.50246972 \\
 5 & -0.32786372-47.812584 i & 10.89643-20.957857 i & 0 \\
 6 & 0.4056661-19.314651 i & -142.72395-9.0982076 i & -0.48919802 \\
 7 & 0.2367434+68.158535 i & -15.241377+30.147026 i & 0 \\
 8 & -0.43162509+19.345375 i & 248.99391+9.1424079 i & 0.48446434 \\
 9 & -0.18567822-87.930128 i & 19.504859-39.032769 i & 0 \\
 10 & 0.44368914-19.35722 i & -383.60261-9.1618631 i & -0.48222893 \\
 \hline
\end{array}
$$
$$
 \begin{array}{|r|r|r|r|r|}
 \hline
 \ell& J_0(P_{-1}) & J_1(P_{-1})  & J_0(\kappa) & J_1(\kappa) \\
 \hline
 2 & -337.06643+85.15083 i & 66.553306-18.416882 i & -26.1142 & 36.026256 \\
 3 & 19.119854-82.676796 i & 15.022065-101.32406 i & 0 & 0 \\
 4 & 524.94212-38.915946 i & -126.27358+48.912959 i & 22.620668 & -107.28865 \\
 5 & -20.504489+85.504877 i & -15.347091+242.82248 i & 0 & 0 \\
 6 & -738.33137+25.959912 i & 184.5521-75.107992 i & -22.02319 & 220.87866 \\
 7 & 20.918971-86.31449 i & 15.435462-446.38189 i & 0 & 0 \\
 8 & 956.254-19.584755 i & -242.41322+100.36684 i & 21.810085 & -375.96945 \\
 9 & -21.097679+86.65854 i & -15.472351+712.39424 i & 0 & 0 \\
 10 & -1175.8634+15.752413 i & 300.09629-125.2532 i & -21.709449 & 572.47272 \\
 \hline
\end{array}
$$
\caption{Jumps in radial functions at $r=r_0=6M$ for $a=0.99M$, $m=2$. Upper: Determined from Weyl scalars and the trace. Lower: determined from regularity on the sphere. Note that $J_0(h) = 0$.}
\label{tbl:jumps-a099}
\end{table}  

\section{Barack-Lousto variables\label{sec:BL}}
In the Schwarzschild case, the projections of the metric perturbation can be written in terms of Barack-Lousto variables $\hb{i}(r)$ as follows:
\begin{subequations}
\begin{align}
 \hrm_{l_+l_+} &= \frac{1}{r f^2} \left( \hb1 + \hb2 \right)  \\
 \hrm_{l_-l_-} &=  \frac{1}{r f^2} \left( \hb1 - \hb2 \right) \\
 \hrm_{m_+m_+} &= \frac{r}{\Lambdah}  \left( \hb7 - i \hb{10} \right) \\
 \hrm_{m_-m_-} &= \frac{r}{\Lambdah} \left( \hb7 + i \hb{10} \right) \\
 \rho \hrm_{l_+m_+} &= -\frac{r}{2 f \lambdah} \left( \hb4 + \hb5 - i(\hb8 + \hb9) \right) \\
 \rhoc \hrm_{l_+m_-} &= \frac{r}{2 f \lambdah} \left( \hb4 + \hb5 + i(\hb8 + \hb9) \right)  \\
 \rhoc \hrm_{l_- m_+} &=  \frac{r}{2 f \lambdah} \left( \hb4 - \hb5 - i(\hb8 - \hb9) \right)   \\
 \rho \hrm_{l_- m_-} &= -\frac{r}{2 f \lambdah} \left( \hb4 - \hb5 + i(\hb8 - \hb9) \right)   \\
 \Sigma \Delta \hrm_{l_+l_-} &= - r^3 \hb6 \\
 \hrm \equiv \frac{1}{\Sigma} \left( \Delta \hrm_{l_+l_-} + \hrm_{m_+m_-} \right) &= \frac{1}{r} \left( \hb3 - \hb6 \right) .
 \label{eq:hBL}
\end{align}
\end{subequations}
Inversely,
\begin{subequations}
\begin{align}
 \hb1 &= \tfrac{1}{2} r f^2 \left(  \hrm_{l_+l_+} +  \hrm_{l_-l_-} \right)  \\
 \hb2 &= \tfrac{1}{2} r f^2 \left(  \hrm_{l_+l_+} -  \hrm_{l_-l_-} \right)  \\
 \hb3 &= r^{-1} \, \hrm_{m_+ m_-} \\
 \hb4 &= -\tfrac{1}{2} \lambdah f \left( \hrm_{l_+ m_+} - \hrm_{l_+ m_-} - \hrm_{l_- m_+} + \hrm_{l_- m_-} \right)  \\
 \hb5 &= -\tfrac{1}{2} \lambdah f \left( \hrm_{l_+ m_+} - \hrm_{l_+ m_-} + \hrm_{l_- m_+} - \hrm_{l_- m_-} \right) \\
 \hb6 &= - r f \, \hrm_{l_+ l_-} \\
 \hb7 &= \frac{\Lambdah}{2r} \left(\hrm_{m_+ m_+} + \hrm_{m_- m_-} \right) \\
 \hb8 &= -\tfrac{i}{2} \lambdah f \left( \hrm_{l_+ m_+} + \hrm_{l_+ m_-} - \hrm_{l_- m_+} - \hrm_{l_- m_-} \right)  \\
 \hb9 &= -\tfrac{i}{2} \lambdah f \left( \hrm_{l_+ m_+} + \hrm_{l_+ m_-} + \hrm_{l_- m_+} + \hrm_{l_- m_-} \right) \\
 \hb{10} &= \frac{i \Lambdah}{2r} \left(\hrm_{m_+ m_+} - \hrm_{m_- m_-} \right)  .
\end{align}
\label{eq:hBL-inverse}
\end{subequations}

\begin{figure}
\begin{center}
 \includegraphics[width=18cm]{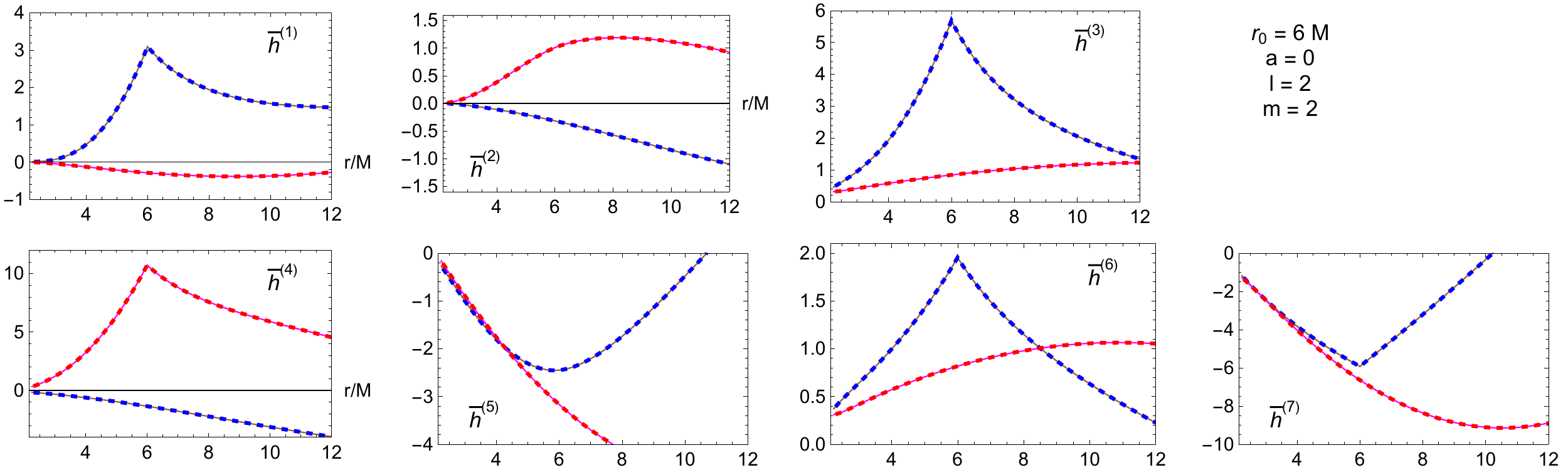} 
 \includegraphics[width=17cm]{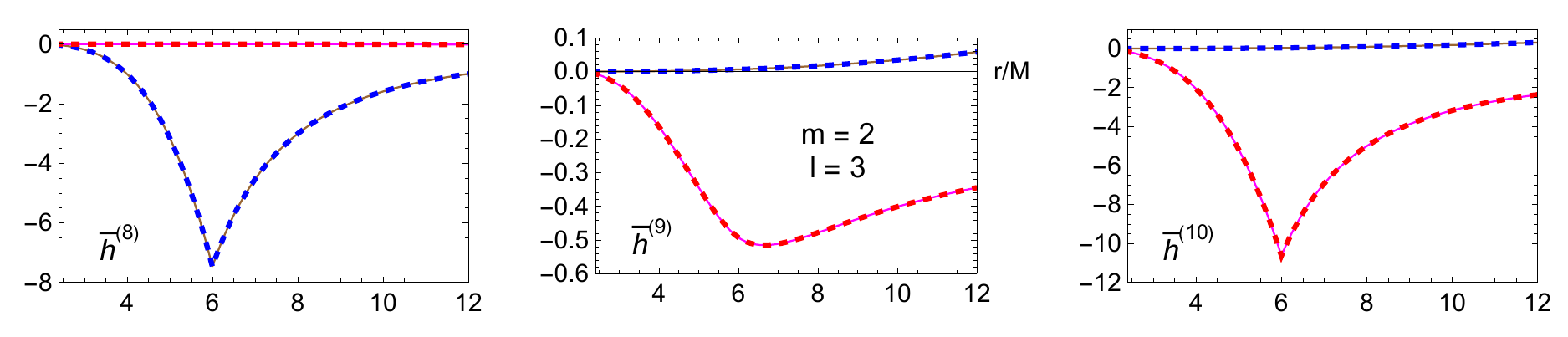} 
\end{center}
 \caption{Modes of the metric perturbation in Barack-Lousto variables $\hb{i}(r)$ (see text) for $a=0$, $r_0 = 6M$ and $m = 2$. The upper  plots show the seven components of the even-parity $\ell=2$ mode, and the lower plots show the three components of the odd-parity $\ell=3$ modes. Labelling: New results [dashed lines], validation data [solid lines], real part [blue/brown] and imaginary part [red/magenta].}
 \label{fig:schw-m2b}
\end{figure}

Figure \ref{fig:schw-m2b} shows the typical radial profile of BL variables $\hb{i}(r)$ for $m=2$, and a comparison with the validation data set (see Sec.~\ref{subsec:schw}). 

\bibliographystyle{apsrev4-1}
\bibliography{LorenzGaugeKerrCirc}

\end{document}